\documentclass[10pt, two column, twoside]{IEEEtran}
\usepackage{color}
\usepackage[colorlinks, linkcolor=color1, anchorcolor=blue, citecolor=color1]{hyperref}

\usepackage{hyperref}

\usepackage{url}
\usepackage{amssymb}
\usepackage{amsmath}
\usepackage[lined,boxed,commentsnumbered, ruled]{algorithm2e}
\usepackage{mathrsfs}
\usepackage{algorithmic}
\usepackage{bm}
\usepackage{array}

\usepackage{pgfplots}
\usepackage{tikz}
\usetikzlibrary{arrows}
\usepackage{subfigure}
\usepackage{graphicx,booktabs,multirow}

\definecolor{colorhkust}{RGB}{20,43,140}
\definecolor{colortsinghua}{RGB}{116,52,129}
\definecolor{color1}{RGB}{128,0,0}

\newcommand{\rev}{\color{black}}

\date{}

\begin{document}

\title{{Edge Artificial Intelligence for 6G: \\Vision, Enabling Technologies, and Applications}}
\author{Khaled B. Letaief, \IEEEmembership{Fellow, IEEE},
Yuanming~Shi,~\IEEEmembership{Senior Member,~IEEE,} 
        Jianmin Lu,\\
        and Jianhua Lu, \IEEEmembership{Fellow, IEEE} 
        \thanks{This work was supported in part by Project No. 20210400L016 under RDC Corporation Ltd. and in part by the Natural Science Foundation of Shanghai under Grant No. 21ZR1442700.}
        \thanks{Khaled B. Letaief is with the Department of Electronic and Computer
Engineering, The Hong Kong University of Science and Technology, Hong Kong, and also with Peng Cheng Laboratory, Shenzhen 518066, China (e-mail: eekhaled@ust.hk).}
         \thanks{Yuanming Shi is with the School
of Information Science and Technology,
ShanghaiTech University, Shanghai 201210, China (e-mail: shiym@shanghaitech.edu.cn). ({\textit{Corresponding Author}})}
\thanks{ Jianmin Lu is with Huawei Technologies Co., Ltd., Shenzhen 518066, China (e-mail: lujianmin@huawei.com).}
\thanks{Jianhua Lu is with the Department of Electronic Engineering and the Beijing National Research Center for Information Science and Technology, Tsinghua University, Beijing 100084, China (e-mail: lhh-dee@mail.tsinghua.edu.cn).}}
                           
\maketitle
\IEEEpeerreviewmaketitle


\maketitle

\begin{abstract}
The thriving of artificial intelligence (AI) applications is driving the further evolution of wireless networks. It has been envisioned that 6G will be transformative and
will revolutionize the evolution of wireless from ``connected things" to ``connected intelligence". However, state-of-the-art deep learning and big data analytics based AI systems require tremendous computation and communication resources, causing significant latency, energy consumption, network congestion, and privacy leakage in both of the training and inference processes. By embedding model training and inference capabilities into the network edge, edge AI stands out as a disruptive technology for 6G to seamlessly integrate sensing, communication, computation, and intelligence, thereby improving the efficiency, effectiveness, privacy, and security of 6G networks. In this paper, we shall provide our vision for scalable and trustworthy edge AI systems with integrated design of wireless communication strategies and decentralized machine learning models. New design principles of wireless networks, service-driven resource allocation optimization methods, as well as a holistic end-to-end system architecture to support edge AI will be described. Standardization, software and hardware platforms, and application scenarios are also discussed to facilitate the industrialization and commercialization of edge AI systems.
\end{abstract}


\section{Introduction}
\label{intro}

\subsection{Roadmap to 6G: Vision and Technologies}

\begin{figure*}[t]
        \centering
        \includegraphics[width=0.9\linewidth]{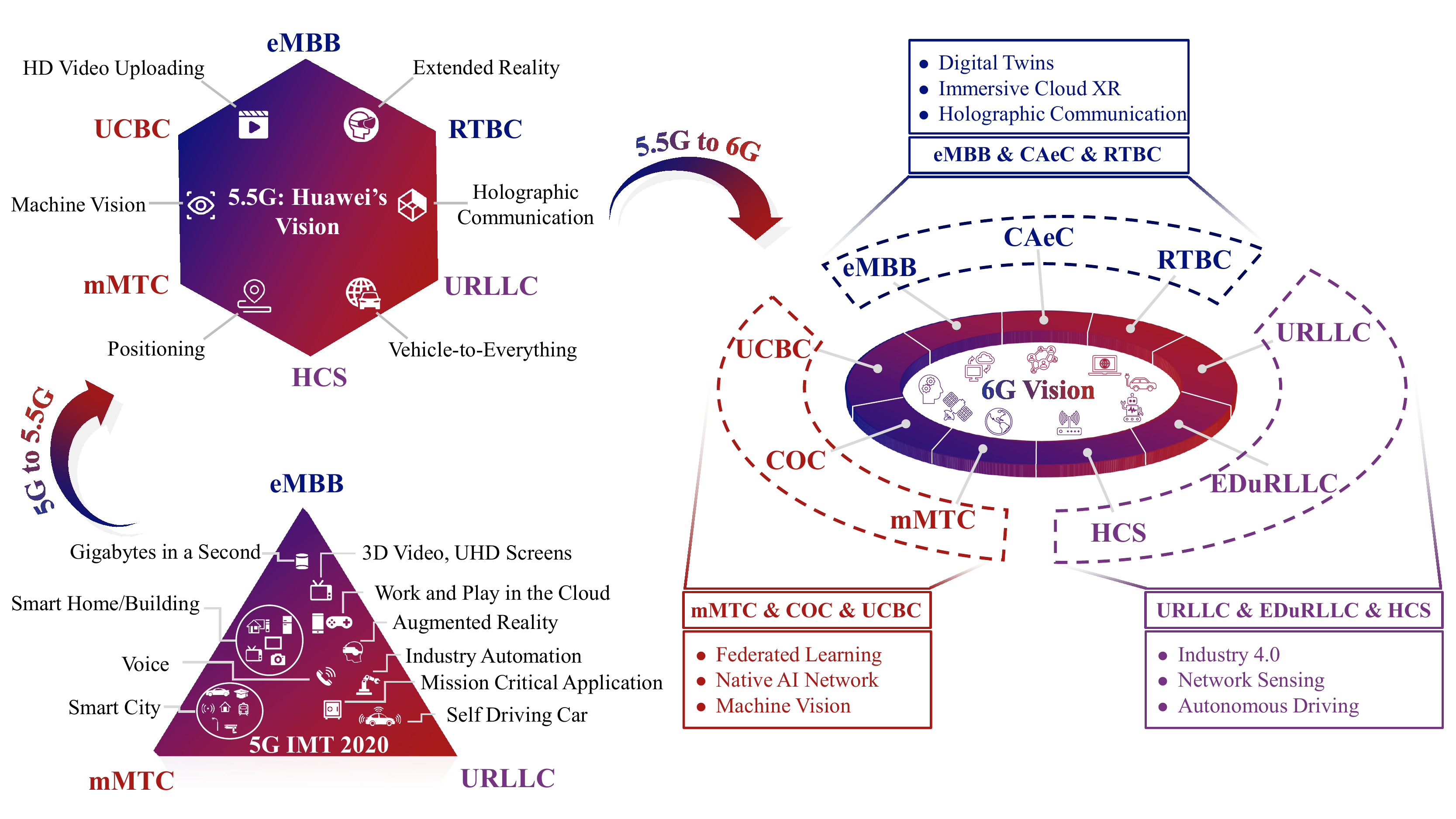} 
        \vspace{-4mm}
        \caption{Towards 6G: the evolution of use cases from 5G to 6G.}\label{6G}
        \vspace{-4mm}
\end{figure*}

With the standardization and worldwide deployment of 5G networks, researchers, companies, and governments have initiated the vision, usage scenarios, and disruptive technologies for future 6G. In particular, the United States \cite{RINGS}, European Union \cite{EU6G}, and China \cite{IMT6G} have recently funded 6G projects with a common goal of enabling connected intelligence. Besides, the International Telecommunication Union (ITU) has  published the system requirements and driving characteristics for Network 2030  \cite{ITU2030}. To improve
a real-time immersive experience and interaction, as well as accelerate
intelligence upgrades for industrial internet-of-things (IoT) and digital twins, multiple companies are now considering
new usage scenarios. For example, based on typical use cases in 5G \cite{Jeff_JSAC5G,JSAC_5GShafi} (i.e., enhanced mobile
broadband (eMBB), ultra-reliable and low-latency
communications (URLLC), and massive machine type communications (mMTC)),  Huawei has recently proposed three
additional application scenarios in the vision of 5.5G. These include uplink centric
broadband communication (UCBC), real-time broadband communication (RTBC),
and harmonized communication and sensing (HCS) \cite{HW55}. It is expected that 6G will go beyond the mobile internet
to support ubiquitous artificial intelligence (AI) services and Internet of Everything (IoE) applications \cite{RINGS, EU6G, IMT6G, ITU2030, IEEEPro21_6Gwireless},
including sustainable cities, connected autonomous systems, brain-computer
interfaces, digital twins, tactile and haptic internet, high-fidelity holographic society, extended reality (XR) and metaverse \cite{Maier_ComMag20}, e-health, etc. Researchers in industry and academia have published
many visionary 6G proposals \cite{you2021towards, Walid_6g, letaief2019roadmap} to provide a better understanding, sensing, controlling, and interacting for a physical world. In particular, three new application
services were envisioned for 6G, including computation oriented communications
(COC), contextually agile eMBB communications (CAeC), and event defined uRLLC
(EDuRLLC) \cite{letaief2019roadmap}. Based on these quoted usage scenarios, we  present the evolution  of visionary use cases for 6G  in Fig. \ref{6G} by integrating intelligence, coordination, sensing,  and computing for a connected cyber-physical world. 

To shape the future of 6G use cases in 2030, multi-disciplinary research and various disruptive technologies are required, including spectrum exploration technologies, devices and circuit technologies, as well as networking, computing, sensing, and learning functionalities. In particular, AI, especially deep learning (DL), provides a revolutionary approach to design and optimize
6G wireless networks across the physical, medium-access, and application layers \cite{letaief2019roadmap, ali2020white}. Specifically, DL provides a  novel way to design 6G air interface by optimizing the radio environment \cite{Tiejun_JSAC21}, communication algorithms \cite{kim2018communication},  hardware, and applications in a unified way \cite{OSheaH17, Jakob_ComMag21}. This has inspired the recent success applications for joint source-channel coding (JSCC)
\cite{SaiduttaAF21}, task-oriented communication \cite{strinati20216g, abs-2102-04170}, semantic
communication \cite{Qin_TSP21Semantic}.  Besides, machine learning (ML) also provides a
paradigm shift for automatically learning high performance and fast optimization algorithms to solve the resource allocation problems in wireless networks  \cite{yifei20lorm, yifei21gnn, liang20learningra, WeiYu_JSAC21}.  The domain knowledge (e.g., optimization models and theoretical tools) was further incorporated into the DL framework for optimizing  ultra-reliable and low-latency communication networks \cite{SheSGLYPV21}. An ML  approach was also developed for addressing the communication, networking, and security challenges for vehicular applications \cite{Kato_IEEEPro20}. With the development of wireless data collection,
learning models and algorithms, as well as software and hardware platforms, we
envision that AI will become a native tool to design disruptive wireless
technologies for accelerating the design, standardization, and commercialization of 6G. 
On the other hand, the evolution
of 6G wireless communication technologies and communication theory will
also inspire the progress and development of AI techniques in terms of novel learning theory, new deep neural network (DNN)\ architectures, customized software and hardware
platforms. 

Given the requirements of emerging 6G, connected intelligence is expected to be the central focus
and an indispensable component in 6G \cite{tong20216g}. This shall revolutionize the evolution of wireless from
``connected things" to ``connected intelligence", thereby  enabling the
interconnections between humans, things,
and intelligence within a hyper-connected cyber-physical world \cite{letaief2019roadmap}. {\textit{Edge AI}} provides a promising solution for connected intelligence by enabling data collection, processing,
transmission, and consumption at the network edge \cite{Yuanming_CST20, zhuo2019edge}. Specifically, by embedding the
training capabilities across the network nodes, edge training is able to preserve privacy and confidentiality, achieve high security and fault-tolerance, as well as reduce network traffic congestion and energy consumption. For instance, over-the-air federated learning (FL) provides a collaborative ML framework to train a global statistical model over wireless networks without accessing edge devices' private raw data \cite{yang2020over}. By directly executing the AI models at the network edge, edge inference can provide low-latency and high-reliability AI services by requiring less computation, communication, storage, and engineering resources. For example, edge device-server co-inference is able to remove the communication and computation bottlenecks by splitting a large DNN model between edge devices and edge servers \cite{shao20edgeinfer}. However, edge AI will cause task-oriented data traffic flows over wireless networks, for which disruptive  wireless techniques, efficient resource allocation
methods and holistic system
architectures need to be developed. To embrace the era of edge AI, wireless communication systems and  edge AI algorithms need to be co-designed  for seamlessly integrating communication, computation, and learning.          

\subsection{Edge AI: Challenges and Solutions}
\begin{figure*}[h]
         \centering
         \subfigure[History of wireless communication and key timeline. In particular, 3GPP Release 17 approved the NWDAF (network data analytics function-5G network AI) federated learning technology standard in 2020.]{
                 \begin{minipage}[b]{1\textwidth}
                   \centering
                         \includegraphics[width=0.9\textwidth]{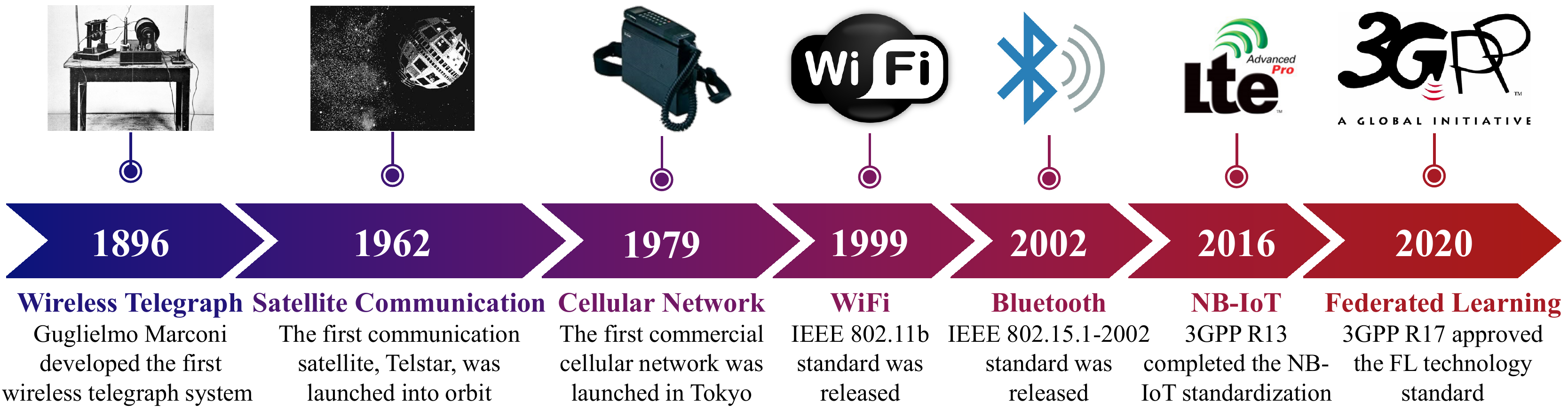}
                 \end{minipage}
         }
         \subfigure[History of artificial intelligence and key timeline. In particular, IEEE 3652.1-2020 approved the first federated machine learning standard in March 2021.]{
                 \begin{minipage}[b]{1\textwidth}
                   \centering
                         \includegraphics[width=0.9\textwidth]{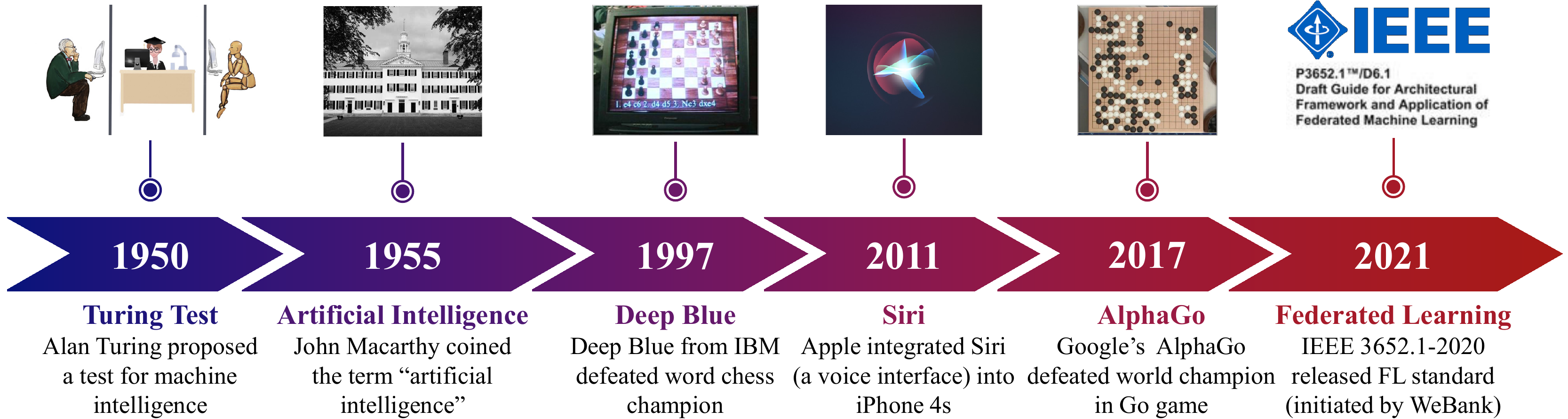}
                 \end{minipage}
         }
      \subfigure[Paper organization and structure: 1) technologies: communication-efficient edge training in Section \ref{edgetraining}, and communication-efficient edge inference in Section \ref{edgeinference}; 2) systems: resource allocation for scalable and trustworthy edge AI systems via theory-driven and data-driven optimizations in Section \ref{resourceall}; 3) architectures: a holistic end-to-end architecture for edge AI systems in Section \ref{archedgeai}; 4) commercializations: standardizations for edge learning and computing, software and hardware platforms, as well as potential applications including autonomous driving, IoT, and smart healthcare in Section \ref{spa}.]{
       \begin{minipage}[b]{1\textwidth}
        \centering
        \includegraphics[width=0.9\textwidth]{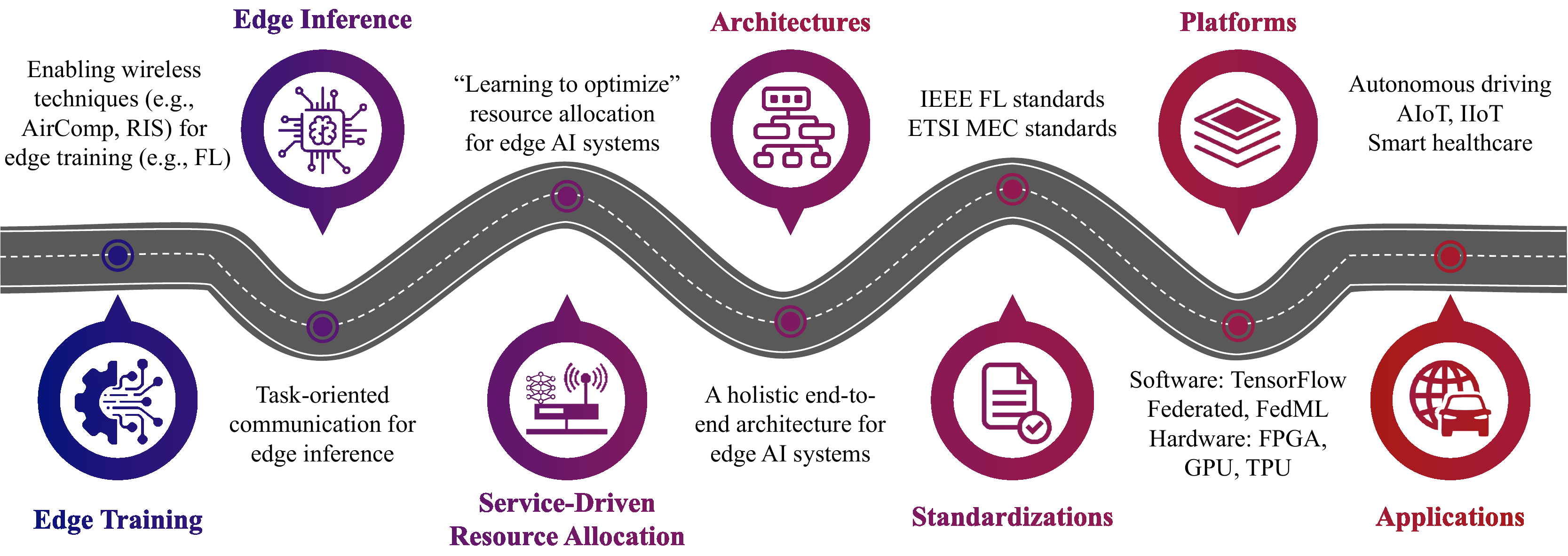}
       \end{minipage}
      }
         \caption{Roadmap to edge AI.} \label{roadmap}
 \end{figure*}
 
Creating a trustworthy and scalable  edge AI system will be of utmost importance for imbuing connected intelligence in 6G. The challenges of trustworthiness and scalability are  multidisciplinary spanning ML, wireless networking, and operation research. Specifically,      \textit{trustworthiness} in terms of privacy and security is  one of
the key requirements for  6G intelligent services and applications, for
which the general data protection regulation (GDPR) needs to be satisfied, and directly transmitting or collecting data from users are forbidden. To tame privacy leakages and adversarial attacks, various edge learning models and architectures have been proposed, including FL (i.e., server-client network architecture with data partition among edge devices) \cite{yang2019federated, kairouz2019advances}, swarm learning (i.e., decentralized device-to-device (D2D) communication architecture without central authority) \cite{warnat2021swarm}, and split learning (i.e., model parameters partitioned among edge devices and edge servers) \cite{gupta2018distributed, park21commdlover}. Distributed reinforcement learning (RL) \cite{Tianyi_TCNS21, zhang2018fully}
and trustworthy learning techniques \cite{Robust1, LiuS21} 
were further proposed to
address the dynamic and adversarial learning environments, respectively. In particular, differential privacy \cite{DP}, lagrange coded computing \cite{pmlr-v89-yu19b}, security multi-party computation, quantum computing, blockchain, and distributed ledger technologies can be further leveraged to build trustworthy edge AI architectures. However, with limited storage, computation, and communication
resources in the wireless edge networks, deploying an edge
AI system  causes a significant {\textit{scalability}} issue in terms of
latency,  energy and accuracy. To address this challenge, a paradigm shift for wireless system design is required from data-oriented communication (i.e., maximizing communication rate or reliability
based on Shannon theory) to {\emph{task-oriented communication}} (i.e., achieving fast and accurate intelligence distillation
at the network edge).   

 \begin{figure*}
 \centering
 \includegraphics[width=1\linewidth]{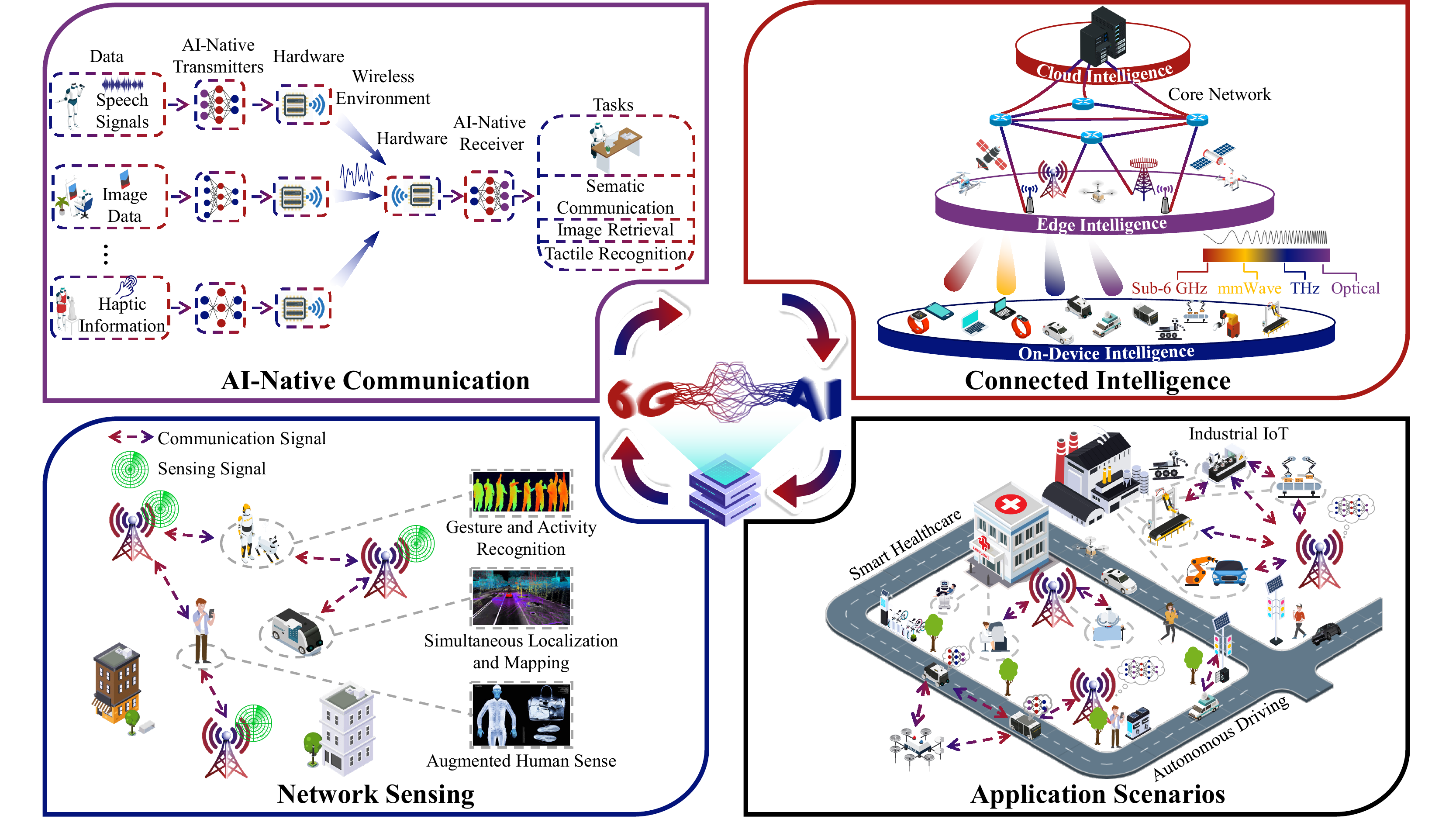} 
 \caption{Edge AI empowered 6G networks: integrated sensing, communication, computation, and intelligence.}\label{AI6G}
\end{figure*}

In this paper, we shall provide a comprehensive picture for the design of scalable and trustworthy edge AI systems by matching the principles and architectures of wireless networks with the task structures of edge AI models and algorithms. The system performance metrics for edge AI are further characterized to facilitate efficient resource allocations based on operation research and ML. Specifically, to design a communication-efficient edge AI training system, we  will provide novel multiple access schemes (e.g., over-the-air computation (AirComp) for model aggregation \cite{yang2020over, Kaibin_TWC20, AmiriG20}) to support massive access for edge devices, new multiple antenna techniques (e.g., cell-free massive MIMO \cite{than2020cellfreemimo, DBLP:journals/twc/NgoAYLM17} and reconfigurable intelligent surface (RIS) \cite{abs-2011-05051, Renzo_jsac20RIS}) to support fast exchange for high-dimensional model updates, and next-generation network architectures (e.g., space-air-ground integrated network (SAGIN)  \cite{seyyedali2020federatedtofog, Jiajia_CST18}) to support diverse edge learning models and topologies. To design a communication-efficient edge inference system with low-latency and reliability guarantees, interference management, cooperative transmission, and task-oriented communication  will be introduced to support edge device distributed inference \cite{Yuanming_TSP19MapReduce}, edge server cooperative inference \cite{yang20iot, hua21serverinference}, and edge device-server co-inference \cite{shao20edgeinfer}, respectively. We then provide a holistic view for mathematically modeling the resource allocation problems in edge training and inference systems, which are categorized as mixed combinatorial optimization, nonconvex optimization and stochastic optimization models. A ``learning to optimize" framework is further introduced to facilitate scalable, real-time, robust, parallel, distributed, and automatic optimization algorithms design for service-driven resource allocation in edge AI systems \cite{yifei20lorm, yifei21gnn, Dongning_JSAC19, WeiYu_JSAC21}. We also provide a holistic  end-to-end  architecture for edge AI systems. Moreover, standardizations, resource allocation optimization solvers, software and hardware platforms, and application scenarios are discussed. The roadmap to edge AI ecosystem is demonstrated in Fig. {\ref{roadmap}} to encourage multidisciplinary collaborations among information science, computer science, operation research, and integrated circuits.

\subsection{Edge AI Empowered 6G Networks}
The developed edge AI  technology will serve as a distributed neural network to  accelerate the evolution of sensing capabilities, communication strategies, network optimizations, and application scenarios in 6G networks. Specifically, edge AI paves the way for network sensing and cooperative perception to understand the network environments and services for an agile and intelligent decision making. For example, edge simultaneous localization and mapping (SLAM) 
\cite{ben2020edge, xu2020edge} has recently been developed to  deploy DL based visual SLAM algorithms on  vehicles by edge inference. Edge AI can also help design AI-native communication strategies  for the physical layer (e.g., task-oriented semantic communication  \cite{Zhijin_JSAC21scl}) and medium access control layer (e.g., random access protocol \cite{JSAC21_DQL}). For instance, edge DL approach has been developed in \cite{Zhijin_JSAC21scl} to deliver low-latency semantic  tasks (e.g., text messages) by learning the communication strategies in an end-to-end fashion based on JSCC. Furthermore, edge AI provides a new paradigm for optimization algorithms design to enable service-driven resource allocation in 6G networks \cite{Yifei_AIem21}. For instances, distributed RL \cite{Dongning_JSAC19},
decentralized  graph neural networks \cite{yifei21gnn}, and distributed
DNN   \cite{Tony_JSAC19distribued}, are able to
automatically learn the distributed resource allocation optimization algorithms. By seamlessly integrating sensing, communication, computation, and intelligence, edge AI shall empower 6G networks to support diversified intelligent applications, including autonomous driving, industrial IoT, smart healthcare, etc. 

To further imbue native intelligence, native trustworthiness, and native sensing in 6G, mimicking nature for  innovating edge AI empowered future networks can be envisioned. Inspired by the dynamic spiking neurons in the human brain, the energy consumption and latency of edge AI can be significantly reduced by processing the learning tasks in an event-driven manner \cite{Jang_SPM19, Jang_CL21}. The brain-inspired stigmergy-based federated collective intelligence mechanism was proposed in \cite{Hongang_WCM20} to accomplish multi-agent tasks (e.g., autonomous driving) through simple indirect communications. By leveraging the prior knowledge of the immune system and brain neurotransmission, a brand-new network security architecture and fully-decoupled radio access network have recently been proposed in \cite{Quan_IWC20}  and \cite{Quan_JCIN19}, respectively. These results on nature-inspired edge AI models and  network architectures provide a strong evidence that one can establish an integrated data-driven and knowledge-guided framework to design and optimize 6G networks. Further details and description of the edge AI empowered 6G network are  provided in Fig. \ref{AI6G}, which highlight the integration of sensing, communication, computation and intelligence in a closed-loop ecosystem.

\subsection{Key Contributions}
We provide extensive discussions, visions, and summaries of wireless techniques, resource allocations, standardizations, platforms, and application scenarios to embrace the era of edge AI for 6G. The major contributions  are summarized as follows:

\begin{itemize}
\item The vision (i.e., connected intelligence for 6G), challenges (i.e., trustworthiness and scalability) and solutions (i.e., wireless techniques, resource allocations and system architectures) for edge AI, as well as edge AI empowered 6G network, are introduced and summarized in Section {\ref{intro}}.
\item  The communication-efficient edge training system is presented in Section {\ref{edgetraining}}, including the edge learning models and algorithms, followed by the promising wireless techniques and architectures to support their deployment.  
\item The communication-efficient edge inference system is introduced in Section {\ref{edgeinference}}. Here, we introduce horizontal edge inference and vertical edge inference by cooperative transmission and  task-oriented communication, respectively.
\item A unified framework for resource allocation in edge AI systems is provided in Section {\ref{resourceall}}. Here, we present operation research based  theory-driven and machine learning based data-driven approaches for designing efficient resource allocation optimization algorithms.
\item A holistic end-to-end architecture for edge AI systems is proposed in Section {\ref{archedgeai}}, including network infrastructure, data governance, edge network function, edge AI management and orchestration.   
\item The standardizations, software and hardware platforms, and application scenarios are discussed in Section \ref{spa}. This will help facilitate the booming market of edge AI in the 6G era.       
\end{itemize}  

{\rev{We summarize the main topics and  relevant technologies as well as highlight the representative 
results in Table \ref{results}.}}

\newcommand{\lencolone}{0.135\textwidth}

\newcommand{\lencoltwo}{0.11\textwidth}

\newcommand{\lencolthree}{0.25\textwidth}

\newcommand{\lencolfour}{0.415\textwidth}

\begin{table*}[htbp]
        
        \centering

        
        \caption{An Overview of the main Topics and Representative
                                Results}
        
        \label{tab_structure}
        
        \resizebox{1\textwidth}{!}{
                
                \begin{tabular}{m{\lencolone}<{\centering}m{\lencoltwo}<{\centering}m{\lencolthree}<{\centering}m{\lencolfour}} 
                        
                        \toprule
                        
                        Sections & Topics & Methods &\hspace{2.4cm} Representative Results \\ \toprule
                        
                        \multicolumn{1}{c}{\multirow{10}[0]{\lencolone}[-1.3cm]{\centering Section \ref{edgetraining}: Communication-Efficient Edge Training}} & \multirow{5}[0]{\lencoltwo}[-0.73cm]{ \centering  Section \ref{edgetraining}-A: Edge Learning Models and Algorithms} &  \multirow{1}[0]{\lencolthree}[0cm]{\centering Federated Learning} &   Horizontal and vertical federated learning \cite{McMahanMRHA17, kairouz2019advances, yang2019federated}; federated optimization \cite{wang2021field}   \\ \cmidrule(r){3-4}
                        
                        &       & \multirow{1}[0]{\lencolthree}[0cm]{\centering Decentralized Learning }    & Swarm learning \cite{warnat2021swarm}; consensus-based methods \cite{Boyd_TIT06}; diffusion strategies \cite{Sayed_SPM13}; decentralized training \cite{lu2021optimal} 
                        \\ \cmidrule(r){3-4} 
                        
                        &       & Model Split Learning & Model parameter partitioned edge learning \cite{li2014communication}; split  learning \cite{gupta2018distributed}  \\ \cmidrule(r){3-4}
                        
                        &       & Distributed Reinforcement Learning & Multi-agent reinforcement learning \cite{lee2020optimization}  \\ \cmidrule(r){3-4}
                        
                        &       & \multirow{1}[0]{\lencolthree}[0cm]{\centering Trustworthy Learning}  & Differential privacy \cite{DP}; secure model aggregation \cite{chen2017distributed};  blockchain smart contract \cite{warnat2021swarm}   \\ \cmidrule(r){2-4}
                        
                        & \multirow{5}[0]{\lencoltwo}[-0.5cm]{\centering Section \ref{edgetraining}-B: Wireless Techniques for Edge Training} & Over-the-Air Computation &  Low-latency analog model aggregation \cite{yang2020over, Kaibin_TWC20, AmiriG20}  \\ \cmidrule(r){3-4}
                        
                        &       &\multirow{1}[0]{\lencolthree}[0cm]{\centering Massive Access Techniques}   & Grant-free random access \cite{liu2018massive, Tao_IoTJ19}; NOMA \cite{Ding_JSAC17NOMA, Dai_ComMag15}; blind demixing \cite{Shi_TSP20}    \\ \cmidrule(r){3-4}
                        
                        &       & Ultra-Massive MIMO &  Cloud-RAN \cite{Yuanming_TWC14gsbf}; cell-free massive MIMO \cite{than2020cellfreemimo}      \\ \cmidrule(r){3-4}
                        
                        &       & Reconfigurable Intelligent Surfaces & RIS-empowered edge training \cite{abs-2011-05051} /edge inference \cite{hua21serverinference}   \\ \cmidrule(r){3-4}
                        
                        &       &  Space-Air-Ground Integrated Networks & UAV-aided model aggregation \cite{fu2021uav}; fog learning \cite{seyyedali2020federatedtofog}  \\ \cmidrule(r){1-4}

                        \multicolumn{1}{c}{\multirow{5}[0]{\lencolone}[-0.56cm]{\centering Section \ref{edgeinference}: Communication-Efficient Edge Inference}} & \multirow{2}[0]{\lencoltwo}[0.03cm]{\centering Section \ref{edgeinference}-A: Horizontal Edge Inference} & Edge Device Distributed Inference & Wireless MapReduce \cite{Yuanming_TSP19MapReduce}  \\ \cmidrule(r){3-4}
                        
                        &       & Edge Server Cooperative Inference & Energy-efficient edge cooperative inference \cite{yang20iot}  \\ \cmidrule(r){2-4}
                        
                        & \multirow{3}[0]{\lencoltwo}[-0.35cm]{\centering Section \ref{edgeinference}-B: Vertical Edge Inference} & Edge Device-Server Co-Inference & DNN inference via edge computing \cite{Xu_TWC20}  \\ \cmidrule(r){3-4}
                        
                        &       & Ultra-Reliable and Low-Latency Communication & Short packet communication \cite{yury10codingrate}; ultra-reliable low-latency edge computing \cite{Mehdi_TWC19, Wei_TWC20}    \\  \cmidrule(r){3-4}
                        
                        &       & Task-Oriented Communication & Single-device/multi-devices edge inference \cite{abs-2102-04170, shao2021taskoriented}  \\   \cmidrule(r){1-4}

                        \multicolumn{1}{c}{\multirow{9}[0]{\lencolone}[-1.49cm]{\centering Section \ref{resourceall}: Resource Allocation for Edge AI Systems}} & \multirow{5}[0]{\lencoltwo}[-0.7cm]{\centering Section \ref{resourceall}-A: Engineering Requirements and Methodologies} &  Accuracy &  Convergence analysis for edge training algorithms   \cite{Xia_arXiv20, AmiriG20, Mingzhe_TWC21}   \\ \cmidrule(r){3-4}
                        
                        &       & \multirow{1}[0]{\lencolthree}[0cm]{\centering Latency}  &  Delay analysis and optimization of wireless FL \cite{li2021delay, mingzhe21convergence};  low-latency edge inference \cite{Yuanming_TSP19MapReduce, shao20edgeinfer}   \\ \cmidrule(r){3-4}
                        
                        &       &  \multirow{1}[0]{\lencolthree}[0cm]{\centering Energy}  & Wireless-powered over-the-air FL \cite{Yuanming_IoT21}; energy-efficient FL \cite{Zhaohui_TWC21};  green edge inference \cite{yang20iot, hua21serverinference}  \\ \cmidrule(r){3-4}
                        
                        &       & Trustworthiness & Privacy \cite{LiuS21}, security \cite{Huang_BF21}, optimality \cite{abs-2104-10095}, interpretability \cite{Samek_IEEEPro21}  \\ \cmidrule(r){3-4}
                        
                        &       &  Service-Driven Resource Orchestration &  Heterogeneous demands for edge AI services \\ \cmidrule(r){2-4}
                        
                        & \multirow{4}[0]{\lencoltwo}[-0.6cm]{\centering Section \ref{resourceall}-B: Optimization Models and Algorithms} &  \multirow{1}[0]{\lencolthree}[0cm]{\centering Mixed-Combinatorial Optimization}    &  Sparse optimization models and algorithms \cite{ShiZCL18}; learning to branch-and-bound \cite{yifei20lorm}; algorithm unrolling \cite{shi2021}  \\ \cmidrule(r){3-4}
                        
                        &       & \multirow{1}[0]{\lencolthree}[0cm]{\centering Nonconvex Optimization}   & Large-scale convex approximation \cite{Yuanming_TSP15}; DC programming \cite{gotoh2018dc, yang2020over};  manifold optimization \cite{boumal2014manopt, Yuanming_TWC16MC}; GNN  \cite{yifei21gnn}   \\ \cmidrule(r){3-4}
                        
                        &       & \multirow{1}[0]{\lencolthree}[0cm]{\centering Stochastic Optimization}  & Robust optimization \cite{Letaief_TSP15}; chance constrained programming \cite{yang20iot}; deep RL \cite{lee2020optimization}; transfer learning \cite{Tomluo_TWC21}     \\ \cmidrule(r){3-4}
                        
                        &       &  End-to-End Optimization  & DNN \cite{WeiYu_TWC21}; GNN  \cite{WeiYu_JSAC21}   \\ \cmidrule(r){1-4}

                        \multicolumn{1}{c}{\multirow{4}[0]{\lencolone}[-0.63cm]{\centering Section \ref{archedgeai}: Architecture for Edge AI Systems}} & \multicolumn{2}{c}{  Section \ref{archedgeai}-A: End-to-End Architecture for Edge AI Systems     }        & Deep integration of sensing, communication, computation and intelligence \cite{tong20216g}  \\ \cmidrule(r){2-4}
                        
                        &\multicolumn{2}{l}{Section \ref{archedgeai}-B: Data Governance}      & Independent data plane \cite{Shencong_JSAC16}; multi-player roles \cite{Dai_TIFS20}  \\ \cmidrule(r){2-4}
                        
                        & \multicolumn{2}{p{0.36\textwidth}}{Section \ref{archedgeai}-C: Deeply Converged Communication and Computing at the Edge}       & \vspace{0.22cm} AI for networks; networks for AI \cite{letaief2019roadmap}  \\ \cmidrule(r){2-4}
                        
                        & \multicolumn{2}{l}{Section \ref{archedgeai}-D: Edge AI Management and Orchestration}       & Planning, deploying, maintaining, and optimizing edge AI models and edge network infrastructures \cite{tong20216g}   \\ \cmidrule(r){1-4}
                        
                        \multicolumn{1}{c}{\multirow{7}[1]{\lencolone}[-0.68cm]{\centering Section \ref{spa}:  Standardizations, Platforms, and Applications}} & \multirow{2}[0]{\lencoltwo}[-0.1cm]{\centering Section \ref{spa}-A: Standardizations} & Learning &  IEEE 3652.1-2020 \cite{9382202}   \\ \cmidrule(r){3-4}
                        
                        &       & Computing & ETSI ISG MEC \cite{mec9}  \\ \cmidrule(r){2-4}
                        
                        & \multirow{2}[0]{\lencoltwo}[-0.35cm]{\centering Section \ref{spa}-B: Platforms} & Software & FedML \cite{He2020FedMLAR}, FATE \cite{webank_fate}, HarmonyOS \cite{Huawei_harmonyOS}  \\ \cmidrule(r){3-4}
                        
                        &       & Solver & CVX \cite{cvx};  SCS \cite{o2016conic}; Open-L2O \cite{chen2021learning}  \\ \cmidrule(r){3-4}
                        
                        &       & Hardware & Edge computing hardware  \cite{CST20_edl}; RF hardware  \cite{amakawa2021white}
                        \\ \cmidrule(r){2-4}
                        
                        & \multirow{3}[1]{\lencoltwo}[-0.1cm]{\centering Section \ref{spa}-C: Applications} & Autonomous Driving & Perception; HD mapping; edge SLAM \cite{zhangjun19iov}  \\ \cmidrule(r){3-4}
                        
                        &       & Internet of Things & AIoT \cite{Yuanming_IEEENet20, Zhijin_JSAC21scl}; IIoT\cite{Tie_CST20, Dinh_WC21}  \\
                        \cmidrule(r){3-4}
                        
                        &       & Smart Healthcare & Clinical disease detection \cite{warnat2021swarm};
                        haptic communication \cite{Anton_CST18}   \\
                        
                        \bottomrule
                        
                \end{tabular}%
                
        }

        \label{results}
\end{table*}

\section{Communication-efficient edge training}
\label{edgetraining}
In this section, we shall present various communication-efficient distributed optimization
algorithms for edge training, followed by promising  enabling wireless techniques to
support the deployment of edge learning models and algorithms.

\subsection{Edge Learning Models and Algorithms}



The training process of edge AI models typically involves minimizing a loss
or empirical risk function  to fit a global model from decentralized data
generated by a massive number of intelligent devices. The goal of the
distributed optimization for edge training is to minimize the  global loss function $\mathcal{L}$, namely,
\setlength\arraycolsep{2pt}
\begin{eqnarray}
\label{elma}
\mathop {\sf{minimize}}_{\bm{\theta}\in\mathbb{R}^{d}}~~\mathcal{L}(\bm{\theta}):=\sum_{k\in\mathcal{S}}w_k\mathcal{L}_k({\bm{\theta}};
\mathcal{D}_k),
\end{eqnarray}
where ${\bm{\theta}}\in\mathbb{R}^d$ are the model parameters, $\mathcal{L}_k$ is the local loss function of device $k$ over local dataset $\mathcal{D}_k$, $\mathcal{S}$ denotes  the set of participating edge nodes,
and $w_k\ge0$ with $\sum w_k=1$ denotes the weight for each local loss function.
Considering the network topology for edge training, the heterogeneous local
dataset $\mathcal{D}_k$, varying device  participation $\mathcal{S}$, dynamic
communication and computation environments, as well as privacy concerns and
adversarial attacks, highly-efficient  and
trustworthy distribution optimization algorithms need to be developed. As shown in Fig. \ref{distribution:dataset:architecture}, based on the data partition and model partition principles \cite{Yuanming_CST20}, we will first introduce various edge training architectures, including FL, decentralized learning, and model split learning. We then present distributed RL and trustworthy learning techniques to accommodate dynamic and adversarial environments, respectively, as shown in Fig. \ref{distribution:dataset:environment}. \\

\subsubsection{Federated Learning}
\begin{figure*}
 \centering
 \subfigure[Federated learning.]{
  \begin{minipage}[b]{0.315\textwidth}
   \centering
   \includegraphics[width=1\textwidth]{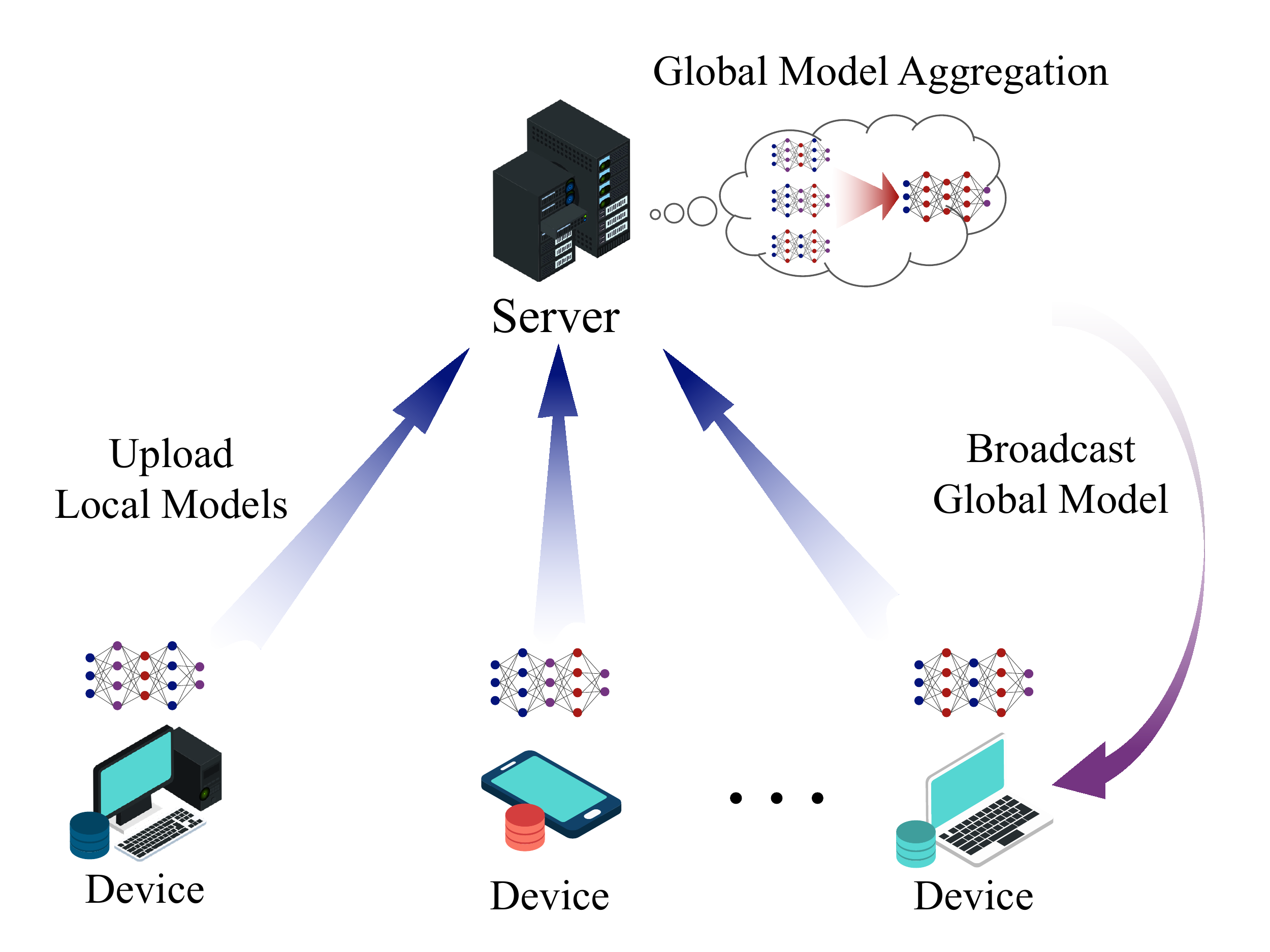}
  \end{minipage}
 }
 \subfigure[Decentralized learning.]{
  \begin{minipage}[b]{0.315\textwidth}
   \centering
   \includegraphics[width=1\textwidth]{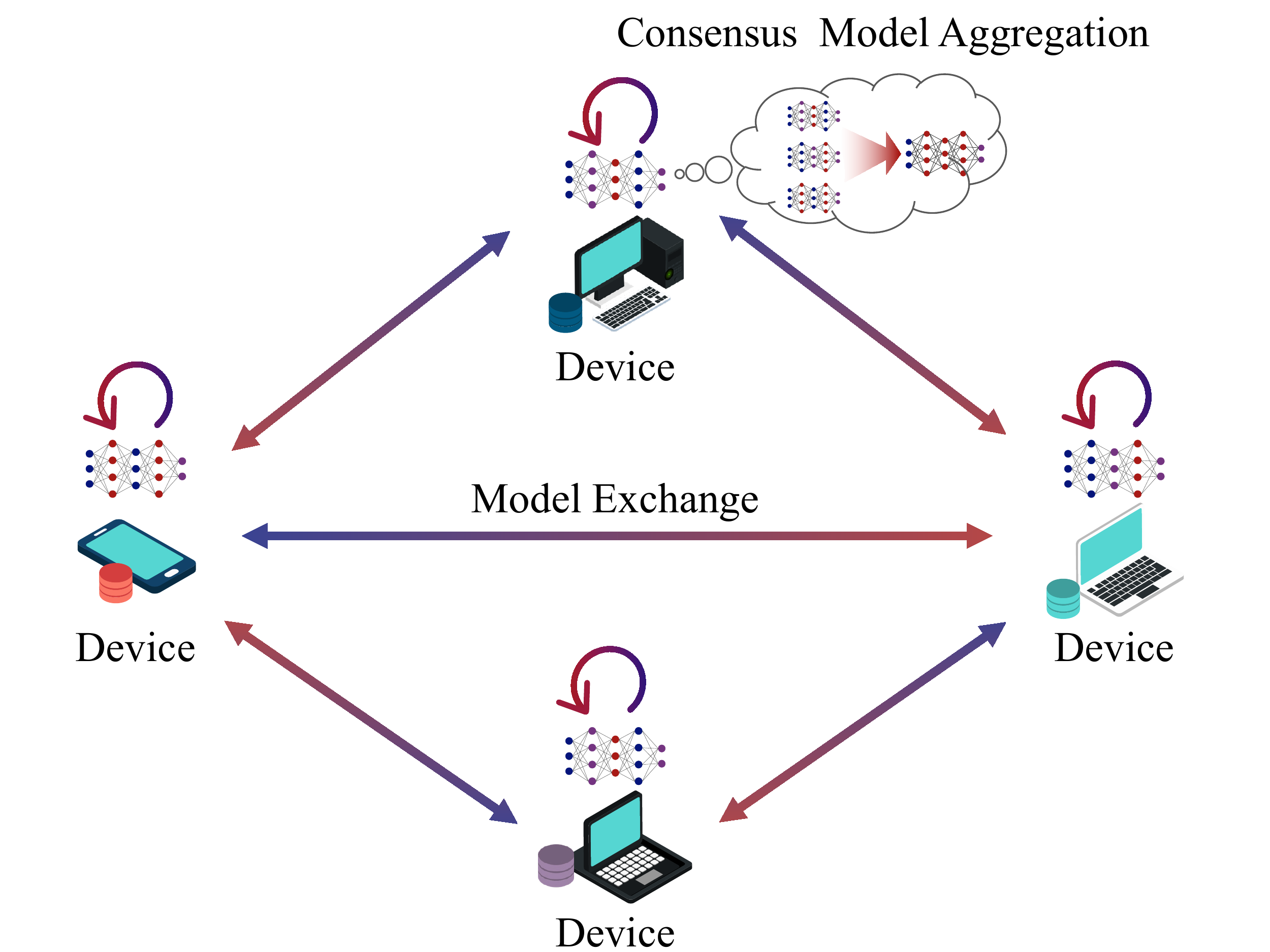}
  \end{minipage}
 }
 \subfigure[Model split learning.]{
  \begin{minipage}[b]{0.315\textwidth}
   \centering
   \includegraphics[width=1\textwidth]{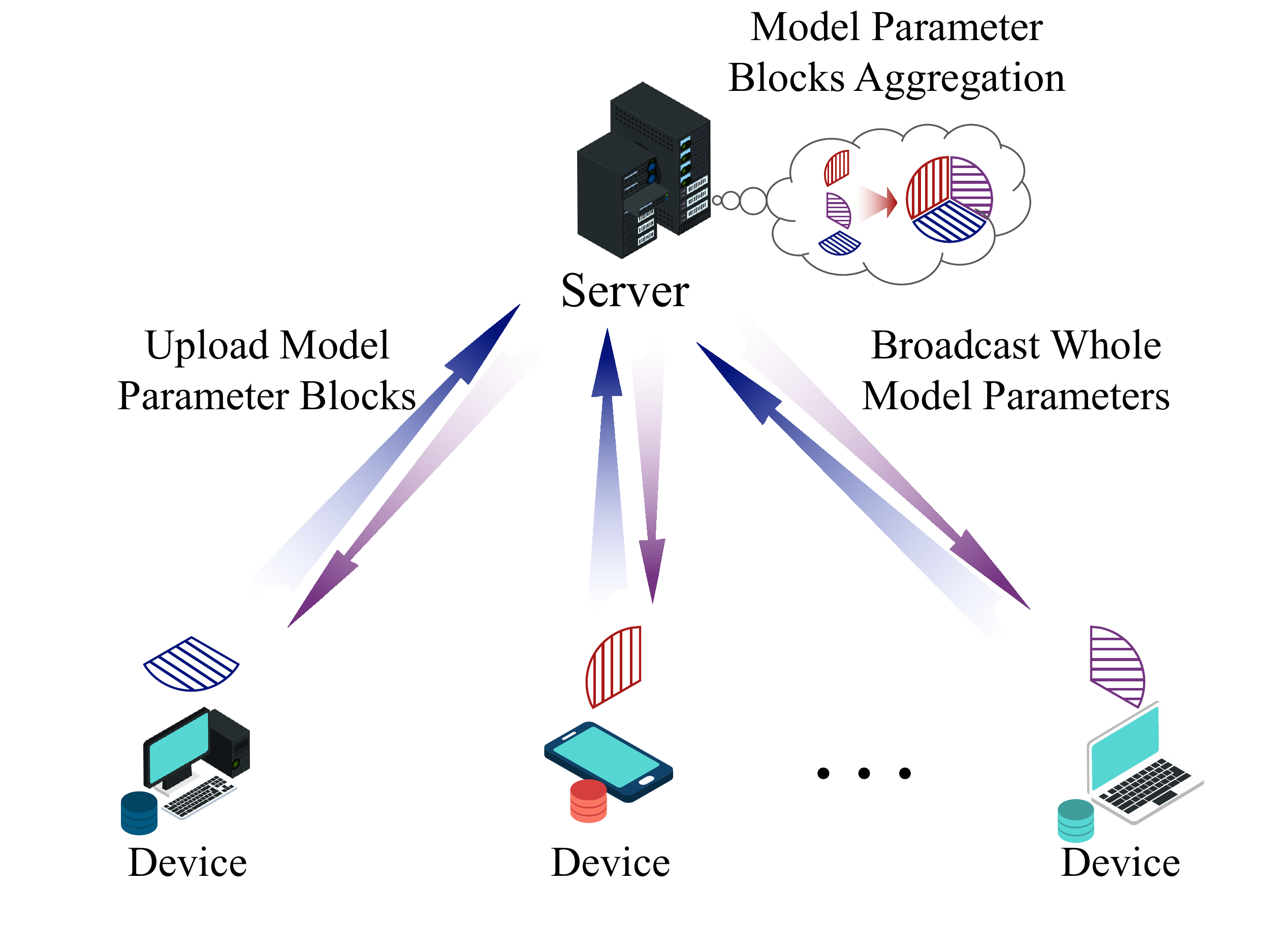}
  \end{minipage}
 }
 \caption{Edge learning models and architectures.} \label{distribution:dataset:architecture}
\end{figure*}

FL is a collaborative ML framework
to train a global statistical model without accessing  edge devices' private
raw data, wherein a dedicated edge  server is responsible for
aggregating local learning model updates and disseminating global
learning model updates \cite{kairouz2019advances}, as shown in Fig. \ref{distribution:dataset:architecture}
(a). {\rev{FL
is being adopted by many industrial practitioners, including Google's Gboard
mobile keyboard for next word prediction and emoji suggestion, Apple's
QuickType keyboard for vocal classifier, NVIDIA for COVID-19 patients 
oxygen needs prediction, and WeBank for money laundering detection \cite{wang2021field}}.} Compared with the cloud data center based distributed learning, cross-device
FL raises unique challenges for solving the distributed training
optimization problems, including high communication costs with a large
model frequently  exchanged over wireless networks, statistical heterogeneity
with non-identical
local data distributions and sizes, system heterogeneity with varied storage,
computation and communication capabilities, as well as dynamic devices participation
\cite{Tian_SPM20}. A growing body of recent works have developed effective methods
to address these unique challenges in FL.

To address the challenge of expensive communication overheads for intermediate
local updates with a central server, federated averaging \cite{McMahanMRHA17}
turns out to be effective to reduce the number of communication rounds by
performing multiple local updates, e.g.,  running multiple  stochastic
gradient descent (SGD)\ iterations  on each edge device. The local updating approach
is able to learn a global model within much fewer communication rounds compared
with the vanilla distributed SGD method, i.e., only running one mini-batch with
SGD at each edge device. Model compression, such as quantization and sparsification,
is another notable way to address the communication bottleneck by reducing
the size of the exchanged messages during each model update round. Scalar quantization
is a typical way to implement lossy compression for the high-dimensional
gradient
vectors
by quantizing each of their entries to a finite-bit low precision value \cite{AlistarhG0TV17,
NIPS2017_89fcd07f, pmlr-v80-bernstein18a}, which was further improved by the
recent proposal of vector quantization \cite{Kaibin_TSP20, Chen_TSP21QFED}.
 Sparsification, on the other hand, proposes to only communicate the informative
elements of the gradient or model vectors among nodes \cite{WangSLCPW18,
AjiH17}.  A set of algorithms combining the local updates method and model
compression have shown the capability of achieving high communication efficiency
\cite{abs-2103-14272, haddadpour2021federated}. In particular, a lazily aggregated
quantized gradient method was further proposed in
\cite{Chen_PAMI20} to reduce both the amount of  exchanged data and  communication rounds by reusing
the outdated gradients for the less informative quantized gradients.

Although the above periodical compressed update methods have shown empirical
or theoretical success  for tackling the communication challenge, the heterogeneity
in systems and local datasets  may slow down or even diverge the convergence
\cite{Li2020On, LiSZSTS20}, for which various algorithms and models have
been proposed to address the statistical and system heterogeneity challenges. To
learn the AI models from statistically heterogeneous local datasets,
various effective and personalized models have been proposed to rectify the
original model (\ref{elma}), including  regularizing local loss functions at each device
\cite{LiSZSTS20, acar2021federated, NEURIPS2020_f4f1f13c},
distributionally robust modeling  \cite{NEURIPS2020_ac450d10,
OPT-026}, multi-task learning  \cite{NIPS2017_6211080f}, as well as
the meta-learning  approaches \cite{NEURIPS2020_24389bfe}. Running a local update
 at  the devices with heterogeneous computation capabilities may yield
objective inconsistency  or client drift, i.e., the learned model can be far from  the desired true model.  
To address this problem,  an operator splitting
method was proposed to avoid the local models drifting apart from the global
model \cite{PathakW20}. A normalized model aggregation method was also developed
to ensure that the global model converges to the desired true model \cite{WangLLJP20}.
A novel federated aggregation scheme was further developed in \cite{pmlr-v130-ruan21a}
to address the system heterogeneity issue concerning the dynamic, sporadic
and partial device participation. To leverage the computation capabilities across the device-edge-cloud heterogenous
network, a hierarchical model aggregation approach was proposed in \cite{abs-2103-14272}
to reduce the latency by controlling the two aggregation intervals.\\    
\subsubsection{Decentralized Learning}
Decentralized ML learns a global model from inherently
decentralized data structures via peer-to-peer communications  over the underlying communication network topology without a central authority
\cite{lian2017can}, as shown in Fig. \ref{distribution:dataset:architecture}
(b). {\rev{It has great potentials for applications in the autonomous
industrial systems, including cooperative automated driving, cooperative
simultaneous localization and mapping, and collaborative robotics in advanced
manufacturing environments \cite{Mehdi_ComMag21}.}} The decentralized learning architecture harnesses the benefits of communication efficiency,
computation scalability and data locality. In particular, swarm learning
\cite{warnat2021swarm} provides a completely decentralized AI solution based
on decentralized ML by keeping local datasets at each edge device. This can achieve high privacy, security, resilience and scalability. Compared with the
sever-client learning architecture in FL, decentralized learning can
accommodate the decentralized D2D communication network architectures
and protocols with arbitrary connectivity graphs (e.g., cooperative driving
and robotics networks). It can also overcome the straggler dilemma with heterogeneous hardware,
as well as improve the robustness to data poisoning attacks and master node fails \cite{warnat2021swarm,
Mehdi_ComMag21}. The convergence behavior of decentralized learning highly depends on the decentralized averaging mechanism and the network topology for data exchange \cite{lu2021optimal}. Typical decentralized aggregation approaches include the consensus-based
methods \cite{Boyd_TIT06} and diffusion strategies \cite{Sayed_SPM13}. 

To improve the communication efficiency for exchanging 
the locally updated models at edge devices within their neighbors, one may  reduce either the number of communication rounds (i.e., improve
convergence rate) or the volume of exchanged data per round. Specifically, the variance
reduction with the gradient tracking method was investigated in \cite{Ran_SPM20}
to achieve a fast convergence rate. Periodic-averaging via running multiple
local updates before decentralized averaging is an effective way to reduce
the number of communication rounds among devices \cite{wang2019cooperative, koloskova2020unified}.
Besides, quantizing or sparsifying the locally updated models can reduce the volume of the exchanged messages to address the communication
bottleneck \cite{koloskova2019decentralized}. A consensus distance controlling
framework was further developed in \cite{kong2021consensus} to achieve the
trade-off between the learning performance and the exactness of  decentralized
averaging for decentralized DL.  Moreover, a communication network
topology design is also critical to improve the communication efficiency
\cite{pmlr-v108-neglia20a}, for which a group alternating direction method
of multipliers \cite{elgabli2020gadmm} was proposed to form a connectivity
chain by dividing the workers into head and tail workers. To address the
heterogeneity issue of local datasets, the momentum-based method \cite{lin2021quasi}
 has recently been developed to achieve good generalization performance.\\

\subsubsection{Model Split Learning} Model split learning enables a collaborative
learning process across the edge devices and edge servers by partitioning the model
parameters across the edge nodes, as shown in Fig. \ref{distribution:dataset:architecture}
(c). That is, each edge node $k$, including edge devices and edge servers, is only responsible for updating ${\bm{\theta}}_k$ with ${\bm{\theta}}=[{\bm{\theta}}_1,{\bm{\theta}}_2,\dots, {\bm{\theta}}_{\mathcal{S}}]$ in (\ref{elma}).  This model splitting architecture can  achieve higher privacy levels and
better trade-offs between communication and computation. It is thus particularly
applicable for DL with a large model parameters size, whereas
the data partition based training method, e.g., FL, normally
requires the local update of a whole copied global model at each involved edge
device. The model parameter partitioned edge learning  approach \cite{li2014communication}
proposed to train only a block of model parameters based
on the  coordinate decent method for the decomposable ML
models \cite{wright2015coordinate} or the alternating minimization approach
for the general DL models \cite{pmlr-v97-choromanska19a}. However, this
approach is prone to data privacy leakage as the datasets need
to be shared across edge devices.  Vertical FL, on the other hand,
can directly learn the global model from the partitioned data features
among different edge devices without sharing them \cite{gu2020federated}. Therefore,
the data features and the associated model parametric blocks are split among
edge devices, for which the asynchronous SGD method can be applied for  vertical
FL \cite{hu2019fdml}. Consensus algorithms were also developed
in \cite{ying2018supervised} to jointly learn a model under the decentralized
network while keeping the distributed data features locally.

Split DL further provides a flexible way to train a DNN
by dividing  it into  lower and upper segments located at the edge device-side
and edge server-side, respectively \cite{gupta2018distributed}. 
{\rev{It
 can be typically applied to the medical diagnosis and millimeter wave channel prediction  \cite{park21commdlover}.
}}Split DL is able to preserve privacy
without sharing raw data and enjoys computation scalability by allowing that only edge
devices  perform simple computation for the lower segments. {\rev{Compared with FL, split DL
can significantly improve  computation efficiency,
reduce communication costs, as well as achieve higher learning accuracy, data security
and system scalability.}} Specifically,
edge devices and edge server collaboratively train the whole neural network, which
involves routing the activations of the edge device-side subnetwork to the edge sever
via forward propagation, and downloading the gradients of the edge server-side
subnetwork to update the lower segment via back propagation. However, exchanging
the instantaneous intermediate values between edge devices and edge server becomes the
communication bottleneck, especially in the case with multiple edge
devices. Therefore, a joint communication strategy and neural network architecture
design is required \cite{park21commdlover} for split training of various
DNNs with heterogeneous edge devices.  Considering the large-scale privacy-sensitive and
delay-sensitive IoT applications, Lyu
\textit{et al.} \cite{lyu2020foreseen} proposed a hybrid fog-based privacy-preserving
DL framework, where a fog-level DNN is partitioned between the edge
device and the fog server side. \\

\subsubsection{Distributed Reinforcement Learning} 
\begin{figure}
 \centering
 \subfigure[Distributed reinforcement learning.]{
  \begin{minipage}[b]{0.45\textwidth}
   \centering
   \includegraphics[width=1\textwidth]{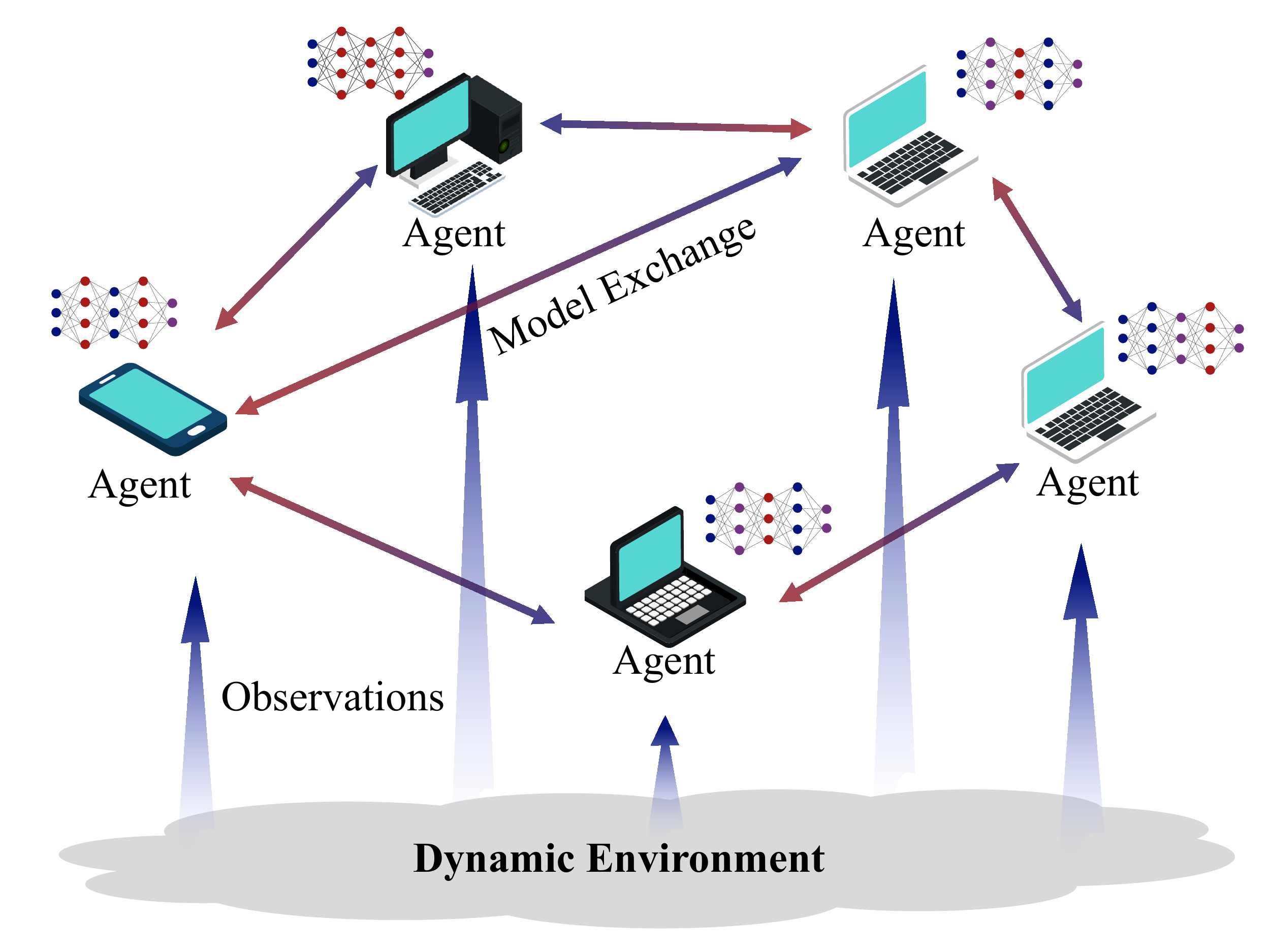}
  \end{minipage}
 }
 \subfigure[Trustworthy learning.]{
  \begin{minipage}[b]{0.45\textwidth}
   \centering
   \includegraphics[width=1\textwidth]{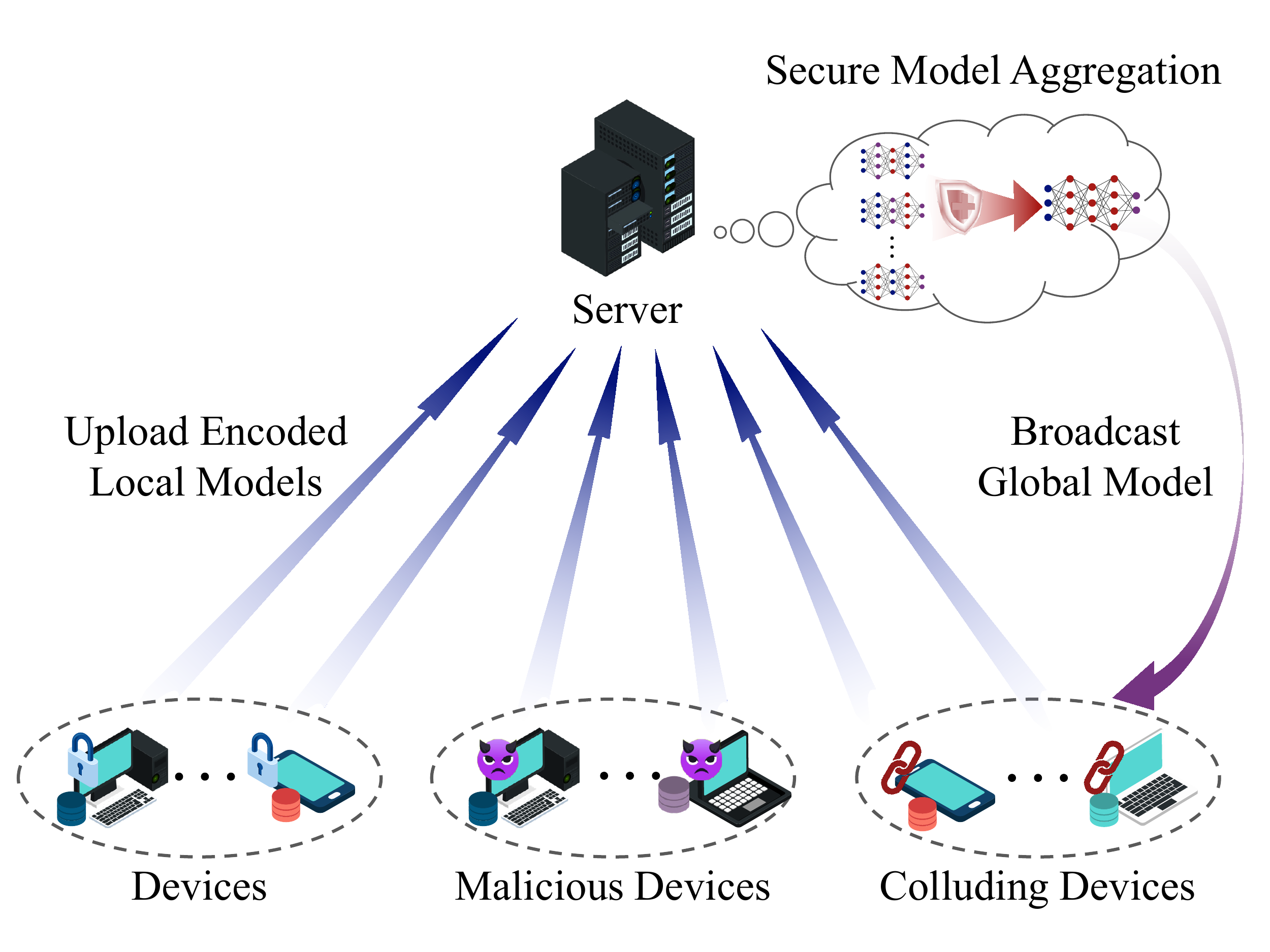}
  \end{minipage}
 }
 \caption{Edge learning modes in dynamic and adversarial environments.} \label{distribution:dataset:environment}
\end{figure}

RL provides a flexible framework for sequential decision making in dynamic
settings by interacting with a dynamic environment, as shown in Fig. \ref{distribution:dataset:environment}
(a). This can be frequently modeled
as decision making and learning in a Markov decision process (MDP) \cite{Kai_SPM17}.
Typical RL algorithms include the model based algorithm, policy-based
algorithm (e.g., natural policy gradient), value based algorithm (e.g., Q-learning),
and actor-critic method. In particular, an asynchronous method, by leveraging
parallel computing, was developed in \cite{mnih2016asynchronous} to solve the
large-scale nonconvex RL problem.  However, in  modern intelligent applications,
e.g., autonomous driving and robotics, it is critical to consider 
multi-agent reinforcement learning (MARL), in which multiple agents collaboratively
interact with a common environment to
complete a common goal and maximize a shared team award with different local
action spaces \cite{lee2020optimization}. Due to the enormous state-action
space, delayed rewards and feedback, as well as the non-stationary and unknown
environments with heterogeneous agents' behaviors, efficient communication strategy among multiple
agents shall play a key role to achieve good and stable performance for MARL.

For the server-client  architecture based MARL, the  edge server
coordinates the learning process for all the edge agents. Ryan \textit{et al.}
\cite{lowe2017multi}  proposed a multi-agent actor-critic
method involving decentralized actors at each agent and a centralized critic
for parameter sharing among the agents. To improve the  communication efficiency
of the distributed policy gradient for MARL, a lazily aggregated policy gradient
was developed in \cite{Tianyi_TCNS21} to reduce the communication rounds by
only communicating informative gradients of partial agents while reusing
the outdated gradients for the remaining agents. For applications without
central coordinators, e.g., autonomous driving, decentralized MARL is essential
wherein the agents only allow the exchange of messages with their neighbors over
a communication connectivity graph \cite{zhang2019multi}.  Zhang \textit{et al.} \cite{zhang2018fully}
proposed decentralized actor-critic
algorithms with function approximation, where each agent makes individual
decisions based on both the information observed locally and the messages
shared through a consensus step over the network. A decentralized entropy-regularized
policy gradient method by only sharing information with neighbor agents
was developed in \cite{Zeng_arXiv20RL} to learn a single policy for multi-task
RL with multiple agents operating different environments.\\

\subsubsection{Trustworthy Learning} To learn and deploy AI models  for  high-stake applications (e.g., autonomous driving) at the network edge,
 it is critical to ensure privacy, security, interpretability, responsibility, robustness,
and fairness for the edge learning processes, as shown in Fig. \ref{distribution:dataset:environment}
(b). However, the
heterogeneity of massive scale edge systems and decentralized datasets raises
unique challenges to design trustworthy edge AI techniques. Although FL addresses the local confidentiality issue by keeping datasets locally,
the shared model updates still cause extreme privacy leakage (e.g., model
inversion attack), the learned global model can be colluded by malicious
attackers \cite{Robust1, Robust3}, and the edge devices may be adversarial
attackers (e.g., data or model poisoning). This calls for rigorous
privacy-preserving mechanisms and secure  aggregation rules \cite{pmlr-v89-yu19b}. Differential
privacy provides a promising lightweight privacy-preserving mechanism to
guarantee a level of privacy disclosure for local datasets by 
adding random perturbations \cite{DP}. The additive noise and signal superposition
properties in the wireless channel can be naturally harnessed as the privacy-preserving
mechanism \cite{LiuS21}. The resulting inherent noisy model aggregation scheme
can limit the privacy disclosure of local datasets at the edge server
for free  while keeping the learning performance unchanged \cite{Anis_TCOM21,
LiuS21}. To improve the communication efficiency for private distributed
learning, Chen {\emph{et al.}} \cite{NEURIPS2020_222afbe0} developed efficient
encoding and decoding mechanisms to simultaneously achieve optimal communication efficiency
and differential privacy  under typical statistical
learning settings. 

Apart from preserving privacy for individual users, edge AI also needs to be
robust to errors and adversarial attackers, as the decentralized nature makes
it easy to be unreliable in the learning process or even completely controlled
by external attackers \cite{chen2017distributed}. To address  Byzantine
attacks (i.e., the faulty edge device can behave arbitrarily badly by modifying
its local updates) in FL with a server-client architecture,
various robust and secure model aggregation schemes (e.g., geometric median
\cite{Zhaoxian_TSP20}, trimmed
mean \cite{Aggregation-trimmed-mean}, and Krum \cite{Aggregation-Krum}) were proposed
to tolerate the Byzantine corrupted edge devices. To simultaneously preserve
privacy for individual users while tolerating Byzantine adversaries, a Byzantine-resilient
secure aggregation framework was developed in \cite{So_JSAC21} to detect adversarial
models without the knowledge of individual local models, as they are masked
for privacy guarantees.  To further avoid malicious edge servers,  blockchain
technology was utilized to provide a decentralized consensus environment to
guarantee
the validity of global models in every learning iteration. This is achieved
by packing the
local models and global model into blocks, which are confirmed under a consensus
mechanism, followed by
linking them into the blockchain \cite{Kim_CL20}. To protect decentralized
learning from attacks, a blockchain based peer-to-peer network was developed
in \cite{warnat2021swarm} to  support swarm learning without a central server.
This high security level in decentralized learning is achieved by securely
enrolling new  nodes via blockchain smart contract to perform local model
training.  

To summarize, the presented edge learning models and algorithms provide a strong evidence that to deploy the edge training process in   wireless networks,
we need to develop new wireless communication techniques and strategies to support massive and flexible edge devices participation, as well as support efficient function
computation for model aggregation (e.g., weighted sum global model aggregation in
FL, consensus model aggregation in decentralized learning,
and robust model aggregation in secure learning). Various edge training architectures
(e.g., server-client, decentralized, and hierarchical network topologies),
as well as high-dimensional model updates exchange motivate us to develop new wireless network principles and architectures to support  edge AI training systems, which will be discussed in the following subsection.

\subsection{Wireless Techniques for Edge Training}

\begin{figure*}
 \centering
 \begin{minipage}{0.65\textwidth}
  \centering
  \subfigure[Over-the-air computation.]{
   \begin{minipage}[b]{0.47\textwidth}
    \centering
    \includegraphics[width=1\textwidth]{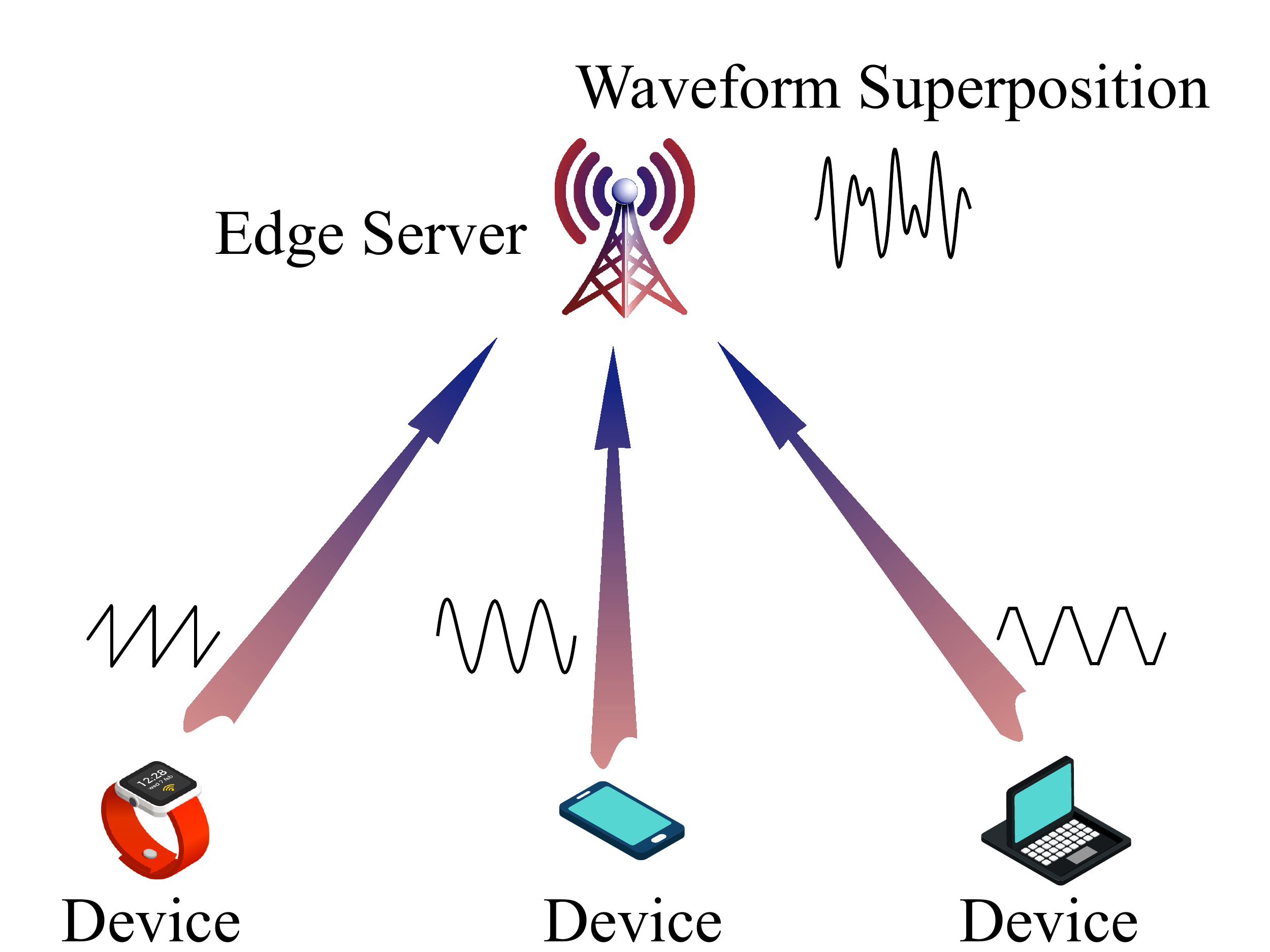}
   \end{minipage}
  }
  \subfigure[Massive access with sporadic traffics.]{
   \begin{minipage}[b]{0.47\textwidth}
    \centering
    \includegraphics[width=1\textwidth]{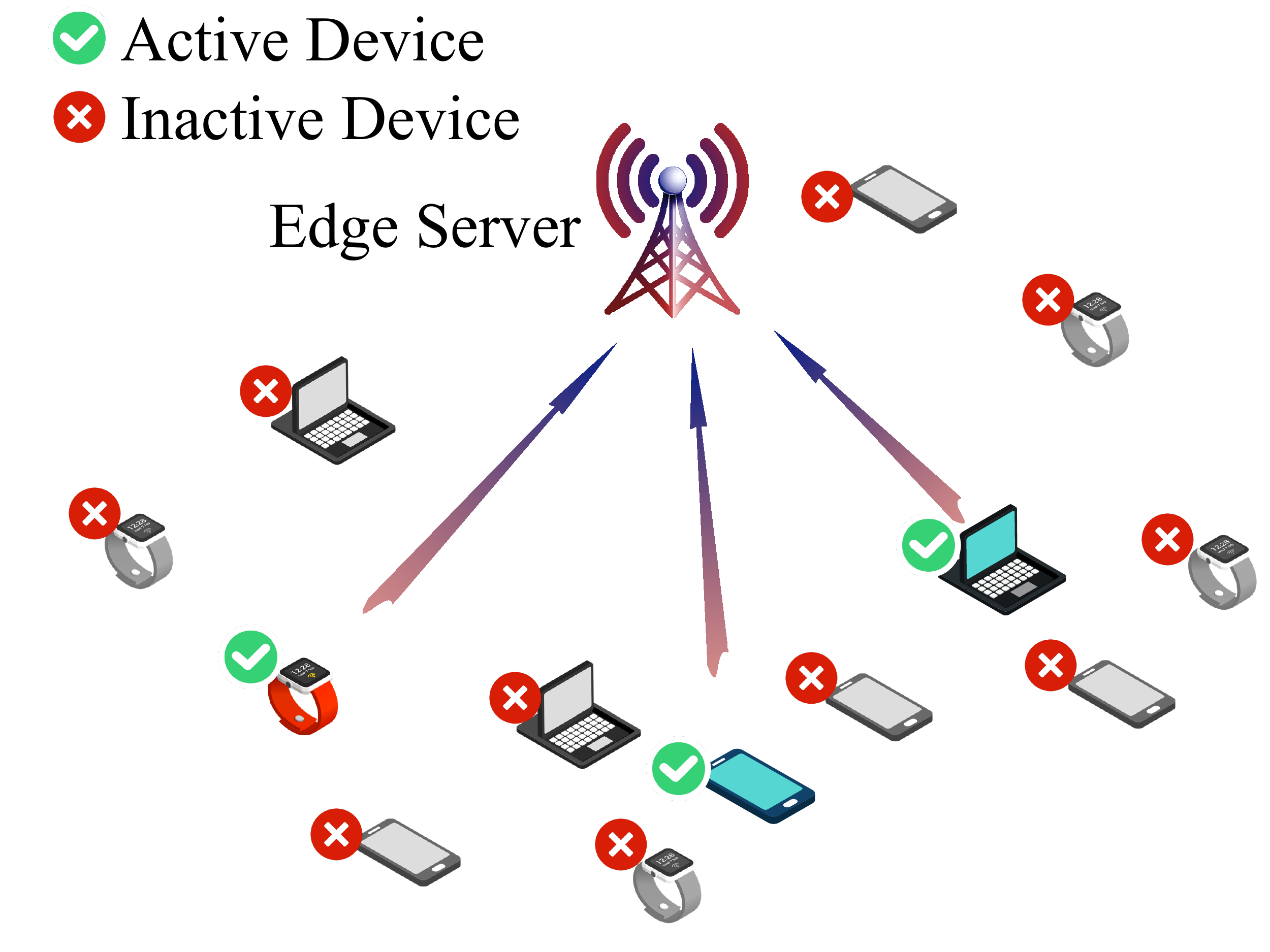}
   \end{minipage}
  }   \\
  \subfigure[Cell-free massive MIMO.]{
   \begin{minipage}[b]{0.47\textwidth}
    \centering
    \includegraphics[width=1\textwidth]{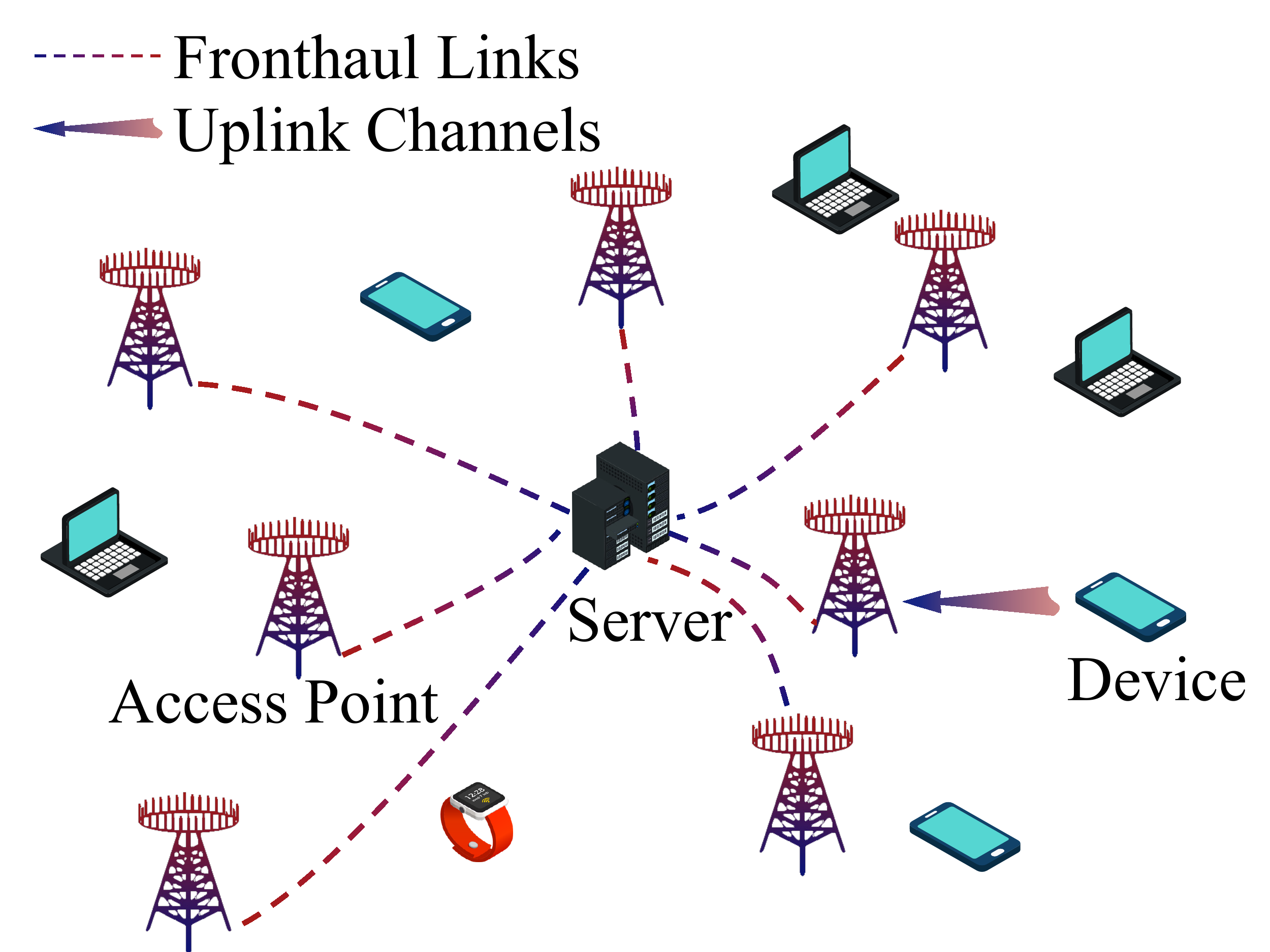}
   \end{minipage}
  }
  \subfigure[Reconfigurable intelligent surface.]{
   \begin{minipage}[b]{0.47\textwidth}
    \centering
    \includegraphics[width=1\textwidth]{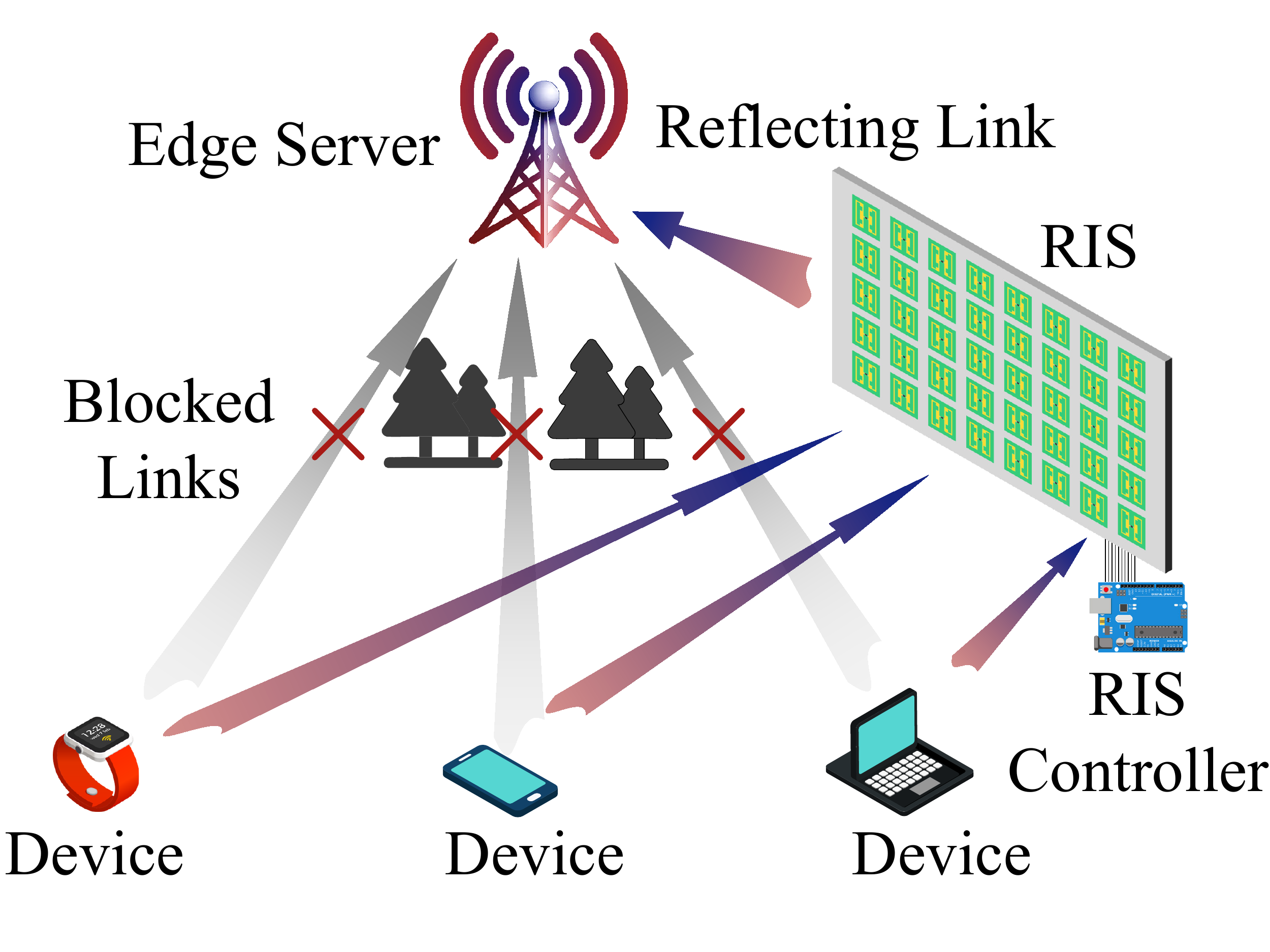}
   \end{minipage}
  }
  
 \end{minipage}%
 \begin{minipage}{0.344\textwidth}
  \subfigure[Space-air-ground integrated network.]{
   \begin{minipage}[b]{1\textwidth}
    \centering
    \includegraphics[width=1\textwidth]{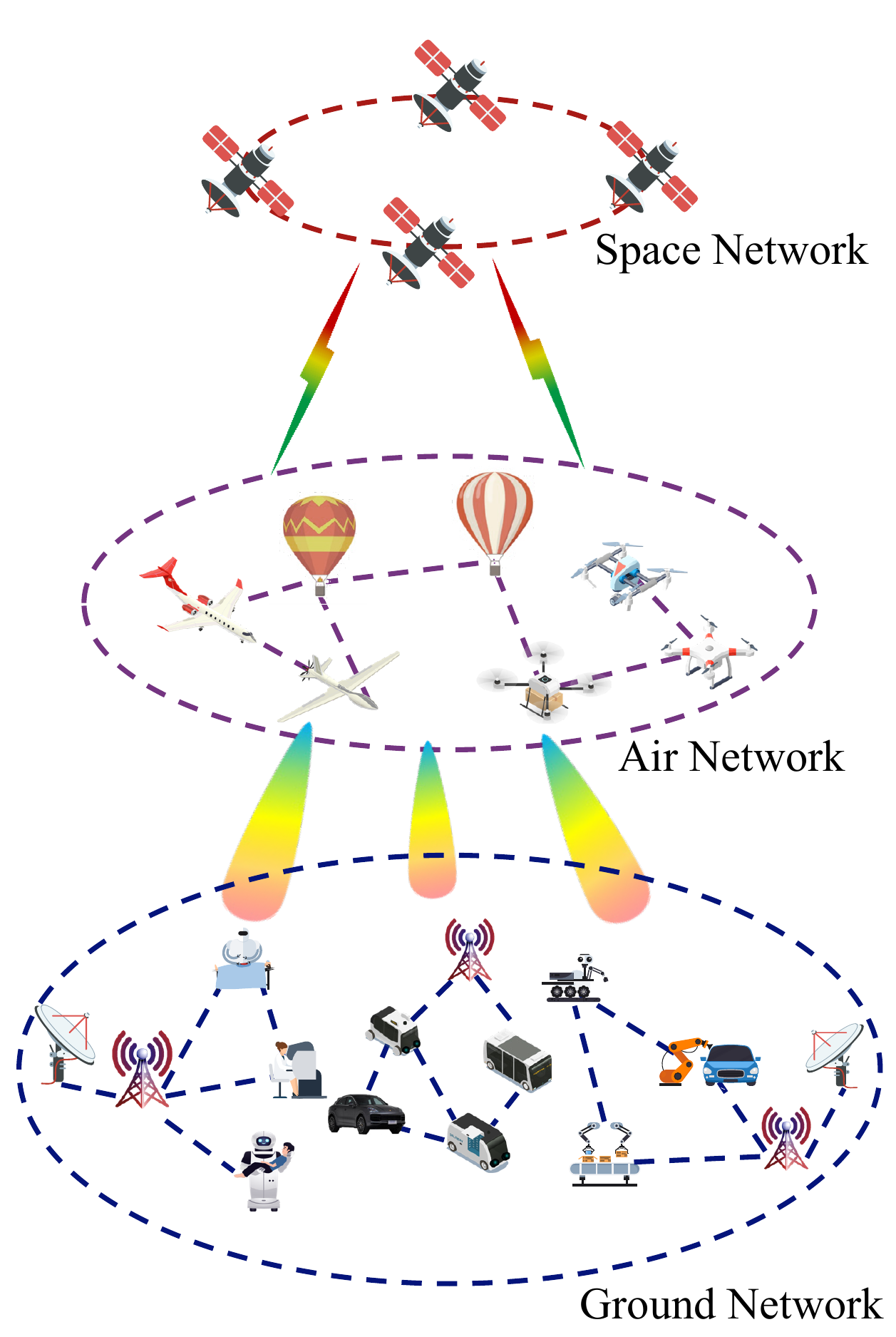}
   \end{minipage}
  }
 \end{minipage}
\caption{Enabling wireless techniques for edge training.} \label{distribution:wireless}
\end{figure*}

As the communication target
for edge AI becomes the learning performance instead of the conventional data rates,
we shall exploit the task structures of edge AI models and algorithms to match  the principles and architectures of wireless networks. 
This helps demystify the efficiency of edge training  in wireless networks,
which yields a learning-communication co-design principle for future 6G wireless
networks to enable AI functionalities sitting natively within 6G. As shown in Fig. \ref{distribution:wireless},
we will introduce next generation multiple access schemes (e.g., AirComp and massive random access) to accommodate a massive number of
edge devices dynamically involved in the training process,
new multiple antenna techniques (e.g., RIS and cell-free massive MIMO) to support high-dimensional model updates exchange,
as well as new network architectures (e.g., SAGIN and unmanned aerial vehicle (UAV) network) to support diversified edge training
models and topologies. \\

\subsubsection{Over-the-Air Computation} Edge training tasks typically involve
computing aggregation functions of multiple local model updates to update a global
model. To accomplish  weighted averaging aggregation in FL,
consensus aggregation in decentralized learning, and robust aggregation (e.g.,
geometric median) in trustworthy learning, the local updates need to be transmitted
from the  edge devices, followed by computing the relevant aggregation
function at the edge sever. However, the limited bandwidth and resource in wireless
networks becomes one of the key bottlenecks to enable a massive number of edge
devices that upload the local model updates for global aggregation. AirComp provides a new  multiple access scheme for
low-latency model aggregation. By concurrently transmitting the  locally updated
models, AirComp can harness interference to reduce communication
bandwidth consumptions. The key idea is that the waveform superposition of a wireless
multiple access can be exploited for computing the nomographic functions
(e.g., the model aggregation function weighted average) over the same channel \cite{Yuanming_IEEENet20}, as shown in Fig. \ref{distribution:wireless}
(a).
Specifically, the transmitted signals at  edge devices are first multiplied
by the fading channels and then superposed over-the-air with additive channel
noise, resulting in a noisy weighted sum of transmitted signals \cite{yang2020over}. This perfectly
matches the structure of model aggregation computation. Note that the robust
aggregation function (i.e., geometric median) does not hold the additive
structure. But we can still approximate it by computing a few number of weighted
averaging functions via AirComp \cite{Huang_BF21}.
The communication latency and bandwidth requirement of AirComp  will not
increase with the number of edge devices, thus relieving the communication
bottleneck in the edge training process.    

Channel fading and noise perturbation in the model aggregation
raise unique challenges for the edge training algorithm design and analysis.
To tackle the channel fading perturbation, a channel inversion method was
proposed in \cite{AmiriG20, Deniz_TWC20, Kaibin_TWC20} by multiplying the
inverse of channel gain for the transmit signal, which may however not satisfy
the power constraint at edge devices. To address these issues, a transceiver
design was provided in \cite{yang2020over} to minimize the distortion for
the perturbed model aggregation, whereas the perturbed model updates are
directly incorporated in the FL algorithm design \cite{Anis_TCOM21}.
Although the analog transmission in AirComp is prone to channel noise, the
additive noise in the model aggregation turns out to be controllable or even
beneficial in the edge training process. Specifically, the channel noise in the model aggregation
yields a new class of noisy FL algorithms. The convergence behavior
demonstrates that the noisy  iterates typically introduce  non-negligible optimality gap
in various FL algorithms, e.g., vanilla gradient method \cite{Tomer_TSP20},
quantized gradient method \cite{Huang_TWC21}, sparsified gradient method
\cite{Deniz_TWC20}, and operator splitting method \cite{Xia_arXiv20}. The
optimality gap  can be  further controlled by transmit power allocation \cite{Shen_arXiv21,
LiuS21, Deniz_TWC20}, model aggregation receiver beamforming design \cite{TWC_21Amiria,
yang2020over, Angela_arXiv20}, and device scheduling \cite{Angela_arXiv20,
yang2020over, Fan_arXiv21}. Besides, channel perturbation in algorithm
iterates can also serve as the mechanism to design saddle points escaping
algorithms \cite{abs-2104-10095}, thereby establishing global optimality
for training the non-convex over-parameterized neural networks in high-dimensional
statistical settings \cite{IEEEProc21_Tong}. The additive channel noise in
model aggregation can also serve as an inherent privacy-preserving mechanism
to guarantee differential-privacy levels for each edge device without sacrificing
learning performance \cite{LiuS21}.  \\

\subsubsection{Massive Access Techniques}
Deploying  cross-devices FL in IoT networks raises
practical challenges, i.e., the IoT devices have sporadic  access to the
wireless network \cite{savazzi2020federated}. It is thus critical to design
practical FL systems to accommodate flexible device participation
with sporadic access to the wireless network \cite{pmlr-v130-ruan21a}, as
shown in Fig. \ref{distribution:wireless}
(b). The grant-free random access
protocol provides a low-latency and low signaling overhead way to detect
the active devices, followed by decoding their corresponding information
data \cite{liu2018massive, Yuwei_JSAC21, Gao_WC20}. In this protocol, active devices can
transmit the data signals directly without waiting for any permission. Sparse signal processing
 provides a promising modeling framework to  simultaneously detect the
active devices and estimate their channels \cite{liu2018massive, Tao_IoTJ19},
which is supported by various efficient algorithms, including the approximate
message passing algorithm \cite{WeiYu_TSP18, Xia_arXiv20RISaircomp} and DNN algorithm unrolling approach \cite{monga2021algorithm, shi2021}.
To  further reduce the latency for data decoding in random access, a sparse
blind demixing framework was developed in \cite{shi2020low} by simultaneously performing  active
device detection, channel estimation and their data decoding.
The key observation is that blind demixing is able to perform low-latency
data decoding for multiple users from the
sum of bilinear measurements without channel estimation at both the transmitters
and receivers \cite{Shi-TWC19, Shi_TSP18}. To enhance the performance, the common sparsity pattern in pilot and user data has been exploited via joint activity detection and data decoding \cite{Jun_ICCsparse, Jun_SPAWC21sparse}.

Random access protocols are promising to support flexible and massive device participation
in the edge training process by identifying active devices with sporadic traffic. It is still  critical to
develop massive access techniques to improve the learning performance by enrolling
more active devices to perform local model update and 
exchange under digital transmission. Nonorthogonal multiple access (NOMA)
\cite{Dai_ComMag15, Ding_JSAC17NOMA} is a key enabling candidate technology
to simultaneously serve massive devices for model aggregation in the same
radio resource block via superposition coding. Typical NOMA schemes
 include the power-domain NOMA with different transmit powers
as weight factors and the code-domain NOMA {\rev{(e.g., sparse code multiple access \cite{Cheng_TVT21} and pattern division multiple
access \cite{Dai_ComMag15})}} with different codes assigned
to users. Therefore, the user's data can be decoded from the simultaneously
transmitted signals via successive interference cancellation. {\rev{In particular, DL provides a powerful method to design and optimize NOMA systems \cite{Neng_TWC20, Chong_JSAC21, Yuxin_JSAC21}.}} Under analog
uncoded transmission, interference can be harnessed via the new massive access
techniques AirComp, for which Dong {\emph{et al.}} further proposed a blind
AirComp for low-latency model aggregation  without channel state information
(CSI) access \cite{Shi_TSP20}. It is thus particularly interesting to integrate
a massive random access protocol (e.g., grant-free random access) and massive
access technique (e.g., AirComp based access technique) with analog uncoded transmission
to simultaneously perform active
device detection, channel estimation and model aggregation, thereby
supporting flexible and low-latency edge devices enrolling for collaboratively training the models. \\ 

\subsubsection{Ultra-Massive MIMO}
Leveraging massive antenna arrays is a key enabling wireless technology
to achieve high spectral and energy efficiency, which is envisioned to be
further scaled up by an order-of-magnitude in 6G  \cite{IEEEPro21_6Gwireless}.
The recent advances in digital beamforming, analog beamforming, as well as
hybrid beamforming have helped the roll-out of massive MIMO into practice by operating
over a wider frequency band. It has been demonstrated that massive MIMO
is able to bring enormous benefits for edge training systems, including high-accuracy
and high-rate for model aggregation, as well as high-reliability for massive device
connectivity. Specifically, massive MIMO can achieve a high computation accuracy
for model aggregation via exploiting spatial diversity \cite{Zhai-TWC21},
and enable ultra-fast model aggregation with simultaneous multi-functions
computation by spatial multiplexing \cite{Zhu-IoTJ19}. Furthermore, for
FL with edge devices sporadically enrolling, the device activity detection
error goes to zero as the number of antenna elements in the BS goes to infinity,
thereby achieving high-reliable devices participation for model updates. 

To scale edge training   to huge physical areas  with massive geographically
distributed edge devices, ultra-dense wireless network  is a promising way to
achieve low-latency, high-reliability and high-performance. This is achieved by  simultaneously uploading massive local model updates
with multiple distributed edge servers\ with abundant communication, computation,
and storage resources, thereby mitigating the stragglers issues (i.e., devices with low communication and computation capabilities may prolong the
training time) and unfavorable channel dynamics. Besides, compared with the
single edge server architecture, distributed
edge servers are robust to server failure issues for reliable edge training. In
particular, cloud radio access network (Cloud-RAN) \cite{shi2015large, Peng_CST16} provides a cost-effective
way to implement  distributed antenna aided edge training systems, for
which reliable model aggregation via AirComp can be achieved by centralized
signal processing and shortening the communication distances between edge
devices and edge servers \cite{Lukuan_ICC21}. The recent proposal of cell-free
massive MIMO \cite{zhang2020ma} serves a promising way to realize the wireless
distributed FL systems  by exploiting the channel hardening
characterization (i.e., the effective channel gain is approximated by its
expectation value) and avoiding sharing instantaneous CSI among edge servers \cite{than2020cellfreemimo}, as shown in Fig. \ref{distribution:wireless}
(c). \\

\subsubsection{Reconfigurable Intelligent Surfaces} To obtain the desired
average function of local model updates for model aggregation via AirComp, magnitude
alignment by scaling the transmit signals (e.g., channel inversion) is normally
required to reduce the channel perturbation \cite{Kaibin_arXiv20aircomp}.
However, due to the resource-limited edge devices and the non-uniform fading
channels, the unfavorable signal propagation environment inevitably leads
to magnitude reduction and misalignment with perturbed model aggregation,
which in turn degrades the learning performance of the edge training process.
Besides, the massive edge devices with sporadic access to the edge servers can
be located at a service dead zone, which makes device activity detection challenging
for weak channel links \cite{Xia_arXiv20RISaircomp}. To enroll multiple
edge devices via simultaneously transmission with NOMA, sufficient
diversified channel gains are normally required for successive interference
cancellation, which however may not always hold in practical scenarios
\cite{Yuanming_TWC21NOMA}. Heterogeneity in terms of computation, communication,
and storage across edge devices is one of major challenges to deploy edge
AI systems. Waiting for the straggler edge devices with slow computation
and communication speeds for model aggregation causes significant delays,
which can be tackled by  computation offloading and task scheduling by
mobile edge computing (MEC) technique \cite{Deniz_TSP19}. However, fully unleashing
the benefits of MEC for straggler mitigation is  limited
by the hostile wireless links \cite{Tong_JSAC20}.

To address the above challenges in terms of  propagation impairments, RIS has been shown to be a cost-effective technology
to support  fast
yet reliable model aggregation with massive edge devices participation by programming
the propagation environment of electromagnetic waves \cite{Yuanming_IEEENet20,
Yuanming_wc21ris}, as shown in Fig. \ref{distribution:wireless}
(d). Specifically, RIS is typically realized by planar or conformal
artificial metamaterials or metasurfaces equipped with a large number of
low-cost passive reflecting elements, which are capable of adjusting the
phase shifts and amplitudes of the incident signals for directional signal
enhancement or nulling,
and thus altering the propagation of the reflected signals \cite{Renzo_jsac20RIS,
Rui_TCOM21, DBLP:journals/wc/HuangHAZYZRD20}. To design an RIS-empowered edge training system, RIS can be leveraged
to align the magnitudes of the transmit signals  by establishing favorable propagation
links in waveform superposition  for AirComp, resulting in boosted received
signal power and accurate aggregated function at the edge server \cite{Yuanming_fang2021over}.
The boosted model aggregation via RIS can support efficient edge devices scheduling
in over-the-air FL, thereby adapting to the time-varying
local model updates and channel dynamics \cite{abs-2011-05051, Angela_arXiv20}.
The reliable sporadic access in edge training can be   developed
by establishing abundant propagation scatters using RIS for accurate activity
detection \cite{Xia_arXiv20RISaircomp}. The latency for local model
updates of the active devices can be further reduced by establishing favorable propagation links via RIS, thereby mitigating stragglers \cite{Tong_JSAC20}. \\

\subsubsection{Space-Air-Ground Integrated Networks}
The typical SAGIN \cite{Jiajia_CST18, Xiangming_JSAC18} provides an integrated
space information platform across the satellite networks (e.g., miniaturized
satellites
\cite{Saeed_CST20}), aerial networks (e.g., UAV
communications \cite{Rui_IEEEPro19}), and terrestrial communications (e.g.,
vehicular communications \cite{Haibo_IEEEPro20}) to provide ubiquitous connectivity
for various edge training architectures, as shown in Fig. \ref{distribution:wireless}
(e). Edge learning over a vehicle-to-everything
network is critical to enable autonomous driving with delay-sensitive
applications \cite{Mehdi_ComMag21}. In this scenario,  the local model updates need
to be fast and reliably aggregated within neighbors via vehicle-to-vehicle
communications \cite{savazzi2020federated}, or to the roadside units via
vehicle-to-infrastructure communication. In particular, radar sensing
provides a promising way to predict the vehicular links \cite{Fan_TWC20} and  holds the potential to provide real-time model aggregation via predictive
beamforming in the model aggregation procedure.  In the scenario with sparsely
deployed edge servers and moving edge devices (e.g., ground vehicles),
UAV, serving as the flying edge servers, can provide a  promising solution to aggregate local
model updates in the whole procedure of edge training by joint UAV trajectory and
transceivers design over dynamic wireless edge networks \cite{fu2021uav}.

To build a scalable edge training system with massive devices
participation for training extremely deep AI models \cite{bonawitz2019towards},
it is  critical to access abundant computation resources across the continuum
of nodes from edge devices, edge servers, to cloud servers \cite{seyyedali2020federatedtofog}. It was shown
in \cite{abs-2103-14272} that the client-server-cloud multi-layer architecture
is able to significantly reduce the training time and energy consumption.
In the scenario without abundant edge and cloud computing infrastructures,
SAGIN provides an ubiquitous computing platform for the  multi-layer hierarchical
edge learning system, where the flying UAVs serve as the proximal edge computing,
and the low earth orbit  satellites serve as the relays to the cloud computing
\cite{Sherman_JSAC19}. To realize SAGIN empowered edge training system, tier-adaptive
aggregation interval management becomes critical to control the local and
global model aggregation intervals \cite{abs-2103-14272} to achieve high
communication efficiency. Besides, the client-edge-satellite association
with dynamic scheduling and offloading is fundamental to tackle the heterogeneity
challenges in terms of system resources and network topologies.

In summary, this section presented multiple access technologies (e.g., AirComp, grant-free random access, NOMA), multiple antenna techniques (e.g., Cloud-RAN, cell-free massive MIMO, RIS), and multiple layer networks (e.g., UAV, SAGIN) that are needed to support low-latency model aggregation and diversified learning architectures and environments. We hope this can inspire more advanced 6G wireless and information techniques (e.g., millimeter-wave and terahertz (THz) communications \cite{Rappaport_IEEEPro14, Rappaport_THz19}, age of information \cite{Roy_JSAC21}) to support  edge AI systems for establishing integrated communication, computation and learning ecosystems.

\section{Communication-Efficient Edge Inference}
\label{edgeinference}

\begin{figure*}
 \centering
 \subfigure[Edge device distributed inference.]{
  \begin{minipage}[b]{0.42\textwidth}
   \centering
   \includegraphics[width=1\textwidth]{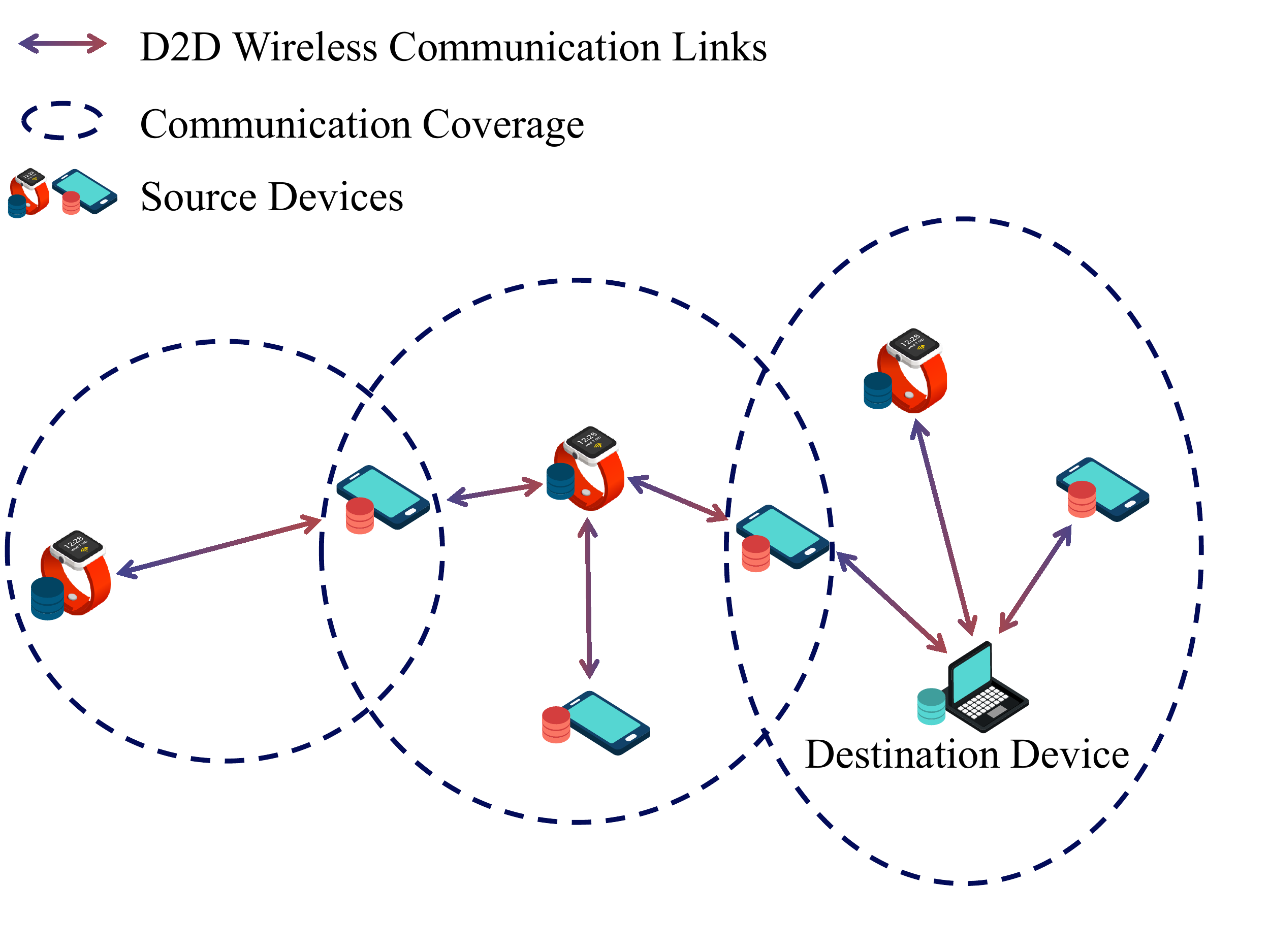}
  \end{minipage}
 }
 \subfigure[Edge server cooperative inference.]{
  \begin{minipage}[b]{0.42\textwidth}
   \centering
   \includegraphics[width=1\textwidth]{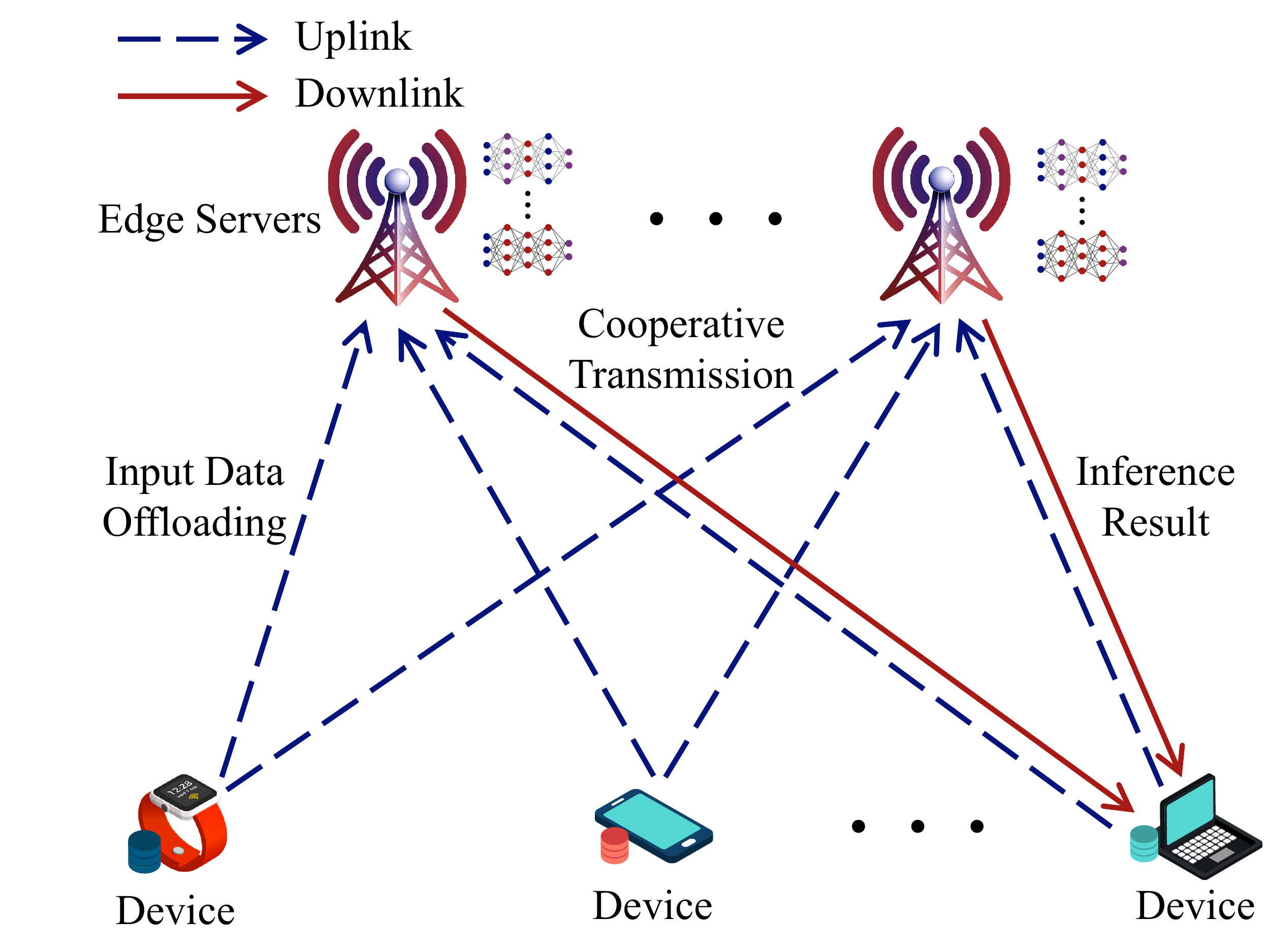}
  \end{minipage}
 }\\
 \subfigure[Edge device-server co-inference with single user.]{
  \begin{minipage}[b]{0.42\textwidth}
   \centering
   \includegraphics[width=1\textwidth]{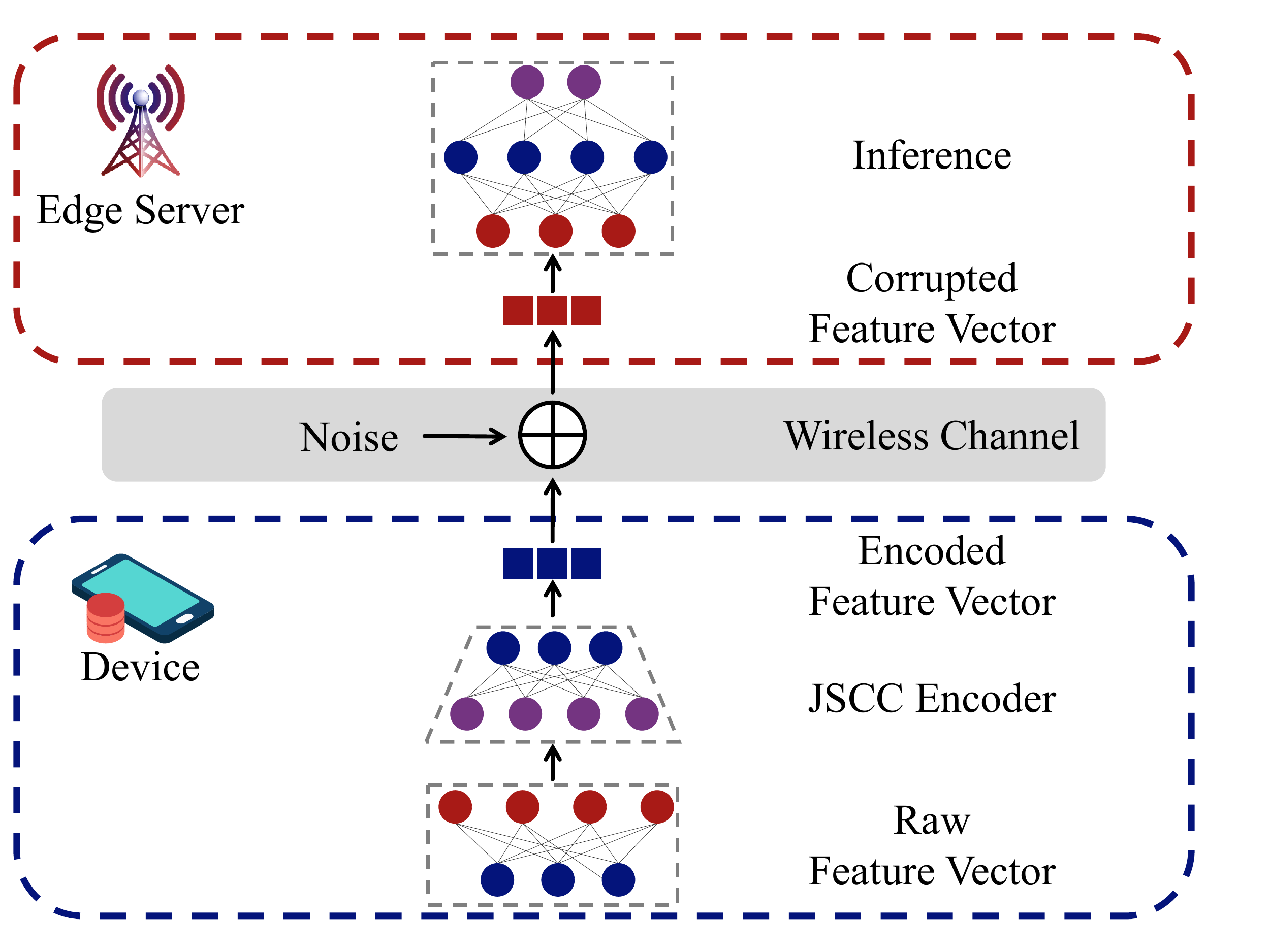}
  \end{minipage}
 }
 \subfigure[Edge device-server co-inference with multiple users.]{
  \begin{minipage}[b]{0.42\textwidth}
   \centering
   \includegraphics[width=1\textwidth]{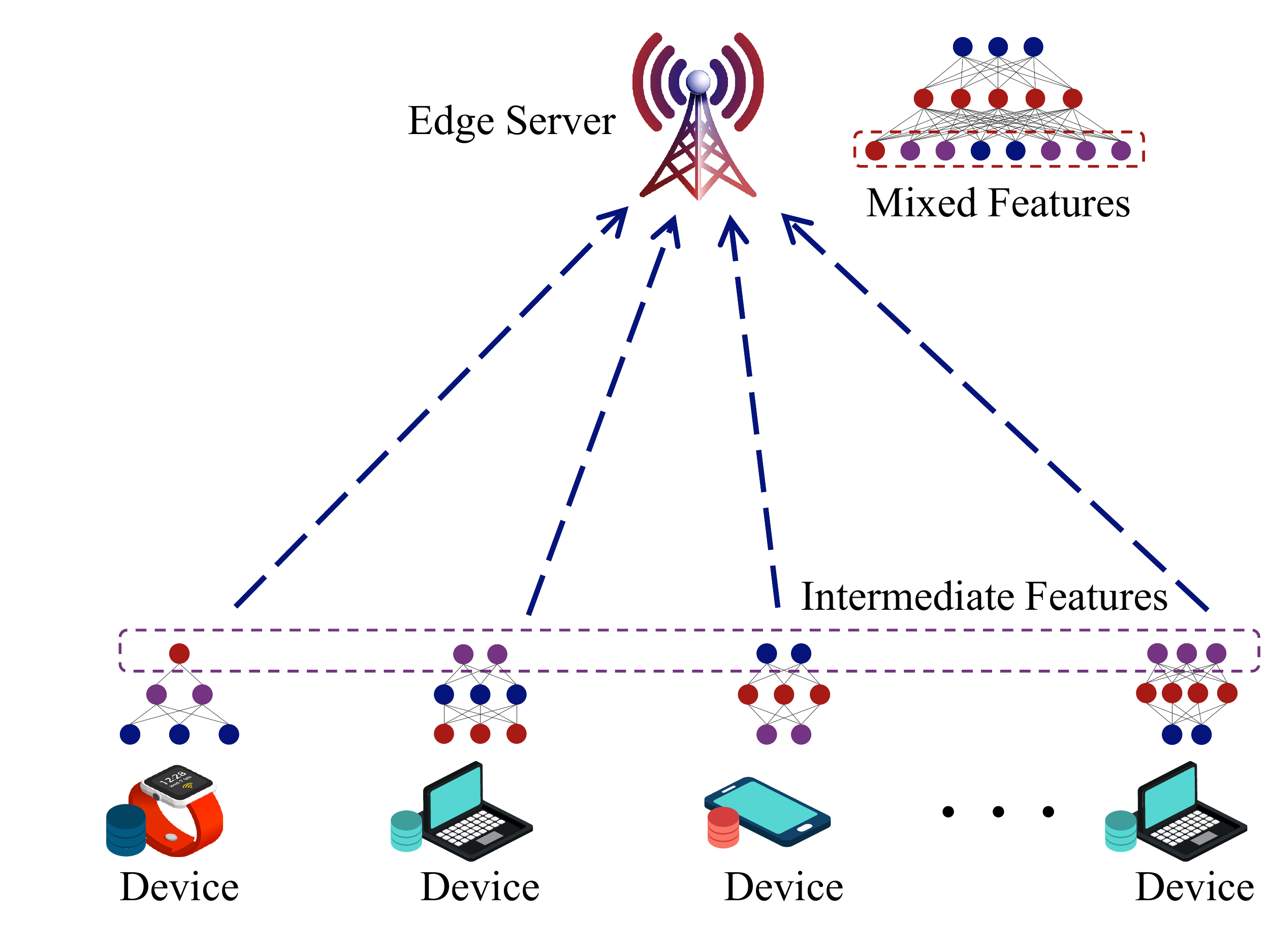}
  \end{minipage}
 }
 \caption{Communication-efficient edge inference systems.} \label{distribution:inference}
\end{figure*}

In this section, we present communication-efficient techniques for edge inference
tasks with latency and reliability guarantees. Based on the dataset distribution characteristics, Yang {\textit{et al.}} \cite{yang2019federated} proposed to categorize FL as horizontal FL (i.e., datasets share the same feature space but different  sample space) and vertical FL (i.e., datasets share the same sample space but differ in the feature space). Hosseinalipour {\textit{et al.}} \cite{seyyedali2020federatedtofog} further proposed a fog learning framework by allowing both vertical communications (i.e., model updates are only exchanged across different network layers) and horizontal communications (model updates can be exchanged between devices in the same network layer). In a similar way, based on different computing
collaboration schemes, we shall propose to categorize edge inference as horizontal
edge inference (i.e., computation resources can only be harvested among
 edge devices, or only be pooled among edge servers), and vertical edge inference (i.e., computation resources can be harnessed between edge devices and edge servers), which are discussed in the following two subsections, respectively.

\subsection{Horizontal Edge Inference}
We consider two different types of horizontal edge inference, as shown in Fig. \ref{distribution:inference}
(a) and Fig. \ref{distribution:inference}
(b).\\
\subsubsection{Edge Device Distributed Inference}Enormous efforts
on TinyML with DL model compression and neural network architecture search
have been conducted to enable low-latency and energy-efficient model inference
on a single  device with limited storage and computation resources \cite{Cheng_SPM18}.
However, due to limited storage capability at edge devices, it becomes extremely difficult to accomplish inference computation
tasks at a single device, for applications such as mobile navigation
with a huge map information dataset \cite{Maddah-TON17}. Edge device distributed
inference based on wireless MapReduce enjoys the advantages of providing
low-latency, high-accurate, scalable,
and resilient services for edge devices  without accessing the cloud data
center \cite{letaief2019roadmap, Yuanming_CST20}. Specifically, edge
device distributed inference involves computing the intermediate values
based on the local input datasets using the map function, followed by sharing
the intermediate values via {{horizontal communication}} among edge devices, thereby
constructing  the desired computation or inference results using the reduce
function \cite{dean2008mapreduce, Yuanming_TSP19MapReduce}.  

To tackle the communication bottleneck for shuffling intermediate values
 in the edge device distributed inference process, a coded distributed computing
approach \cite{Songze-TIT18} was adopted in \cite{Maddah-TON17} to improve
the scalability of wireless MapReduce by inducing the coded multicasting
transmission opportunities. This, however, sacrifices computation efficiency
as computation replication  of the local dataset is needed. To further improve
the spectral efficiency, instead of reducing the volume of communication bits \cite{Maddah-TON17},
a joint uplink and downlink design approach based on the interference alignment
principle was developed in \cite{Yuanming_TSP19MapReduce} to improve the communication
data rates for local intermediate values shuffling. In particular, to compute
the nomographic function \cite{Mario_TWC15} for edge device distributed inference based on  the MapReduce decomposition, a multi-layer hierarchical
AirComp approach was proposed in \cite{Vincent_arXiv21aircomp} to improve the spectral efficiency over the
multi-hop D2D communication network, as shown in Fig. \ref{distribution:inference} (a).\\

\subsubsection{Edge Server Cooperative Inference}
\label{esci} 
DL with high-dimensional
model parameters  is able to provide high accurate intelligent
services. However, it is  challenging to directly deploy such large AI models on IoT
devices  due to very limited onboard computation, storage and energy resources.
Deploying and executing
DL models on edge servers turns out to be a promising solution. However, the limited wireless bandwidth
between edge devices and edge servers becomes the key bottleneck \cite{yang20iot,
hua21serverinference} for edge server cooperative inference. Compressing and encoding the input source data at
edge devices are essential to reduce the uplink communication overheads, for which
various data dimensionality reduction approaches have been proposed by exploiting
the specific computation tasks and communication environments \cite{Yuanming_CST20}.
Besides, for the applications with high-dimension output inference results
(e.g., the output of the NVIDIA's AI system GauGAN is a large-sized photorealistic
landscape image), it is equally
important to design highly efficient downlink communication solutions for
delivering the output inference results for the edge devices \cite{yang20iot,
hua21serverinference}.  

Computation replication has been shown to be effective for reducing the communication
latency in computation offloading when the output size is large \cite{Tao_TWC20}.
This is achieved by executing each inference task at multiple edge servers,
followed by delivering the inference results for multiple edge devices via
downlink cooperative transmission \cite{Wei-JSAC10}. Although edge server
cooperative inference via downlink transmission cooperation is able to significantly
improve communication efficiency by mitigating interference and alleviating
channel uncertainties, it causes extra energy consumption to execute  the
same inference tasks at multiple edge servers. To design a green edge server
cooperative inference system,  joint inference task selection and downlink
coordinated beamforming framework was proposed in \cite{yang20iot} to minimize
the overall computation and communication energy  consumption, as shown in Fig. \ref{distribution:inference} (b). RIS was further leveraged in \cite{hua21serverinference}
to design  green edge server cooperative inference systems by considering both
 uplink and downlink transmit power consumption. The rate splitting method
is also anticipated to be able to further improve the energy-efficiency for edge
server cooperative inference by partially decoding the inference result and partially treating
it as noise in a flexible way \cite{Bruno_ComMag16}.

\subsection{Vertical Edge Inference}
\label{edsc}
We consider two different cases of vertical  edge inference, as shown in Fig. \ref{distribution:inference}
(c) and Fig. \ref{distribution:inference}
(d), with a single edge device and multiple edge devices, respectively. In the following, we shall first present effective techniques for communication-efficient vertical edge inference for these two cases, and then present a new general design principle for resource-constrained vertical edge inference, named \emph{task-oriented communication}.\\
\subsubsection{Edge Device-Server Co-Inference}
Edge device distributed inference enjoys low-latency whereas it has limited
accuracy due to limited processing capabilities and limited bandwidth. Although
edge server cooperative inference is able to achieve high accuracy with DL models, it may raise data leakage issue and excessive communication delay. It
thus becomes inapplicable for  privacy-sensitive and delay-sensitive applications.
To provide ubiquitous AI services across diversified application scenarios,
edge device-server co-inference, as a complementary solution to horizontal edge
inference, is promising to alleviate the communication overheads
while achieving high accuracy and privacy for inferring the DNN models. This is achieved by dividing the DNN model into a computational
friendly segment at the edge device, and the remaining computational heavily
segment  at the edge server \cite{shao20edgeinfer}, as shown in Fig. \ref{distribution:inference} (c). By adaptively partitioning
the computation burdens between the edge devices and edge server,  model split
selection for the neural network is essential to achieve optimal computation-communication
trade-off in the vertical edge inference system via edge device-server synergy and
collaboration  \cite{Xu_TWC20}. To further reduce the communication overheads,
a communication-aware model compression approach  was proposed in \cite{shao20edgeinfer}
to limit the number of the activated neurons at the last layer of neural
network deployed at the edge device. However, the short message transmission
\cite{park21commdlover} and data amplification effect \cite{Jun_ICASSP21}
of the output features extracted by the on-device split model raise unique
challenges to realize real-time vertical edge inference. \\

\subsubsection{Ultra-Reliable and Low-Latency Communication} The packet length
of the extracted output features transmitted from the edge devices can be very
short \cite{yury10codingrate, Mehdi_ComMag21}, for
which the achievable data rate in such a finite block length regime is penalized
by a non-vanishing decoding error probability \cite{yury10codingrate}.
 Besides, the output inference results from the edge server should be delivered
to the edge devices with latency and reliability
guarantees for  mission-critical applications. Considering the system
dynamics, including task arrival dynamics  in the network layer and the  wireless
channel dynamics in the physical layer, cross-layer optimization is needed
 to minimize the end-to-end delay for edge device-server
co-inference \cite{Mehdi_TWC19, Wei_TWC20}.  In particular, MDP supported by linear programming was adopted  in \cite{Wei_TWC20}
to jointly schedule the transmission at edge devices and computation at the edge
server for achieving the optimal power-latency tradeoff for edge device-server co-inference
via MEC. The random delay characteristics were also investigated in \cite{Jianhua_IoTJ21}  by modeling the coupled transmission and computation process as a discrete-time two-stage tandem queueing system. To support multiple edge devices for uploading intermediate
features using short packet transmission, massive MIMO  can
be adopted to combat channel fast fading and provide a nearly deterministic
communication environment due to channel hardening \cite{JSAC20_Poor}. The
received multiple intermediate features can be further aggregated via the
mixup augmentation technique \cite{verma2019manifold} to enable scalable
and cooperative inference  at the edge server, as shown in Fig. \ref{distribution:inference} (d).  \\

\subsubsection{Task-Oriented Communication} As revealed in \cite{shao20edgeinfer}, there exists an intrinsic communication-computation trade-off in resource-constrained vertical edge inference. This is mainly caused by the data amplification issue in DL based inference, namely the dimension of the intermediate feature may be larger than the input data size. Thus, if only a few layers of the neural network were deployed on the edge device, the output feature would have a size larger than the input data, yielding too much communication overhead. To reduce the intermediate feature size, more layers have to be deployed on the edge device, which however will lead to high local computation burden. To resolve this tension between local computation and communication overhead, it is of critical importance to effectively compress and transmit the intermediate feature. Such a communication task is fundamentally different from \emph{data-oriented communication} in current wireless networks, i.e., to transmit a binary sequence at the highest data rate for reliable reconstruction at the receiver. In vertical edge inference, the feature transmission is for the inference task, not for reconstructing the feature vector with high fidelity.  Thus, as advocated in \cite{shao20edgeinfer}, we should rather design the communication scheme for feature transmission in a task-oriented manner, i.e., only transmitting the informative messages for the downstream inference
task at the edge server. Instead of decoding the intermediate features, the
received signal corrupted by channel fading and noise is directly processed
at the edge server to obtain the inference results. 

This \emph{task-oriented communication} principle
constitutes a  paradigm shift for the communication system design from data
recovery to task accomplishment. It was first tested in vertical edge inference via end-to-end training with joint source-channel coding in \cite{ShaoZ20-1}, which helps to reduce both the communication overhead and on-device computation cost. Such design principle has also been applied in other tasks. For example, the DL based end-to-end semantic
communication system was developed in \cite{Qin_TSP21Semantic} via joint semantic
source and wireless channel coding for recovering the meaning of sentences
instead of the original transmitted data samples. The analog JSCC approach was presented in \cite{Deniz-JSAC_2021} to compress
and then code the feature vectors, followed by leveraging the received perturbed
signal directly for wireless image retrieval at the edge server via a fully-connected
neural network. Recently, a novel and generic design framework for task-oriented communication was developed
in \cite{abs-2102-04170}, which is based on the information bottleneck formulation \cite{physics-0004057}. This framework provides a principled way to extract informative and concise representation from the intermediate feature, which is made mathematically tractable via variational approximation. 
 Furthermore, it has been extended to the cooperative inference scenario  with multiple edge devices in \cite{shao2021taskoriented} based on  distributed information bottleneck \cite{aguerri2019distributed} and distributed  source coding theory. 

In summary, this section presented  interference coordination techniques and task-oriented low-latency communication principles for horizontal edge inference and vertical edge inference, respectively. We hope this can motivate the co-design of wireless communication networks and deep learning models to deliver low-latency, energy-efficient and trustworthy edge AI inference services.

\section{Resource Allocation for Edge AI Systems}
\label{resourceall}
In this section, we shall characterize the engineering requirements for designing  communication-efficient edge AI systems, including accuracy, latency, energy, privacy and security. Effective service-driven resource allocation methods based on mathematical programming and ML are then provided to achieve scalability and trustworthiness for edge AI systems.

\subsection{Engineering Requirements and Methodologies}
We identify the engineering requirements for designing scalable and trustworthy edge AI systems. Resource allocation strategies must cater to the needs of edge AI
systems for achieving accurate intelligence distillation into the edge network
at an ultra-low power and low-latency cost. \\ 
\subsubsection{Accuracy}
The edge training process involves
designing the global iterates  ${\bm{\theta}}^{[t]}$ with $t$ as the iteration
index to minimize the empirical loss function while achieving fast convergence
rates with negligible optimality gap for problem (\ref{elma}). To design efficient resource allocation schemes in edge
training systems, it is particularly important to characterize the convergence
behaviors for the global iterates ${\bm{\theta}}^{[t]}$, which typically depend
on the scheduled devices, local updates,
aggregation behaviors, network topologies, propagation environments,  function
landscapes, and underlying algorithms. Specifically, for  edge training systems via AirComp, the global model aggregation errors due to the wireless channel
fading and noise will cause learning performance degradation  \cite{AmiriG20,
Xia_arXiv20}. The optimality gap (i.e., the distance between the current
iterate and the desired solution), characterized by the convergence behavior of the global iterate, can be further controlled by various resource
allocation schemes, including edge devices transmit power control \cite{LiuS21,
xiaowen2021optimized}, edge server receive beamforming \cite{yang2020over,
Fan_arXiv21}, passive beamforming at RIS \cite{abs-2011-05051, Angela_arXiv20},
as well as device scheduling policy \cite{yang2020over, abs-2011-05051}. For
digital design of the edge training system, the optimality basically
depends on the edge devices selection,   packet errors in the uplink transmission,  and model parameter partition, for which user scheduling \cite{yang2020selection}, power control \cite{Mingzhe_TWC21}, batchsize selection \cite{Yu_JSAC21dnn}, aggregation frequency control \cite{Tuor-JSAC19},  and bandwidth allocation \cite{wen20joint} were provided to improve the accuracy in the edge training process.

For edge inference, the accuracy indicates the quality of the inference results for a given task. It is typically measured by
the number of correct predictions from inference, e.g., the classification tasks. For computer vision
applications in autonomous driving, ultra-high accuracy for the DNN model inference
is demanded. For applications in radio resource
allocation via distributed ML, the accuracy of inferring a DNN model can be moderate. The
accuracy of edge inference  depends on the  difficulty of the tasks and datasets, the quality of
the trained
model, the dynamics of wireless communication and edge computation
environments, as well as the methods for processing the models, datasets and features. In particular, for horizontal edge inference via AirComp aided wireless MapReduce, the accuracy for computing a nomographic function is fundamentally limited by  the channel fading and noise, for which various transceivers were designed to minimize the mean square error for inference computation tasks \cite{Vincent_arXiv21aircomp}. The accuracy of vertical edge inference  depends on the informativeness and reliability of the intermediate features transmitted from edge devices, as well as the dynamic wireless environments, for which an ultra-reliable communication and  adaptive JSCC approach need to be developed to improve the inference performance. In particular,  information bottleneck was adopted in \cite{abs-2102-04170} to characterize the relationship between the accuracy of the vertical edge inference and the communication overhead of the intermediate features. \\

\subsubsection{Latency}
     For edge training, the latency consists of computation latency and communication
latency. The computation latency  highly depends on the computation capability
of the edge devices and servers, as well as the size of the models and datasets.
The communication latency is the sum of the  transmission latency of one round with respective to the total learning rounds until convergence for training the global model. In one typical training round, the communication delay in the uplink and downlink transmissions for model updates, is mainly affected by the wireless
communication techniques, bandwidth and power budgets, wireless channel conditions, as well as the scheduled edge devices. Li {\textit{et al.}} characterized the delay distribution for FL over arbitrary fading channels via the saddle point approximation method and
large deviation theory \cite{li2021delay}. The trade-off between the convergence speed and the per-round latency was revealed in \cite{Niu-TWC21} based on the key observation that more scheduled devices yield faster convergence rate while prolonging the time of uploading the local updates at each iteration due to limited radio resources. A probabilistic device scheduling policy  was further proposed in \cite{mingzhe21convergence, chen2021communication} to minimize the overall training time in wireless FL. Besides, the trade-off between the local computation rounds for local model updates and the global communication rounds for global model updates is characterized to guide the resource allocation for minimizing the total learning time and energy consumption \cite{Dinh_TON21}. The convergence speeds of FL algorithms were characterized in \cite{Shen_JSAC21}  by considering non-identical dataset distributions, partial edge devices participation, and quantized model updates in both uplink and downlink communications.

In the case of edge inference, the latency measures the time between the data arrival to the generation of the inference results through the edge AI system.  It  consists of the data pre-processing,
data transmission, model inference, and result post-processing, which highly depend on the computation hardware, communication schemes, DL models and tasks. For the real-time
mobile computer vision application of AR/VR, stringent latency requirements are required,
e.g., 100ms. For  scalable
radio resource allocation application via DL, the inference latency must be within the channel
coherence time (e.g., 10ms) to yield a meaningful resource allocation decision \cite{yifei21gnn}. A low-rank matrix optimization  based transceiver design approach was proposed in \cite{Yuanming_TSP19MapReduce} for fast shuffling intermediate values   in wireless distributed computing, thereby reducing the latency for horizontal edge inference via edge devices collaboration. For vertical edge inference, the dynamic computation partition and early existing scheme was proposed in \cite{Xu_TWC20} to accelerate the inference speed via edge device-server synergy. The cross-layer design approach was adopted in \cite{Wei_TWC20} to reduce the communication and computation latency for the time-sensitive edge inference computing applications. In particular, the DL enabled task-oriented communication framework was developed  to achieve low-latency edge device-server co-inference by merging feature compression, source coding and channel coding for the specific inference tasks \cite{abs-2102-04170, Deniz-JSAC_2021}.  \\

\subsubsection{Energy}
     For edge training, the energy consumption consists of the computation
and communication process. For AlphaGo, it may cost 280 GPUs and a \$3000 electric
bill per game \cite{de2019machine}. It is therefore critical to design energy-efficient edge training
systems to minimize carbon dioxide footprint for contributing the carbon neutrality target. Such a design is mainly dictated by the size of training models, model training
algorithms, and wireless transmission strategies and hardware (e.g., the scaled SiGe bipolar technology \cite{SiGe_18}), and edge computing architectures and hardware. Both  computation energy consumption for local model updates and communication energy consumption for uploading local updates are simultaneously minimized in \cite{Zhaohui_TWC21} by considering the the learning latency and accuracy constraints for wireless FL. The wireless power transfer approach was further adopted in \cite{Kaibin_arXiv21} to power the edge devices for local model computation and communication, for which the active devices with enough harvested energy will contribute to accelerate the learning procedure. To deploy AirComp-assisted FL across massive IoT devices with a limited battery capability,  microwave based wireless power transfer supported by RIS was adopted in \cite{Yuanming_IoT21} to recharge the IoT devices  via energy beamforming at edge server and passive beamforming at RIS.

In the case of edge inference, it becomes particularly important to achieve high energy efficiency for processing the DNN models at the network edge with battery-limited devices. The energy consumption
of executing a DNN model is highly dictated by the computation architecture and
methods (e.g., ultra-low power compute-in-memory  AI accelerator) at the edge
computation nodes \cite{Sze_SSCM20}, the architecture of DNN models \cite{Sze_IEEEPro}, and the wireless transmission for data
exchange during the model inference procedure. For horizontal edge inference via wireless cooperative transmission at multiple edge servers, the sum of the computation and  transmission power consumption for generating and delivering the inference results were minimized via downlink coordinated beamforming \cite{yang20iot}. Energy consumption at the edge devices can be minimized in the cross-layer design for delay-sensitive edge device-server co-inference by computation offloading \cite{Wei_TWC20}. Besides, energy harvesting  becomes a promising technology for the edge computing based vertical edge inference by providing renewable energy resources for  edge devices \cite{Khaled_JSAC16}.\\       
     
\subsubsection{Trustworthiness}
     Trustworthiness is one of the main drivers for developing the next
generation AI technologies. Specifically, the developed AI models and algorithms must be privacy-preserving, adversarial-resilient, robust, fair, optimal and interpretable \cite{Samek_IEEEPro21}.  
For edge training, privacy mainly depends on
the offloading or coding of the raw data and intermediate features. Keeping datasets at devices is
a direct and effective way to preserve user's privacy in FL. Besides, the wireless channel noise yields a noisy model aggregation procedure via AirComp, which provides an inherent
privacy-preserving mechanism to enhance differential-privacy for
each edge device. An adaptive power control method was further developed in \cite{LiuS21} to control the differential-privacy levels in this over-the-air FL system, while avoiding the learning performance degradation.
To address the adversarial attacks, the blockchain based decentralized learning was proposed in \cite{Qu_IoTJ20} to enable secure global model aggregation  by using a consensus mechanism of blockchain. The block generation rate was optimized by considering the communication, computation and consensus delays in the blockchain enabled secure edge learning systems \cite{Choi_TCOM20fl, Qu_IoTJ20}.      
For edge inference, privacy
and security are mainly dictated by the way of processing the input data, of transmitting the inference results, as well as the computation methods
for model inference (e.g., secure multi-party computation).

Establishing optimality for ML algorithms is important to deliver reliable and responsible AI  services. However,  empirical risk minimization for training the models is usually nonconvex, which poses significant challenges to guarantee global optimality for the learning algorithms and models \cite{IEEEProc21_Tong}. Fortunately, under the high-dimensional statistical setting, the local strong convexity and smoothness of the nonconvex loss functions can be exploited to tame the nonconvexity for various learning models, e.g., blind demixing \cite{Shi_TSP18}, phase retrieval \cite{Yuxin_TSP19}, and shallow neural networks \cite{Yuanming_ICASSP19}. Besides, with high-dimensional datasets, the nonconvex loss functions of certain statistical learning models, including over-parameterized neural networks \cite{IEEEProc21_Tong} and dictionary learning \cite{Ju-TIT17}, can enjoy benign global geometric landscape such that all the local minima are  global minima, and all the saddle points can be escaped efficiently using the algorithms including trust region method and perturbed gradient descent method \cite{jin2017escape}. In particular, for edge training, the channel noise yields a perturbed stochastic gradient descent method to  escape saddle points for distributed principal component analysis via AirComp \cite{abs-2104-10095}. Therefore, channel noise can provide a mechanism for both preserving differential privacy \cite{LiuS21} and achieving global optimality \cite{abs-2104-10095}. These evidences indicate that we should embrace channel fading and noise for achieving trustworthy edge AI.    \\

\subsubsection{Service-Driven Resource Orchestration} Edge AI systems  need to incorporate various wireless network architectures and communication strategies by integrating communication and computation.  This will result in a  highly complex and dynamic network, which  requires innovative technologies and solutions. Various use
cases (e.g., autonomous driving, industrial IoT, and smart healthcare) and heterogeneous requirements in terms of accuracy, latency, energy and trustworthiness, would further aggravate the complexity for resource allocation in edge AI systems. Besides, the complex edge servers and base stations will be quite energy-consuming, which brings formidable
challenges for achieving high energy efficiency. To enable efficient resource allocation, it is thus critical to precisely model the heterogeneous demands for edge AI services, and reversely matching them with proper network resource orchestration. This, however, relies on the quantitative relationship between network resources and user requirements for edge AI tasks.  To pave the way for this paradigm shift for service-driven resource allocation in edge AI systems, in the next subsection, we shall provide various intelligent optimization models and algorithms to adapt to diversified network environments and services.                  

\subsection{Optimization Models and Algorithms}
\begin{figure}[t]
 \centering
 \includegraphics[width=0.95\linewidth]{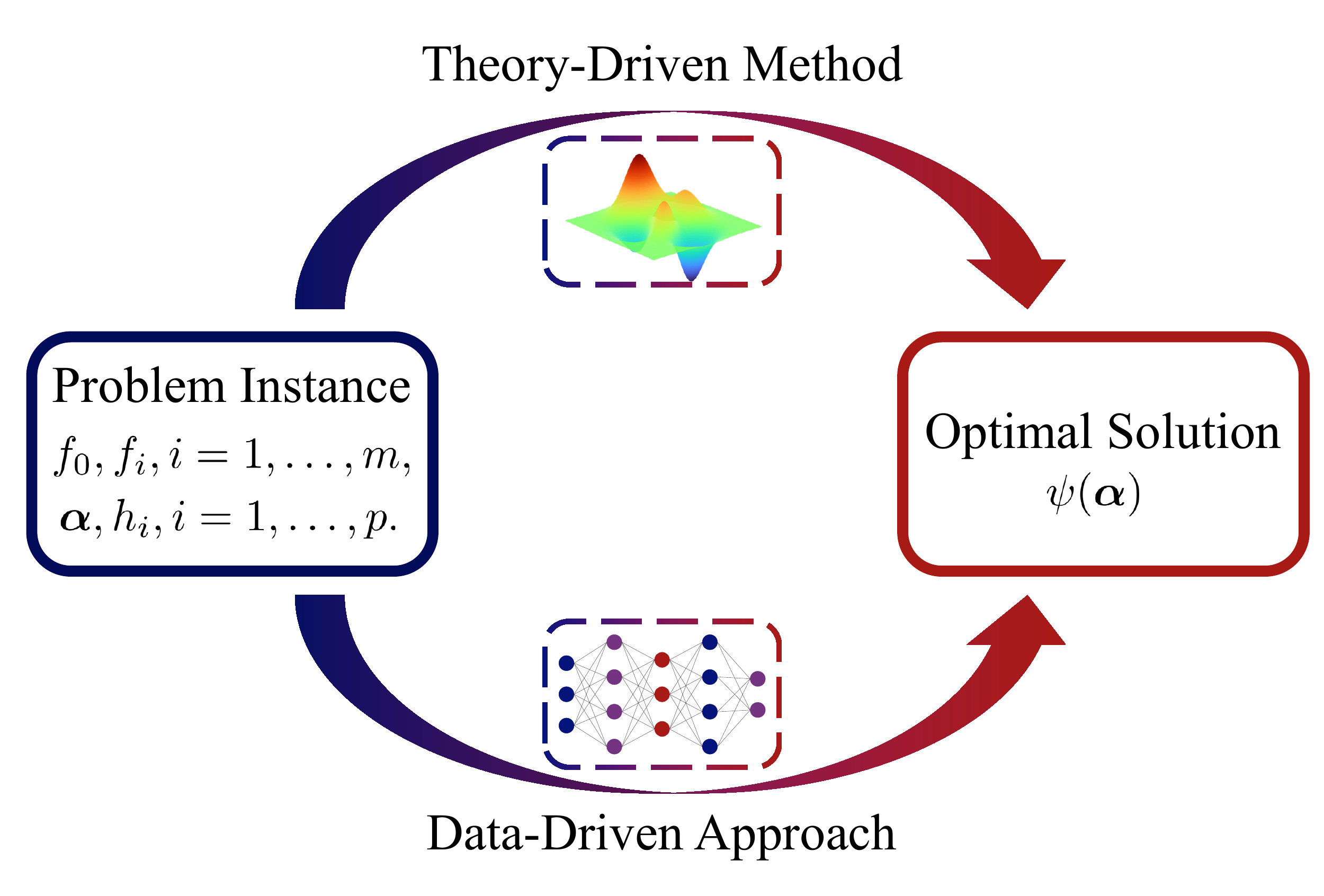} 
 \vspace{-4mm}
 \caption{Resource allocation optimization methods for edge AI systems.}\label{learning2optimize}
 \vspace{-4mm}
\end{figure}
The service-driven network resource management problems for edge AI systems can be classified as a parametric family of mathematical optimization problems:
\setlength\arraycolsep{2pt}
\begin{eqnarray}
\label{opt}
\mathop {\sf{minimize}}_{{\bm{z}}}&& f_0({\bm{z}}; {\bm{\alpha}})\nonumber\\
{\sf{subject~to}}&& g_i({\bm{z}}; {\bm{\alpha}})\le 0,
i=1,\dots, m,\nonumber\\
&& h_i({\bm{z}}; {\bm{\alpha}})=0, i=1,\dots,
p,
\end{eqnarray}
where ${\bm{z}}\in\mathbb{R}^{n}$ is the optimization variable vector consisting of both discrete and continuous variables, ${\bm{\alpha}}\in\mathcal{A}$  is the problem parameter vector with $\mathcal{A}$ denoted as the parameter space (e.g., CSI). For each fixed ${\bm{\alpha}}\in\mathcal{A}$, $f_0:\mathbb{R}^n\rightarrow\mathbb{R}$ is the objective function (e.g., optimality gap in edge training), $g_i:\mathbb{R}^n\rightarrow\mathbb{R},i=1,\dots, m$ are the inequality constraint functions (e.g., latency requirements in edge inference), and $h_i:\mathbb{R}^n\rightarrow\mathbb{R},i=1,\dots, p$ are the equality constraint functions. The resource allocation optimization problems  are typically categorized as mixed-combinatorial optimization, nonconvex continuous optimization, stochastic optimization, and end-to-end optimization. To provide scalable, real-time, parallel, distributed and automatic resource allocation schemes, we shall propose to exploit the landscape of the underling optimization problems (\ref{opt}) by the theory-driven method based on mathematical programming, followed by developing the novel data-driven approach based on machine learning to achieve real-time and distributed implementations, as well as improved and robust performance, as shown in Fig. \ref{learning2optimize}. Here, $\psi({\bm{\alpha}})$ is a mapping function to map the problem
parameter ${\bm{\alpha}}$ to the optimal solution of problem (\ref{opt}).\\

\subsubsection{Mixed-Combinatorial Optimization} The resource allocation problems in edge AI systems involve optimizing across learning, computation and communication. Specifically, for edge training systems, we need to jointly optimize the subcarrier and bandwidth allocation \cite{wen20joint, Mingzhe_TWC21, mingzhe21convergence}, transmit power and receive beamforming  \cite{yang2020over,
abs-2011-05051, Angela_arXiv20}, passive beamforming at RIS \cite{abs-2011-05051, Angela_arXiv20},
device  selection \cite{yang2020over, Niu-TWC21} and activity detection \cite{Tao_IoTJ19}, local updates computation \cite{Zhaohui_TWC21}, and global aggregation
frequency control \cite{abs-2103-14272}, thereby reducing the optimality gap and energy consumption in the distributed learning procedure. For edge  inference via collaboration among edge servers, task selection, coordinated downlink beamforming among edge servers, as well as passive beamforming at RISs were jointly optimized to achieve green edge inference \cite{hua21serverinference, yang20iot}. All of these resource allocation schemes can be formulated as a mixed combinatorial optimization problem, which needs to jointly optimize continuous-valued variables (e.g., beamforming and power control) and discrete-valued variables (e.g., device selection and subcarrier allocation). In particular, sparse optimization provides a powerful modeling approach to solve the mixed combinatorial resource allocation problems by exploiting the sparsity structures in the optimal solutions \cite{ShiZCL18}. For instance, the group sparsity can represent the combinatorial variables for edge devices selection in FL \cite{yang2020over}, edge devices activity detection   \cite{Tao_IoTJ19}, and inference tasks selection \cite{yang20iot}. The algorithmic advantages of the sparse optimization modeling approach are supported by various convex relaxation algorithms \cite{shi2015large, Yuanming_JSAC16sgsbf}, e.g., mixed $\ell_1/\ell_2$-norm
minimization \cite{Yuanming_TWC14gsbf}. {\rev{A typical sparse and low-rank optimization modeling and algorithmic framework was developed in \cite{yang2020over} to support the joint device selection and transceiver design  for improving learning
performance in over-the-air FL systems.}}

Although  operation research provides a theory-driven approach for solving the mixed combinatorial optimization problem or its equivalent sparse optimization problem,  the existing algorithms are either heuristic with noticeable performance loss or optimal with intolerably high computation complexity. To address these challenges, ``learning to optimize" provides a data-driven design paradigm to improve the computation efficiency and system performance for resource allocation \cite{chen2021learning, bengio2020machine}. This is achieved by developing computationally efficient optimization methods by learning from the sampled problem instances using training models and methods. The learned algorithms can be furthered executed online and distributed for real-time resource allocations in edge AI systems. To solve the large-scale mixed combinatorial optimization problem efficiently, imitation learning  was adopted in \cite{yifei20lorm} to learn an aggressive pruning policy in the globally optimal-achieving branch-and-bound algorithm. This learning based brand-and-bound method can significantly save the time for pruning the nodes in the search tree, achieve near-optimal performance with few training samples, as well as guarantee feasibility of constraints without performance degradation. To further speed up the sparse optimization method for the mixed combinatorial optimization problem in edge device activity detection, the DNN based algorithm unrolling framework  was developed in \cite{shi2021} to  achieve theoretical guarantees, performance improvements, interpretability and robustness for the learned sparse optimization algorithms \cite{monga2021algorithm}. This is achieved by mapping the theory-driven iterate operations, i.e., iterative shrinkage thresholding algorithm, into an unrolled recurrent neural network, followed by training the model parameters based on  supervised learning. {\rev{Besides, a multi-agent RL approach was developed in \cite{Zhang_TWC21} to solve the distributed mixed combinatorial optimization problem for task offloading and resource allocation in multi-layer edge inference systems.}} \\

\subsubsection{Nonconvex Optimization} Most of the resource allocation problems in edge AI need to solve a series of nonconvex optimization problems, e.g., nonconvex sparse optimization for device selection in wireless FL, nonconvex quadratic programming for transceiver design in over-the-air FL \cite{yang2020over}, low-rank matrix optimization for interference management in edge device distributed inference \cite{Yuanming_TSP19MapReduce}, and unit modulus constrained phase shifts optimization \cite{feng2021optimization} in RIS-empowered edge AI systems \cite{abs-2011-05051, hua21serverinference}. Convex approximation provides a natural way    to design polynomial time complexity algorithms for nonconvex programs based on the principle of  majorization-minimization \cite{SunBP17} or successive convex approximation \cite{marks1978general}. A two-stage
 framework was provided in \cite{Yuanming_TSP15} for solving general large-scale convex programs with infeasibility detection and  scalable computation. This is achieved by  matrix stuffing technique for fast conic program modeling in the first stage, and operator splitting method for scalable conic program solving in the second stage \cite{Yuanming_TSP15}. Although the semidefinite relaxation approach \cite{Luo_SPM10} is able to convexify the general quadratic programs by matrix lifting and dropping the resulting rank constraints, it fails to return high quality solutions in the high-dimensional settings. This issue was addressed by a difference-of-convex-functions (DC) programm  \cite{gotoh2018dc, yang2020over, Yuanming_TSP19MapReduce} by representing the rank function via an equivalent DC function. {\rev{This DC optimization modeling and algorithmic framework was typically applied to solve the nonconvex passive beamforming
problem in the RIS-empowered FL systems \cite{abs-2011-05051} and edge inference
systems \cite{hua21serverinference}.}}
To solve the large-scale rank constrained matrix optimization problems, Riemannian manifold optimization was proposed to optimize such nonconvex programs directly  by exploiting the manifold geometric structures of fixed-rank matrices \cite{boumal2014manopt, Yuanming_TWC16MC}.

To further enable real-time, automatic and distributed design of nonconvex optimization algorithms for resource allocation in edge AI systems, DL was shown to have great potentials for achieving this goal. A multi-layer perceptron  was adopted in \cite{Mingyi_TSP18} to directly learn the mapping from the problem instance to the output solution generated by the  weighted minimum mean square error (WMMSE) algorithm for nonconvex precoding design \cite{TomLuo_TSP11}. Instead of running the iterates, the learned algorithm via  deep learning can be executed in real-time, as neural networks only involve computationally cheap operations, e.g., matrix-vector multiplication. To reduce the model and sample complexity, as well as improve the performance and interpretability, unfolded neural networks  were developed in \cite{Chowdhury_TWC21,
Ding_JSAC21unfolding, Ding_TWC21unfolding} to parameterize  the iterative policy  via unfolding one iteration of the existing structured algorithm into one layer of a neural network. Graph neural network (GNN) has recently been shown to be able to harness the benefits of generalizability, interpretability, robustness, scalability, superior  performance, real-time and distributed implementation for   learning to optimize nonconvex problems, including power control \cite{Eisen_TSP20}, beamforming \cite{yifei21gnn}, and phase shift design \cite{WeiYu_JSAC21}. This is achieved by modeling wireless network as a graph, followed by using a GNN to parameterize the mapping function $\psi({\bm{\alpha}})$ for the optimal solution.  \\

\subsubsection{Stochastic Optimization} In large-scale edge AI systems, the estimated CSI will be inevitably imperfect
or  partially available \cite{Wadu_TCOM21, TWC_21Amiria}. It is thus critical to design practical resource allocation schemes by considering the CSI uncertainty, for which robust optimization and stochastic optimization are two typical approaches. Specifically, robust optimization approach aims at guaranteeing the worst-case but conservative performance over the uncertainty set. The robust optimization method can usually yield computationally tractable optimization models \cite{Letaief_TSP15}. The stochastic optimization approach, e.g., chance constrained programming, only relies on the probabilistic description of the uncertainty of the problem parameter $\bm{\alpha}$ in problem (\ref{opt}) and is able to provide a trade-off between conservativeness and probabilistic guarantees for the achievable performance \cite{Yuanming_TSP15ccp}. In particular, a statistical learning approach was presented in \cite{yang20iot} to learn a tractable uncertainty set to approximate the chance constrained programming for achieving high  computation efficiency and system performance {\rev{in the energy-efficient edge inference systems}}. However, due to the limited historical samples, it is difficult to characterize the true probability distribution for the CSI uncertainty. Distributionally robust optimization \cite{esfahani2018data}  provides a promising way to achieve worst-case probabilistic performance by incorporating all sample-generating distributions into an ambiguity set. However, finding the globally optimal solution for this method is often computationally intractable.   

DL provides an alternative way to address the uncertainty and dynamics of environment parameters to achieve modeling flexibility and computational efficiency for resource allocation in the complicated edge AI systems. Specifically, DL can provide acceptable performance for resource allocation based only  on geographic locations information of the transmitters and receivers \cite{Wei_JSAC19}.  By considering CSI variations  \cite{Julian_TWC20, Dongning_JSAC19, Yuen_JSAC20} and stochastic task arrivals   \cite{Wei_TWC20, Mehdi_TWC19}, the dynamic communication and computation resource allocation problem can be formulated as a MDP, for which deep RL, a model-free approach,  can provide  efficient and robust solutions \cite{lee2020optimization}. Besides, the learned algorithms can be distributively executed in the multi-agent edge AI systems. However, due to the distribution
shift  for system parameters in episodically dynamic environment, the trained model may suffer
from performance deterioration when the dataset follows a different distribution
in the inference stage \cite{yifei20lorm}. Transfer learning \cite{Tomluo_TWC21} and continual
learning \cite{Mingyi_Hong} have recently been adopted to address such task
mismatch issue in the ``learning to optimize"  framework considering the system distribution dynamics. \\

\subsubsection{End-to-End Optimization} Channel estimation plays a  pivotal role to support effective resource allocation in large-scale edge AI systems \cite{shi2015large, Yuanming_wc21ris}. In particular, exploiting the low-dimensional structures of wireless channels becomes a promising way to address the curse of dimensionality for CSI acquisition in various networks.  
Specifically, in ultra-dense Cloud-RAN,  a high-dimensional structured channel estimation framework was proposed  in \cite{Yuanming_TWC18csi} by inducing the spatial sparsity and temporal correlation prior information using a convex regularizer. Sparsity structures of a massive MIMO channel was exploited in \cite{Jun_TWC16csi} to reduce the training overheads for CSI acquisition.
The signal superposition property of a wireless multiple access channel was exploited to directly obtain the weighted sum of   channels for receive beamformer design, thereby avoiding global CSI estimation \cite{Yu_TCOM20AirComp}. The sparsity in the activity pattern  was leveraged to develop the sparse signal processing framework for joint activity detection and channel estimation in grant-free massive access \cite{liu2018massive}. Due to the passive nature of RIS, it becomes infeasible to directly perform signal processing for channel estimation at RIS and the  cascaded channel can only be estimated either at the edge servers or edge devices \cite{Yuanming_wc21ris}. To address this unique challenge, the common reflective channels among all edge devices \cite{Cui_TWC20}, quasi-static property between RIS and edge server channel links \cite{Xiaojun_jsac20, Dai_TWC21}, {\rev{spatial features of noisy channels and additive nature of noises \cite{Chang_TWC21}}}, as well as  channel sparsity \cite{Xiaojun_CL20} and device activity sparsity \cite{Xia_arXiv20RISaircomp}, were exploited to reduce the training overhead. 

However, all of the above works follow the ``estimate-then-optimize" framework by  first performing pilots-based channel estimation, followed by allocating resources based on the estimated CSI. However, this two-stage approach fails to achieve a low  signaling overhead and superior system performance. Although the low-dimensional structures have been exploited for designing efficient channel estimation methods, the additional information (e.g., user location and mobility), are difficult to be modelled and incorporated into a unified mathematical model for CSI acquisition overhead reduction, which may exceed latency. Besides, the artificially  defined criterion (e.g., mean square error) for channel
estimation may not be aligned with the ultimate  goal
for resource allocation in edge AI systems. To address this challenge, a DL approach has
recently been proposed to merge the two stages into an ``end-to-end optimization"
framework for resource allocation \cite{WeiYu_TWC21}. This is achieved by
directly mapping the received pilots (i.e., the problem parameters $\bm{\alpha}$ in (\ref{opt}) can be the received pilots) into the resource allocation policy
without explicit channel estimation. This mapping function is further parameterized
by a DNN to capture the inherent structures of the resource allocation problems.
For instance, the GNN was adopted to model the permutation invariant and equivalent
properties  of the mapping function for resource allocation in the RIS empowered
TDD wireless networks \cite{WeiYu_JSAC21}. The neural calibration approach \cite{ Letaief_ICC21nc} was developed in FDD massive MIMO systems to map the received pilots at edge devices  into feedback bits, followed by directly mapping the feedback bits  into the downlink beamformers \cite{ WeiYu_TWC21}. 

In summary, this section presented the  operation
research based theory-driven  and ML based data-driven methods for designing effective, real-time, distributed and robust resource allocation strategies in edge AI systems. We hope these results can stimulate more service-driven resource allocation methods (e.g., network slicing  \cite{Afolabi_CST18}) and optimization approaches (e.g., multi-objective optimization \cite{Emil_SPM14}). The presented ``learning to optimize" framework is also promising for resource allocation in  various future wireless networks.

\section{Architecture for Edge AI Systems} 
\label{archedgeai}
\begin{figure*}[t]
 \centering
 \includegraphics[width=0.95\linewidth]{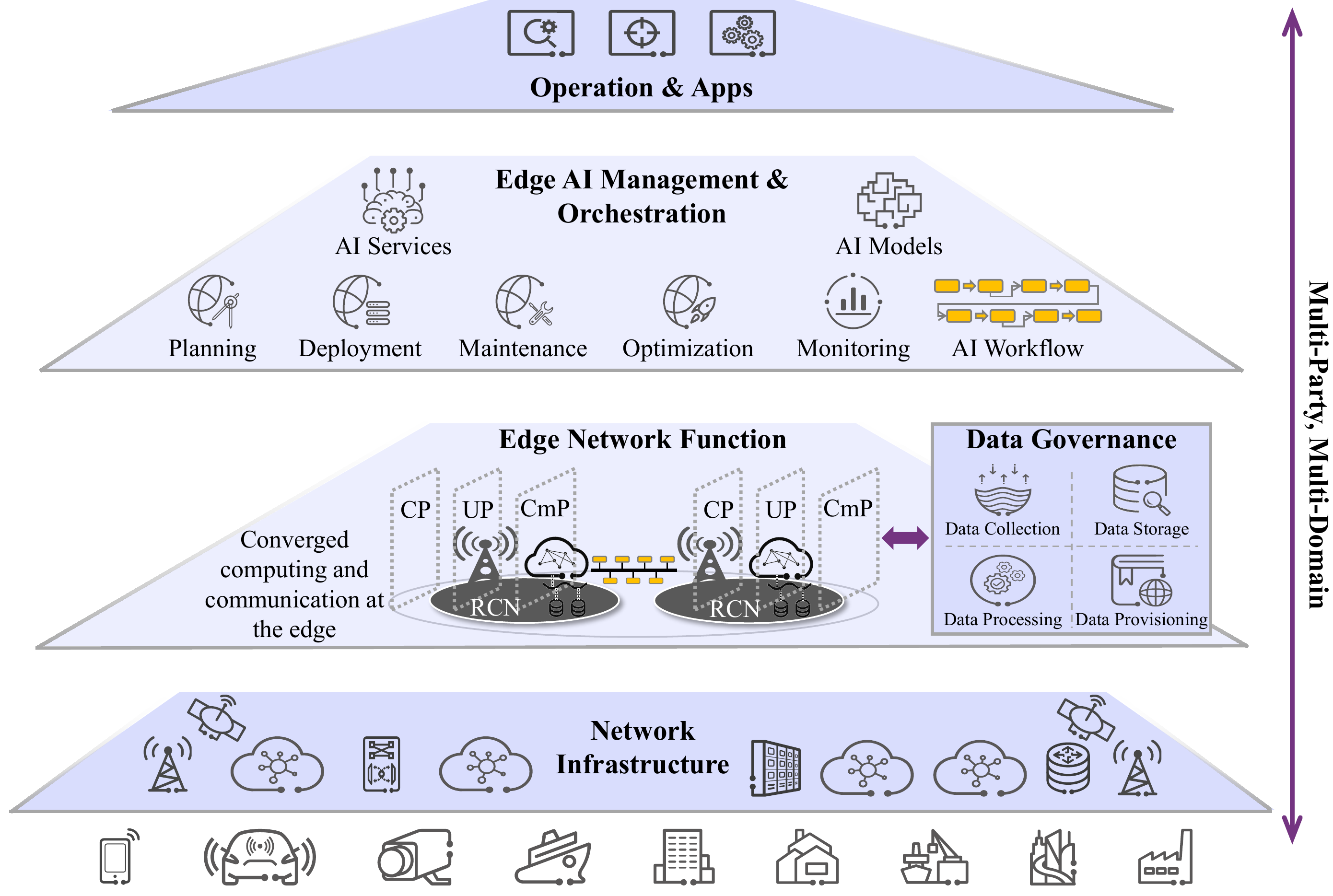} 
 \caption{E2E architecture for edge AI systems with radio computing nodes (RCN) to allow seamless integration of communication and computing capabilities. New independent computing planes (CmP) in RCN will also be used to host AI tasks and collaborate with communication functions in the control $\&$ user planes (CP and UP).}\label{architecture}
\end{figure*}
In this section, we present a new mobile network architecture for edge AI systems, supported by the wireless network infrastructures  in  Section {\ref{edgetraining}} and Section {\ref{edgeinference}}, as well as the service-driven resource allocations in Section \ref{resourceall}. We will provide an end-to-end (E2E) architecture design  across the network infrastructure, data governance, network function, network management,  as well as operations and applications.

\subsection{End-to-End Architecture for Edge AI Systems}
For each new generation of mobile networks, new services and capabilities have been introduced at the architecture level in order to meet more and typically more stringent demands. The mobile network was originally designed to deliver voice services. Since then, both the architecture and deployment of mobile networks have followed a centralized and hierarchical paradigm that reflects the nature of voice traffic and packet traffic of the mobile internet. To realize the vision of ``connected intelligence", 6G will break and shift these traditional paradigms towards a novel architecture and design that meet new requirements for the deep integration of communication, AI, computing, and sensing at the network edge with new integrated  capabilities empowered by  evolutionary, as well as, revolutionary enabling technologies.

 Under this new design philosophy, we introduce a holistic E2E architecture for scalable and trustworthy 6G edge AI systems, as illustrated in Fig. {\ref{architecture}}. By providing new wireless network infrastructures,
enabling efficient data governance, integrating communication and computation
at the network edge, as well as performing automated and scalable edge AI
management and orchestration, the proposed E2E architecture will provide a scalable and flexible platform to support diversified edge AI applications with heterogeneous service requirements.      

\subsection{Data Governance} 
Due to the expected huge energy consumption, as well as, security and privacy concerns, we envisioned that data in future 6G networks need to be collected, processed, stored and consumed at the the network edge. Since data and AI applications in 6G are expected to be much more diverse than ever before, it is incentive that there will be a provision for  a unified and efficient data governance framework at the architecture level. Data governance goes far beyond conventional data collection and storage, which will also consider the data availability and quality, data sovereignty, knowledge management and legal implication. Data governance  also must consider the mechanism to comply with the regional or national data protection policies and regulations of the data source in terms of usage rights and obligations such as GDPR.\\ 

\subsubsection{Independent Data Plane} 5G has introduced a new network data analytics function (NWDAF) in the core network  to implement AI-based network automation, optimize the related network functions (e.g., AI-based mobility management \cite{Shencong_JSAC16}), and improve user service experience, etc. One of its main goals is to collect and analyze data from other 5G network elements to train AI models and implement AI inference for automated and scalable network optimization. Meanwhile, similar mechanisms such as collecting   and analyzing data based on the existing SON/MDT (self-organizing networks and minimization of drive tests), was adopted for 5G radio access networks (RAN). In 6G, such a separated data collection and analytics mechanism needs to evolve to a unified and more efficient paradigm.   
An independent data plane in 6G could contribute to organizing and managing data efficiently while also considering privacy protection  \cite{ tong20216g}. This paves the way for natively embedding edge AI into the 6G networks by leveraging multi-domain data.    \\ 

\subsubsection{Multi-Player Roles} The data governance ecosystem includes different roles: data customer, data provider, data owner and data steward, etc. These could be taken by the same or different business entities, including individual users. Hence, data governance is a typical scenario that involves multiple players. It  thus becomes essential to establish a multi-party data trading platform to negotiate data rights and prices among different business entities while achieving trustworthiness, fairness and efficiency.  This can be achieved using decentralized technologies such as blockchain with smart contracts design \cite{tong20216g, Dai_TIFS20}. This will improve data efficiency and business ecosystem for the deployment of edge AI.

\subsection{Deeply Converged Communication and Computing at the Edge} 

In 5G, the superior performance has been achieved by leveraging the AI capability into RAN \cite{qIAN_CST18, Liang_CST19}. For instance, we can optimize radio resource scheduling and mitigate interference using machine learning methods \cite{yifei21gnn}. Such utilization of AI in 5G can be refereed to as  {\textit{AI for networks}}.
The targets of edge AI are not only AI for networks, but also {\textit{networks for AI}} \cite{letaief2019roadmap}, as presented in Section {\ref{edgetraining}} and Section {\ref{edgeinference}}. This will depend on the new functional capabilities of future networks, including how to make  computing as a foundational capability of future 6G networks. A new type of radio equipment may emerge,  which we refer to   as a radio computing node (RCN), which allows the computing resources to be seamlessly converged with the communication capability.   This will  require the introduction of a new independent computing plane (CmP) in RCN to host AI tasks and collaborate with the communication functions in the control plane (CP) and user plane (UP) \cite{tong20216g}. This will also enable the flexible integration of computation, communication and intelligence for edge AI.    

\subsection{Edge AI Management and Orchestration}
Edge AI involves a diverse set of learning models and algorithms, network infrastructures, as well as complicated collaboration for communication, computation and intelligence.  
Developing a  framework for edge AI management and orchestration thus becomes an essential aspect for the design of the native AI support at the architecture level. This framework needs to be designed so as to facilitate the seamless integration and deployment of AI services, especially from third-parties. This can be achieved by planning, deploying, maintaining, and optimizing the decentralized machine learning models and algorithms, as well as the edge network infrastructures and functions. The edge AI management and orchestration shall also include AI workflow, distributed and streaming
data, along with heterogeneous network resources, etc. Scale and cross-domain issues will be huge challenges for such a framework and this may involve complicated standardization efforts. Hence, building such new framework which will fully rely on standardization may not be feasible. We may instead leverage the open-source approach \cite{tong20216g}
to commercialize some of the components in this framework.

In summary, this section presented the edge AI system architecture from an E2E perspective detailing its network infrastructure, data governance, edge network function, as well as edge AI management and orchestration. The standardization efforts, hardware and software platform, and application scenarios will be further discussed in the next section. We hope this novel E2E architecture can stimulate more innovative and out-of-the-box ideas for the evolution of edge AI system architectures.

\section{Standardizations, Platforms, and Applications}
\label{spa}
In this section, we will first discuss the standardization for edge learning models
and algorithms, as well as integrated computing functionalities at the network edge. The research-oriented and production-oriented
platforms are then provided, including distributed optimization based FL
software, large-scale optimization based resource allocation solvers, as
well as edge AI computing and communicating hardware. To accelerate commercialization
for edge AI, the  application scenarios are also investigated,
including autonomous driving, industrial IoT, and smart healthcare.

\subsection{Standardizations}
The standardization of 6G will not be limited to the communications part, but also to the deep integration of communications, intelligence, and computing. The 3rd Generation Partnership Project (3GPP) may start an overall study into 6G systems around the end of 2025 (3GPP Release 20), while starting research into technical specifications around the end of 2027 \cite{tong20216g}.
In this subsection, we will introduce the standardizations on  trustworthy edge learning models
and algorithms, as well as wireless computing functionalities implemented in  digital or analog communication systems.  \\

\subsubsection{Learning} 

The first technical standard for FL was approved
on March 2021 as  IEEE 3652.1-2020 \cite{9382202}, {\textit{IEEE Guide for
Architectural Framework and Application of Federated Machine Learning}}.
 This IEEE standard for FL is developed by the Learning Technology Standards
Committee of the IEEE Computer Society with participants from the  shared machine
learning working group, including 4Paradigm, AI Singapore,
Alipay, Huawei, JD iCity, Tencent, WeBank, and Xiaomi, etc. Specifically,
the IEEE 3652.1-2020 standard provides the guidelines for architectures and
categories of FL from the perspectives of data, user and system, followed by identifying  the
associated application scenarios, performance  evaluations, and regulatory
requirements. Standardization plays a vital role in creating a private
and secure FL ecosystems at large-scale to provide consumer products and
services in the market. Besides, various standards
for data privacy and security
have been developed by the information security, cybersecurity and privacy
protection technical committee from the International Organization for Standardization
(IOS) and the International Electrotechnical Commission (IEC). For example,
ISO/IEC TS 27570 \cite{ISOTS27570}, {\textit{Privacy Protection - Privacy
Guidelines for Smart Cities}}, provides guidelines and recommendations for
the management of privacy  and the usage of standards. ISO/IEC DIS 27400
\cite{ISO27030},
{\textit{Guidelines for Security and Privacy in Internet of Things (IoT)}},
provides guidance for principles and controls to provide private and secure
IoT systems, services and solutions. The technical committee on cybersecurity
of the European Telecommunications Standards  Institute (ETSI) has recently unveiled
 ETSI EN 303 645 \cite{ETSIEN303645}, {\textit{Cyber Security for Consumer
Internet of Things:
Baseline Requirements}}, to provide cybersecurity standard and baseline for
IoT consumer products and certification schemes. All these standards are
applicable for developing private and secure edge AI models and algorithms to provide trustworthy
products and services. \\        
\subsubsection{Computing} The computing functionality can be implemented in wireless networks by either digital modulation or analog modulation. Specifically,  MEC provides a promising
solution for deploying  edge AI systems in current wireless systems with digital modulation \cite{CST20_edl}. The standardization
activities on MEC thus pave a way to integrate  edge AI into mobile
networks at a maturity level. Specifically, the  ETSI ISG MEC (Industry Specification
Group for Multi-access Edge Computing) has established a standardized and
open ecosystem for  both edge-aware and edge-unaware applications at
the network edge. It has published a set of white  papers and specifications
covering across user equipment application, service application, as well
as management, mobility, and orchestration related application programming
interfaces (APIs). Besides, 3GPP
5G specifications define the key enablers and architectures for edge computing
to allow traffic routing, policy control, and network management for 
collaboration in a MEC system and a 5G system \cite{3gpp1}. The collaboration between two independent systems of MEC and 5G can be further optimized in 6G, where communication and computing can be converged into one system by adding a computing plane \cite{tong20216g}. In  particular,
ETSI ISG MEC has recently   developed  a synergized mobile
edge cloud architecture by leveraging and harmonizing the existing and ongoing
standards (including 3GPP, ETSI ISG MEC, GSMA, and 5GAA) \cite{mec9}. Although
5G is rolled out globally, the modern mobile systems are widely deployed
based on digital modulation instead of  analog modulation \cite{haykin1994introduction}.
To support analog communication based AirComp for edge training in current wireless networks  \cite{JSAC_5GShafi},
one may either directly leverage the existing digital modulator with quantized
analog signals or introduce an additional analog modulator with a matched
filter for decoding the received signals \cite{park21commdlover}. It is obvious
that  more efforts are needed to incorporate AirComp functionalities into
the future 6G standards to mature edge AI systems.

\subsection{Platforms}
We present the software and hardware platforms for deploying edge AI models and algorithms, as well as the  optimization solvers for resource allocation in edge AI systems. \\

\subsubsection{Software} There is a rapidly growing body of software platforms
for simulations and  productization of edge AI algorithms and models. FL
library, TensorFlow Federated, Leaf, and PySyft have provided excellent open
software frameworks for FL simulations and evaluations. To further accelerate
research
progress and facilitate algorithmic innovation and performance comparison
in realistic FL environments, FedML \cite{He2020FedMLAR}, a research-oriented
open FL library, has recently been established to support diverse FL computing
environments and  topological architectures with standardized FL algorithm
implementations and benchmarks. As a  production-oriented software project,
FATE \cite{webank_fate} has been developed in the Webank's AI Department for
financial industry by supporting various secure computing protocols and 
FL architectures. Besides, existing edge computing frameworks (e.g., Baidu
``Baetyl" and Huawei ``KubeEdge") provide promising solutions to deliver
edge AI services. For  edge AI empowered IoT applications, Microsoft ``Azure
IoT Edge", Google ``Cloud IoT",  Amazon ``Web Services (AWS) IoT" and NVIDIA
``EGX" provide edge AI platform to bring real-time AI services across a wide
range of applications, including smart retial, home, manufacturing, and 
healthcare. Huawei has recently released   a next-generation
operating system, HarmonyOS \cite{Huawei_harmonyOS}, to enable seamless collaboration
and interconnection among smart edge devices across diverse platforms. This
empowers connected intelligence by deploying edge AI in the operating systems.\\

\subsubsection{Solver} Resource allocations for edge AI systems and wireless
networks are booming through the development of  various  large-scale optimization models
and algorithms. General-purpose large-scale optimization software solvers
are important to enable rapid prototyping and deploying  resource allocation
optimization algorithms for edge AI systems. Specifically, CVX \cite{cvx}
provides a two-stage software framework for modeling and solving general
large-scale convex optimization problems. This is achieved by automatically
transforming the original problem instances into
standard conic programming forms, followed by calling the advanced off-the-shelf
conic solvers, e.g.,  MOSEK \cite{mosek} and SCS \cite{o2016conic}. To further
speed up the modeling phase and avoid repeatedly parsing and re-generating conic
forms, a matrix stuffing technique was presented in \cite{Yuanming_TSP15}
to generate the mapping function between the original problem and the
conic form in a symbolic way instead of the time-consuming numerical way
using CVX. It is thus particularly interesting to develop a solver  to automatically
generate the mapping functions for conic transformation in a symbolic forms.
Besides, Gurobi \cite{gurobi} and MOSEK \cite{mosek}  are among  the fastest
solvers for solving the general mixed-integer second-order conic programs.
Chen {\textit{et al.}} recently released the software package ``Open-L2O"
\cite{chen2021learning} to implement the ``learning to optimize" framework
for benchmarking performance fairly  and designing algorithm automatically.\\

\subsubsection{Hardware} The achievable performance and benefits of edge
AI systems are conditioned upon the availability of edge AI computing hardware
and  radio frequency (RF) hardware technologies. Specifically,
edge AI computing hardware can
be categorized as graphic processing unit (GPU)-based hardware (e.g., NVIDIA's
GPUs), field programmable gate array (FPGA)-based hardware (e.g., Xilinx's
SDSoC), and application specific integrated circuit (ASIC)-based hardware
(e.g., Google's TPU). The detailed comparisons for various edge AI computing
hardware can be found in \cite{CST20_edl}. In particular, the chip design procedure for edge AI hardware can be significantly accelerated by the recent proposal of deep RL assisted fast chip floorplanning  \cite{mirhoseini2021graph}. Besides, the
massive broadband connectivity requirements for edge AI systems motivate
the innovations in RF hardware technologies. The benefits
of RIS-empowered FL systems highly depend on the capabilities of manipulating
electromagnetic waves at the metasurfaces \cite{Renzo_jsac20RIS}, whose reconfigurability
is typically enabled by switches, tunable material, topological metasurfaces,
and hybrid metasurfaces \cite{Qiu-NL21}. THz communication with frequency
band 0.1-10 THz, is envisioned as a promising enabler for achieving sensing,
communication, and learning in an integrated edge AI system. To approach
this THz region,  RF hardware technologies and solutions were thoroughly investigated
in \cite{amakawa2021white}, including semiconductor circuits, antenna forms,
packaging and testing of transceivers.                

\subsection{Applications}
We discuss edge AI enabled application scenarios by inspiring new communication algorithms, resource allocation optimization algorithms, as well as data processing methods.  \\

\subsubsection{Autonomous Driving} Autonomous driving basically refers to
self-driving vehicles  that move without
the intervention of human drivers. Self-driving vehicle  integrates
various  innovative technologies, including advanced sensor technologies, new
energy automobiles, next generation AI technologies, as well as future vehicular
networks. Autonomous driving can  significantly improve the safety,
passenger comfort, travel and logistics efficiency, collision avoidance,
and energy  efficiency. Edge AI shall provide a pivotal role
for achieving ultra-low latency communication, intelligent networking, real-time
data analytics, as well as high security for intelligent vehicles \cite{zhangjun19iov,
Kato_IEEEPro20}. A general DL framework was proposed in \cite{SheSGLYPV21}
to enable
ultra-reliable and low-latency vehicular communication, by incorporating
the domain knowledge including information theoretical tools and cross-layer
optimization design. To minimize the vehicles' queuing latency,
a FL approach was developed in \cite{SamarakoonBSD20} to 
learn the tail distribution of the queue lengths. To cope with the high mobility
and heterogeneous structures in vehicular networks, DL becomes
powerful for dynamic resource allocation \cite{liang20learningra} and network
traffic control \cite{Kato_IEEEPro20}. In particular, edge AI techniques,
including distributed RL \cite{Dongning_JSAC19, Mingzhe_JSAC21uav},
decentralized  GNN \cite{yifei21gnn}, as well as distributed
DNN with binarized output layer  \cite{Tony_JSAC19distribued}, are able to
learn and execute the distributed resource allocation polices in an automatic
and real-time manner.
 
The data processing tasks for autonomous driving mainly include perception,
high-definition (HD) mapping, as well as SLAM \cite{zhangjun19iov, liu2019edge}. Specifically, to understand the
 environments for intelligent decision making, various sensory data  from onboard
sensors (e.g., light detection and ranging (LiDAR), cameras, radar and sonar)
need to be processed for the perception tasks, including localization, object
detection and tracking. The perception capability can be enhanced by edge AI systems, e.g., edge device-server co-inference of DNN models for
vision based perception tasks \cite{Davide_RAM11}. HD mapping aims at constructing
a representation of  the vehicles operating environments,
e.g., obstacles, landmarks position, curvature and slope. This is imperative
to achieve high accurate localization for autonomous driving. The edge server cooperative inference method in Section {\ref{esci}} can be adopted to reduce the
storage and communication overheads for updating the HD map by collecting
fresh data from the vehicles in the dynamic environments \cite{zhangjun19iov}.
SLAM comprises simultaneously estimating the state of a vehicle and constructing
a map of the environment \cite{Cadena_TOR16}, which paves the way for achieving
full autonomy in autonomous driving \cite{bresson2017simultaneous}. Edge SLAM
\cite{ben2020edge, xu2020edge} has recently been developed to execute 
DL based visual SLAM algorithms on edge vehicles. This is achieved
by deploying the tracking computation parts on the edge vehicles while offloading
the remaining parts (e.g., local mapping and loop closure) to the roadside edge server
via  vertical edge inference in Section \ref{edsc}. \\

\subsubsection{Internet of Things} Artificial Intelligence of Things (AIoT)
leverages AI technologies and IoT infrastructures to improve the human-machine
interactions and enable multi-agent communications and collaborations. AIoT
goes beyond the conventional communication paradigm for audio, video
and data delivery. It will enable semantic communication \cite{Zhijin_JSAC21scl}
to exchange semantic information among agents. Shannon and Weaver categorize
communication into three levels, including transmission level (i.e., transmit
symbols accurately), semantic level (i.e., convey the desired meanings precisely),
and effectiveness level (i.e., produce the desired actions effectively) \cite{shannon1949mathematical}.
Sematic communication is able to significantly improve the communication
efficiency by only transmitting the extracted relevant information for sematic
information delivery tasks with the semation error as the performance metric.
A distributed edge DL approach has been recently  developed in
\cite{Zhijin_JSAC21scl} to enable low-latency semantic communication over
IoT networks. This is achieved by jointly optimizing the  compressed DNN
based transmitters at the edge IoT devices and the  quantized DNN based receivers
at the edge server over the wireless fading channels.      

Industrial IoT (IIoT) is a production-oriented industrial network for connecting
industrial devices and equipments, processing and exchanging generated data,
as well as optimizing the production system  \cite{Sisinni_TII18IIoT}. Besides,
digital twin is becoming a key technology for  smart manufacturing in industrial
4.0 by connecting physical machines and digital representations in a
cyber-physical system \cite{Nee_TII19dd, Greyce-IEEEPro21}. This is achieved
by providing a virtual representation of the industrial entities and products'
life-cycle to predict and optimize the behaviors of the manufacturing process.
Edge AI provides a promising way to model and deploy digital twins for IIoT
networks to process the high volume of industrial streaming data with low-latency
and high-security guarantees. Specifically, edge computing provides a
general platform for inferring DNN models via computation offloading to reduce
network latency and operation cost in IIoT \cite{Tie_CST20}. {\rev{FL becomes a key enabling technology to support intelligent IIoT applications
(e.g., smart grid and smart manufacturing) and provide IIoT services (e.g., data offloading and mobile crowdsensing) \cite{Dinh_WC21}.}} In particular, blockchain empowered FL  was proposed in \cite{Yunlong_TIT20} to provide
secure communication and private data sharing schemes for constructing digital
twin IIoT networks, followed by reducing
communication overheads via asynchronous model  aggregation. \\

\subsubsection{Smart Healthcare} 
Smart healthcare aims to realize a common platform for efficient and personalized healthcare, intelligent health monitoring, and  precision medicine development via collaboration among multiple participants
(e.g.,
doctors, patients, hospitals, and research institutions). This is achieved
by emerging advanced technologies, including DL \cite{obermeyer2016predicting,zhou2021review},
Tactile Internet, IoT, edge AI, and wireless communications. In particular,
edge AI with distributed and secure DL has been demonstrated to be able
to significantly improve the reliability, accuracy, scalability, privacy
and security for precision medicine and Internet of Medical Things \cite{Sun_IOMT19}, including medical imaging, drug development, and chronic disease
management \cite{subramanian2020precision}.
Specifically, Kaissis {\textit{et al.}} in  \cite{kaissis2020secure} presented
a FL approach
 for medical imaging to preserve privacy and avoid potential attacks against
the
datasets or learning algorithms. 
Besides, swarm learning has recently been developed in \cite{warnat2021swarm}
to provide a  decentralized and confidential
clinical disease detection solution for diseases (e.g., COVID-19, tuberculosis,
and leukaemia). This is achieved by leveraging the blockchain and edge computing
techniques to develop a secure and private decentralized learning architecture
while keeping the medical data locally. MIT Media Lab established a split
learning project to allow health entities
collaboration for training patient diagnostic models without sharing sensitive
raw data \cite{vepakomma2018split}. An RL approach for
decisions
making in patient treatment was introduced in \cite{gottesman2019guidelines}
to realize safe and risk-conscious
healthcare practice.

Haptic communication \cite{Anton_CST18} aims at delivering
the skill set (e.g., the manipulation skills representation learned from
the multisensory tactile and visual  data
\cite{fazeli2019see}, and the signatures of the human grasp learned using a tactile glove \cite{sundaram2019learning}) over the Tactile Internet in an ultra-reliable and low-latency
manner. It has potentials  in healthcare applications including tele-diagnosis,
 tele-rehabilitation, and tele-surgery, which turns out to be essential during the ongoing COVID-19 pandemic. Edge AI  becomes a key enabling technique
for the Tactile Internet with human-in-the loop to facilitate ultra-responsive
 and truly immersive tactile actuation in the tele-operation systems \cite{Simsek-JSAC16}.
This is achieved by enabling the network edge with  intelligent prediction
capability for haptic information (e.g., tactile feedback and control
traffics) \cite{IoT21_traffice}, as well as the  intelligent resource allocations
across the whole network layers \cite{Nattakorn_CST21}. Specifically, a distributed
optimization framework was developed in \cite{JSAC_18fogtactile} to design
an edge computing assisted Tactile Internet for achieving both the ultra-low
latency and high energy efficiency. Such a distributed optimization algorithm
can be further learned via the distributed DL techniques \cite{Tony_JSAC19distribued}.
Besides, a variational optimization framework was proposed in \cite{TII19_DLNOAM}
to enjoy low-latency and high-reliability for massive access in the Tactile Internet.
The variational decision function can be further parameterized via DNNs with the capability of distributed training and inference for practical
deployments \cite{TII19_DLNOAM, yifei21gnn}.

In summary, this section presented standardizations, platforms, and applications for practical deployment of edge AI systems. Combining the presentations of edge training in Section \ref{edgetraining}, edge inference in Section \ref{edgeinference}, resource allocation in Section {\ref{resourceall}}, and system architecture in Section \ref{archedgeai}, we complete the roadmap for edge AI ecosystem, as shown in Fig. \ref{roadmap}. We hope these results can encourage more communities and stakeholders to engage in industrializing and commercializing edge AI in the era of 6G.

\section{Conclusion}
Embedding low-power, low-latency, reliable, and trustworthy intelligence into the network edge is an inevitable trend and disruptive shift in both academia and industry. Edge AI serves as a distributed neural network to imbue connected intelligence in 6G, thereby enabling intelligent and seamless interactions among the human world, physical world, and digital world. The challenges for building edge AI ecosystems are multidisciplinary spanning wireless communications, machine learning, operation research, domain applications, regulations and ethics.  In this paper, we have investigated the key wireless communication techniques, effective resource management approaches and holistic network architectures to design scalable and trustworthy edge AI systems. The standardizations, platforms, and applications were  also discussed for productization and commercialization of edge AI.  We hope that this article will serve as a valuable reference and guideline for further considering  edge AI opportunities across theoretical, algorithmic, systematic, and entrepreneurial considerations to embrace the exciting era of edge AI.

\bibliographystyle{IEEEtran}
\bibliography{reference}
\begin{IEEEbiography}[{\includegraphics[width=1in,height=1.25in,clip,keepaspectratio]{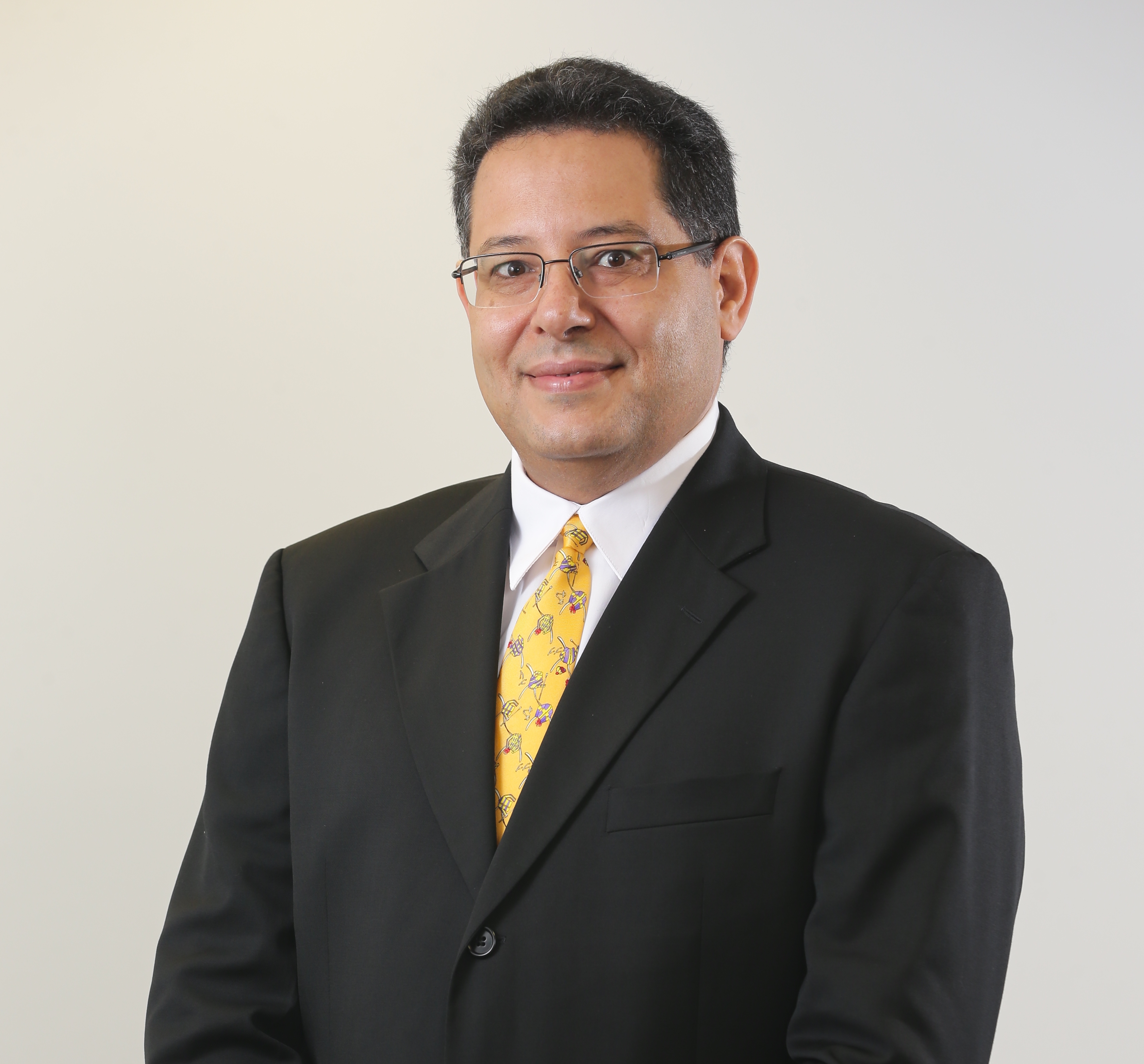}}]{Khaled. B. Letaief}
	 (S'85-M'86-SM'97-F'03) is an internationally recognized leader in wireless communications and networks with research interest in artificial intelligence, big data analytics systems, mobile cloud and edge computing, tactile Internet, 5G systems and beyond.  In these areas, he has over 720 papers along with 15 patents, including 11 US inventions.
	 
	He is a Member of the United States National Academy of Engineering, Fellow of IEEE, Fellow of Hong Kong Institution of Engineers, and Member of the Hong Kong Academy of Engineering Sciences. He is also recognized by Thomson Reuters as an ISI Highly Cited Researcher and was listed among the 2020 top 30 of AI 2000 Internet of Things Most Influential Scholars.
	
	Dr. Letaief is the recipient of many distinguished awards and honors including the 2021 IEEE Communications Society Best Survey Paper Award; 2019 Distinguished Research Excellence Award by HKUST School of Engineering (Highest research award and only one recipient/3 years is honored for his/her contributions); 2019 IEEE Communications Society and Information Theory Society Joint Paper Award; 2018 IEEE Signal Processing Society Young Author Best Paper Award; 2017 IEEE Cognitive Networks Technical Committee Publication Award; 2016 IEEE Signal Processing Society Young Author Best Paper Award; 2016 IEEE Marconi Prize Paper Award in Wireless Communications; 2011 IEEE Wireless Communications Technical Committee Recognition Award; 2011 IEEE Communications Society Harold Sobol Award; 2010 Purdue University Outstanding Electrical and Computer Engineer Award; 2009 IEEE Marconi Prize Award in Wireless Communications; 2007 IEEE Communications Society Joseph LoCicero Publications Exemplary Award; and 19 IEEE Best Paper Awards.
	
	He served as consultants for different organizations including Huawei, ASTRI, ZTE, Nortel, PricewaterhouseCoopers, and Motorola.  He is the founding Editor-in-Chief of the prestigious IEEE Transactions on Wireless Communications and been involved in organizing many flagship international conferences.
	
	From 1990 to 1993, he was a faculty member at the University of Melbourne, Australia.  Since 1993, he has been with the Hong Kong University of Science \& Technology (HKUST) where he is currently the New Bright Professor of Engineering.  While at HKUST, he has held many administrative positions, including the Dean of Engineering, Head of the Electronic and Computer Engineering department, Director of the Wireless IC Design Center, founding Director of Huawei Innovation Laboratory, and Director of the Hong Kong Telecom Institute of Information Technology.  
	
	Dr. Letaief is well recognized for his dedicated service to professional societies and IEEE where he has served in many leadership positions.  These include IEEE Communications Society Vice-President for Conferences, elected member of IEEE Product Services and Publications Board, and IEEE Communications Society Vice-President for Technical Activities. He also served as President of the IEEE Communications Society (2018-19), the world's leading organization for communications professionals with headquarter in New York City and members in 162 countries.   In 2022-23, Dr. Letaief will serve as member of the IEEE Board of Directors.
	
	Dr. Letaief received the BS degree with distinction in Electrical Engineering from Purdue University at West Lafayette, Indiana, USA, in December 1984. He received the MS and Ph.D. Degrees in Electrical Engineering from Purdue University, in Aug. 1986, and May 1990, respectively.
\end{IEEEbiography}

\begin{IEEEbiography}[{\includegraphics[width=1in,height=1.25in,clip,keepaspectratio]{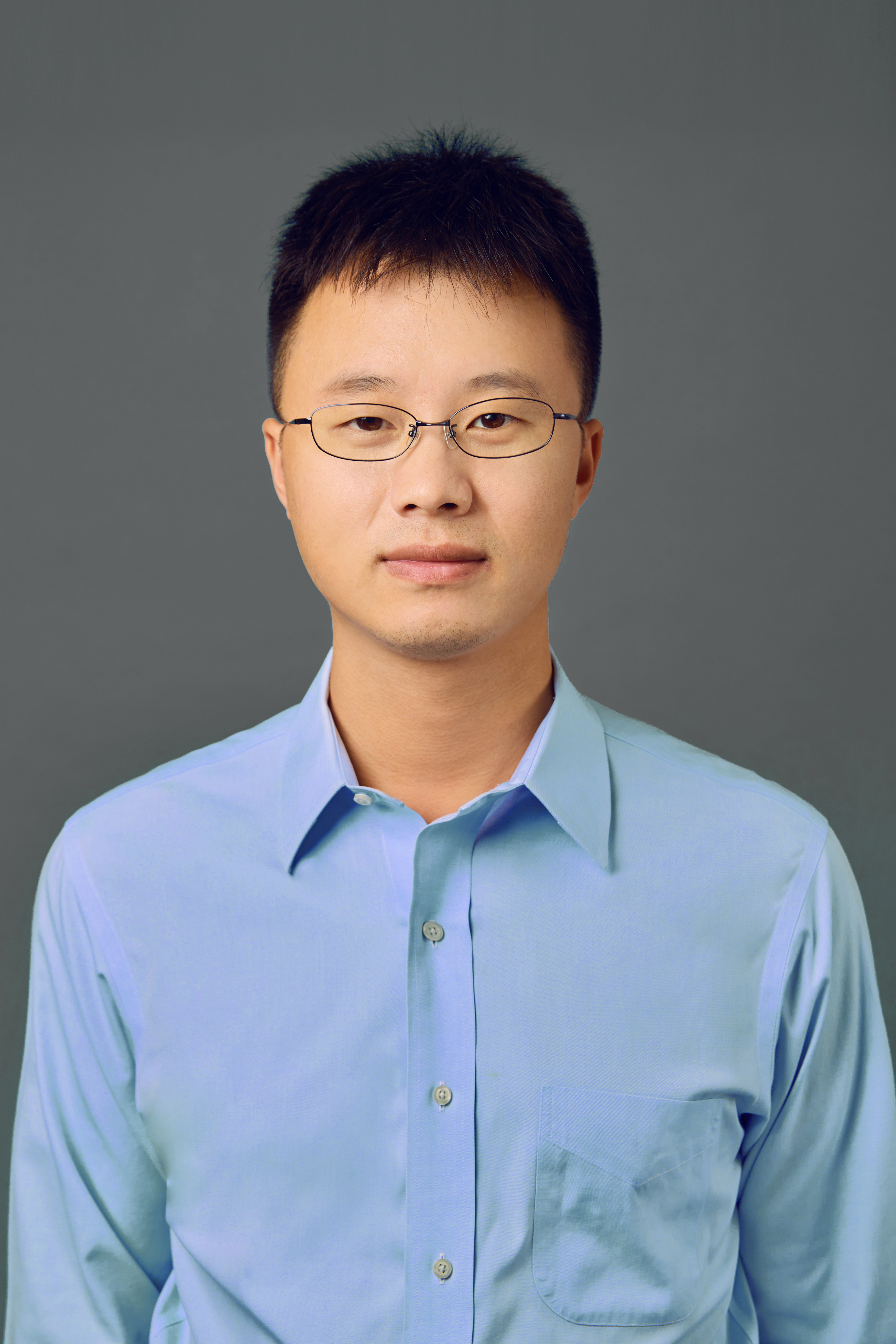}}]{Yuanming Shi}
	(S'13-M'15-SM'20) received the B.S. degree in electronic engineering from Tsinghua University, Beijing, China, in 2011. He received the Ph.D. degree in electronic and computer engineering from The Hong Kong University of Science and Technology (HKUST), in 2015. Since September 2015, he has been with the School of Information Science and Technology in ShanghaiTech University, where he is currently a tenured Associate Professor. He visited University of California, Berkeley, CA, USA, from October 2016 to February 2017. His research areas include optimization, statistics, machine learning, wireless communications, and their applications to 6G, IoT, and edge AI. He was a recipient of the 2016 IEEE Marconi Prize Paper Award in Wireless Communications, the 2016 Young Author Best Paper Award by the IEEE Signal Processing Society, and the 2021 IEEE ComSoc Asia-Pacific Outstanding Young Researcher Award. He is also an editor of IEEE Transactions on Wireless Communications and IEEE Journal on Selected Areas in Communications.
\end{IEEEbiography}
\begin{IEEEbiography}[{\includegraphics[width=1in,height=1.25in,clip,keepaspectratio]{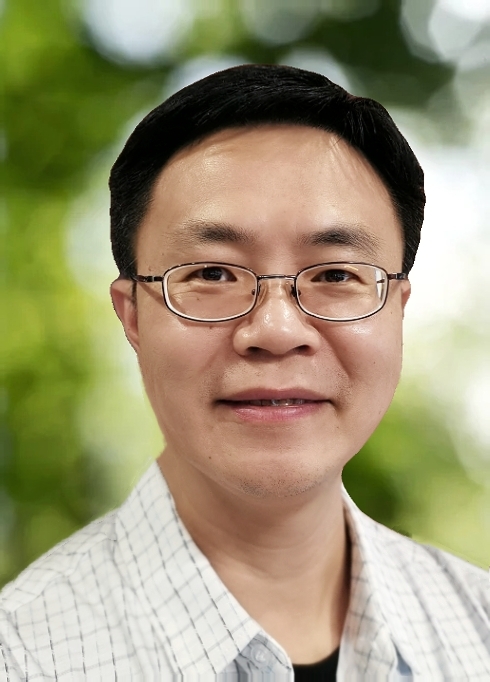}}]{Jianmin Lu}
	 joined the Huawei Technologies in 1999. During the last two decades, he conducted various researches on wireless communications especially on physic layer and MAC layer and developed 3G, 4G and 5G products. He received more than 50 patents during the research. He was deeply involved in 3GPP2 (EVDO/UMB), WiMAX/802.16m and 3GPP(LTE/NR) standardization and contributed several key technologies such as flexible radio frame structure, radio resource management and MIMO. His current research interest is in the area of signal processing, protocol and networking for the next generation wireless communication. He is currently Executive Director of Huawei Wireless Technology Lab.
\end{IEEEbiography}

\begin{IEEEbiography}[{\includegraphics[width=1in,height=1.25in,clip,keepaspectratio]{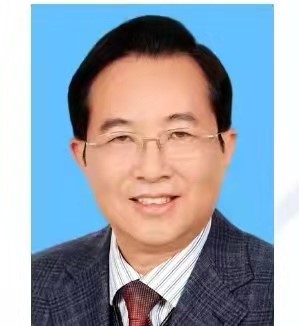}}]{Jianhua Lu}
	(M'98-SM'07-F'15) received his B.S. and M.S. degrees from Tsinghua University, Beijing, China, in 1986 and 1989, respectively, and his Ph.D. degree in Electrical \& Electronic Engineering from the Hong Kong University of Science \& Technology, Hong Kong, China, in 1998. Since 1989, he has been with the Department of Electronic Engineering, Tsinghua University, where he currently serves as a professor. He is now a vice president of the National Natural Science Foundation of China. 
	
	His research interests include broadband wireless communications, multimedia signal processing, and satellite communications. He has authored/co-authored over 300 referred technical papers published in international renowned journals and conferences and over 80 Chinese invention patents. He was also a recipient of the Best Paper Awards at the IEEE ICCCS 2002, China Comm. 2006, IEEE Embedded-Com 2012, IEEE WCSP 2015, IEEE IWCMC 2017 and IEEE ICNC 2019.
	
	Prof. Lu served as an Editor for IEEE Transactions on Wireless Communications from 2008 to 2011, and Program Committee Co-Chair, as well as, TPC member of many international conferences. He is now the Editor-in-Chief of China Communications.
	
	He is a member of Chinese Academy of Sciences, and a Fellow of IEEE.
\end{IEEEbiography}

\end{document}